*Published in Shape Memory and Superelasticity, 2026*

# Plastic Deformation of B19‘ Martensite by Kwinking: Where it Matters in NiTi Technology

*P. Šittner[1], H. Seiner[2], P. Sedlák[2], O. Molnárová[1], L. Kadeřávek[1], O. Tyc[1], E. Iaparova[1], L. Heller[1]*

[1]Institute of Physics of the CAS, Na Slovance 1992/2, 18200 Prague, Czech Republic
[2]Institute of Thermomechanics of the CAS, Dolejškova 5, Prague, Czech Republic

## Abstract

Nitinol technology, besides utilizing the functional thermomechanical properties derived from the B2 cubic to B19' monoclinic martensitic transformation, also exploits the excellent plastic deformability of NiTi in the martensite state. It originates from the unique mechanism of plastic deformation of the B19' martensite by kwinking involving dislocation slip based kinking assisted by deformation twinning. Although the mechanism of plastic deformation of martensite by kwinking was revealed only very recently, various unusual phenomena that can only be rationalized by kwinking, have been reported in literature in the last 50 years. These phenomena include: 1) cold working with a high degree of reduction without introducing cracks, 2) excellent plastic deformability in the martensite state (plastic deformation up to~80% strain at stresses >1GPa), 3) refinement of austenitic microstructure to a quasi-amorphous state by tensile deformation, 4) observation of high density of {114} deformation bands in austenitic microstructures, 5) systematic ruptures of strengthened NiTi wires in tensile tests via necking at the onset of plastic yielding, 6) localized plastic deformation in tensile tests via propagation of Lüders band fronts with very large localized strain (~40%), 7) unusually long upper stress plateaus in superelastic tensile tests (>8% strain), 8) large plastic strains (> 20 %) generated in a single closed-loop cooling/heating cycle under constant stress, 9) shape setting of already annealed NiTi by heating under external constraint. Finally, we discuss how kwinking deformation was considered in constitutive modelling of thermomechanical behaviors of NiTi and, particularly, what is the role of the kwinking deformation in NiTi technology.

## 1. Introduction

NiTi is by far the most important and widespread shape memory alloy (SMA). Research and development in the last few decades enabled superelastic NiTi technology currently used in medical devices [1] and NiTi SMA based actuators currently used in automotive, aerospace, and robotics [2]. The key material property, which enabled this success, is the B2-B19' martensitic transformation (MT) providing large recoverable strain in thermomechanical load cycles [3]. When this MT proceeds under external stress it is accompanied by incremental plastic deformation that helps to maintain strain compatibility among polycrystal grains and prevents preliminary intergranular fractures [4,5]. Because of this, a properly cold worked/annealed nanocrystalline NiTi wire displays recoverable strain as large as ~12% in closed-loop thermomechanical loading involving stresses as high as 1 GPa without generating excessive plastic strains and without a danger of unexpected fracture at high tensile stresses. This is unique, other polycrystalline SMAs are not capable of that. They typically display either small recoverable strains (~2-4%), as in Cu-based alloys, or recoverable strains that are accompanied by large plastic strains (Fe-based, nickel free Ti-based alloys) or intergranular cracking and fracture in just few thermomechanical load cycles (NiMnGa or Co-based alloys). Since development of engineering applications of SMAs has been constrained by these shortcomings, NiTi (and some NiTi based multicomponent alloys) turned out to be the most successful shape memory and superelastic alloy.

Traditionally, martensite has been considered in the SMA field to be the hard phase, which is highly resistant to plastic deformation, while austenite has been viewed as the soft phase that deforms plastically by dislocation slip [6,7]. In NiTi, however, the situation is reversed. At low temperatures (< 200 °C), the B19' martensite tends to deform via dislocation slip but the B2 austenite phase is highly resistant to dislocation slip when deformed in the absence of MT [8]. At high temperatures (> 300 °C), the austenite deforms plastically via dislocation slip, as clearly evidenced by strain softening and the rate dependence of the stress-strain responses in tensile tests [8]. In the intermediate temperature range of 200-300 °C, stress induced MT proceeding at high stresses is accompanied by dislocation slip in both the austenite and martensite phases by a kind of TRIP-like deformation mechanism [9,10]. However, due to the lack of reliable experimental data, this deformation mechanism is still poorly understood.

NiTi alloys tend to deform plastically via dislocation slip in martensite while the forward and/or reverse MT proceed under external stress above certain thresholds but below the yield stress [11,12,13,14]. Dislocation slip in martensite during the MT proceeding under stress helps to relieve the internal stress concentrations arising due to strain incompatibilities at grain boundaries caused by the mismatch of transformation strains in neighboring grains. However, the same dislocation slip leads to incremental plastic strains generated during the forward and reverse MT in cyclic thermomechanical loads [11,13] resulting in functional fatigue and thereby limiting many promising engineering applications of NiTi. Fortunately, generation of incremental plastic strains by MT proceeding under stress can be suppressed by strengthening the NiTi alloy against dislocation slip in martensite [13,14], either through careful alloy design and/or via postprocessing thermomechanical treatments.

Besides the functional thermomechanical properties derived from the B2-B19' MT involving small tensile strains (<12%), NiTi displays another unique property that is also heavily used in superelastic and actuator technologies. It is its excellent plastic deformability in the low temperature martensite state that overcomes the strength-ductility trade-off paradigm of material engineering. A properly cold worked/heat treated NiTi wire deforms in tensile tests at low temperatures (<100 °C) under high stresses (~ 1 GPa) up to very large plastic strain (~80%) [15]. The excellent plastic deformability in the martensite state enables high degrees of cold work reduction (~90%) during cold rolling [19], cold drawing [20], or any severe plastic deformation [17,18] without introducing cracks. The only requirement is that the alloy is martensitic while it deforms plastically [16]. Plastic deformation of martensite refines the austenitic microstructure and turns the alloy into a near amorphous state. However, this amorphization is not due to severe plastic deformation, but rather due to the unique mechanism of plastic deformation which refines the alloy microstructure. However, near-amorphous microstructures were observed by transmission electron microscopy (TEM) even in NiTi wires deformed in tensile test at low temperatures (<100 °C) up to ~60% elongation [15]. These near-amorphous microstructures transform upon subsequent heating into a nanocrystalline microstructures providing NiTi with excellent functional thermomechanical properties [21].

Although plastic deformation of superelastic and shape memory NiTi alloys has been investigated since the onset of NiTi technology, its mechanism remains poorly understood. Researchers commonly refer to "plastic deformation of NiTi by dislocation slip and deformation twinning" without specifying the phase in which dislocation slip occurs and which slip systems are activated. For information on the state-of-the-art view on deformation mechanisms activated during tensile deformation of martensitic NiTi, see Fig. 7 and related text in recent review article by Chowdhury and Sehitoglu [22].

Deformation mechanism responsible for plastic deformation of B19' martensite which enables the excellent plastic deformability of NiTi was revealed only very recently [23,24]. It was termed "**kwinking**", as it combines dislocation slip based **kinking** and deformation t**win**ning. It was revealed based on the results of: i) theoretical considerations [23], ii) TEM analysis of martensite variant microstructures and intermartensitic interfaces [24, 25, 27] in grains of plastically deformed NiTi shape memory wires, and iii) studies of the evolution of martensite texture during tensile deformation of NiTi wires up to fracture [8,26]. The kwinking deformation is triggered when the yield stress for plastic deformation of oriented martensite is reached in general thermomechanical loading tests on NiTi. The activity of kwinking is experimentally evidenced by the observation of unique martensite variant microstructures consisting of deformation bands arranged in a specific deformation geometry characterized by lattice rotations around common 010]M direction as well as by the presence of intermartensitic interfaces on (20-1), (10-1), (100) planes created by kwinking deformation in these microstructures. The kwinking concept [23] explains the origin of these deformation-induced microstructures and intermartensitic interfaces by considering that the observed deformation bands were created by coordinated [100](001) dislocation slip based kinking combined with (100) deformation twinning.

Following reverse MT of plastically deformed NiTi wires on unloading and heating, these martensitic microstructures are converted into austenitic microstructures consisting of deformation bands separated from the austenite matrix by symmetrical tilt grain boundaries (STGBs) arranged in another specific deformation geometry characterized by lattice rotations around common <011>A direction (lattice correspondent to [010]M direction) [14,15,30]. Majority of the observed deformation bands were evaluated to be [114} austenite twin bands. These peculiar deformation bands and interfaces in austenite have been observed in deformed NiTi alloys since the onset of NiTi research [29-39]. However, as researchers were not familiar with the kwinking deformation mechanism, they naturally considered them to be deformation twins in austenite [29,31,35,36,39,41,42].

We performed systematic analyses of austenitic and martensitic microstructures and permanent lattice defects in deformed nanocrystalline NiTi wires as part of experimental research carried out in the last 15 years [9, 10,11,12,14,15,21,26,43-50]. As other researchers, we have also originally considered the (20-1) twin bands in plastically deformed martensite to be created by unique deformation twinning in martensite [39] which reorients the martensite lattice out of the Eriksen-Pitteri neighborhood [51] in such a way that the imposed strains are not recoverable upon unloading and heating (called non-transformation pathway by Gao et al. [52]). However, the problem with this view is that the mechanisms explaining how the (20-1) twinning in B19' martensite proceeds are either unclear [52] or unrealistic [42]. The kwinking mechanism [23] explains that that the (20-1) deformation bands experimentally observed in plastically deformed NiTi are not deformation twins but (20-1) kwink bands created by coordinated dislocation slip-based kinking combined with (100) deformation twinning.

The story of the research that led to the discovery of kwinking is, however, long and confusing since it spreads over many articles published before the kwinking was discovered. The purpose of this work is to overview the phenomena reported in author's earlier works on nanocrystalline superelastic and shape memory NiTi wires that originate from the kwinking deformation and to discuss them based on the new understanding of the mechanics of the kwinking deformation. After a brief description of the kwinking mechanism (Chapter 3) and introduction of the [100](001) dislocation slip and (001) compound twinning in the B19' martensite that enables it (Chapter 4), we discuss individual aspects of plastic deformation of B19' martensite by kwinking, particularly the temperature dependent kwinking stress (Chapter 5). We discus results of TEM analysis of martensite variant microstructures and permanent lattice defects in whole grains created by the kwinking deformation (Chapter 6), discuss the macroscopic localization of kwinking deformation in necks and Lüders bands (Chapter 7), explain the experimentally observed evolution of martensite texture and refinement of microstructure in tensile tests as being due to activation of kwinking deformation (Chapter 8), discuss several unique phenomena observed in cyclic thermomechanical loading tests that originate from the activation of kwinking (Chapter 9), discuss potential consideration of kwinking deformation in constitutive modelling of thermomechanical behaviors of NiTi (Chapter 10), and finally, we discuss the role of kwinking in further development of NiTi technology (Chapter 11). Each chapter contains last paragraph summarizing key information on the kwinking deformation provided in it.

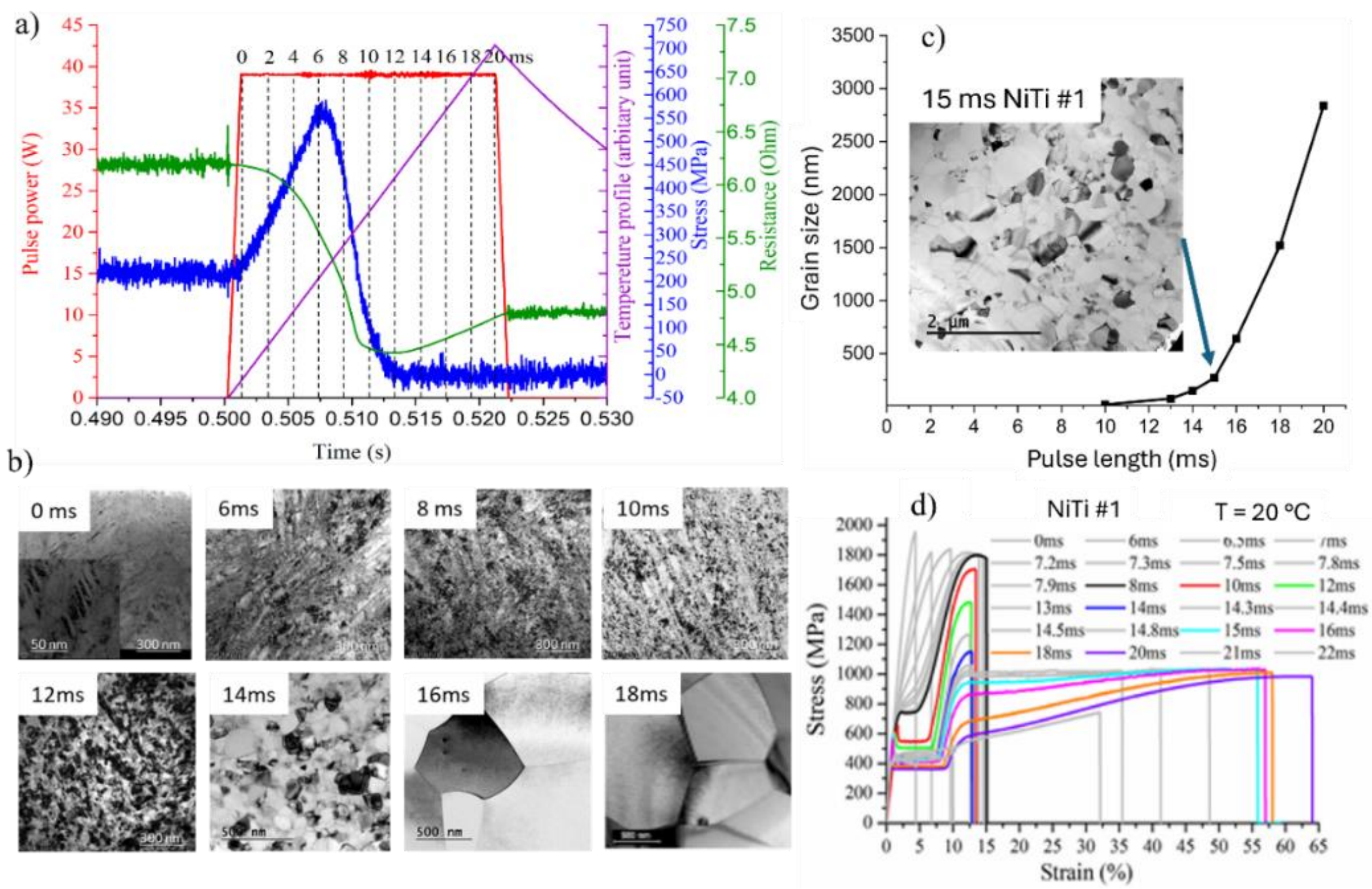

**Figure 1: Electro-pulse heat treatment of cold-worked NiTi wires [21].** The 30 mm long segment of cold drawn NiTi wire was loaded up to 300 MPa stress, its length was constrained and heated by a short pulse of controlled electric power. a) variation of tensile stress, electric resistance, and temperature of the wire during the electric power pulse (P =37.8 W, pulse time t = 20 ms). The tensile stress and electric resistance were measured, temperature was calculated using the heat equation. b) microstructures in NiTi wires heat treated with pulse lengths 0-18 ms observed by TEM. c) correlation between the grain size of the heat treated NiTi wires and pulse length (maximum temperature reached in the pulse treatment). d) tensile stress-strain curves of NiTi #1 wires heat treated with pulse lengths 0-22 ms.

## 2. Experiments

As introduced above, selected results of the ~15 years of research on nanocrystalline NiTi wires related to kwinking deformation are overviewed and discussed in the work. In these earlier experiments, we subjected purposely heat-treated nanocrystalline NiTi wires to various thermomechanical loadings to investigate the involved deformation mechanisms. The experiments were frequently performed with in-situ evaluation of electrical resistance, texture and phase fractions via in-situ synchrotron x-ray diffraction, with analyses of local surface strains or spatially resolved thermography. In addition, we have analyzed martensite variant microstructures and permanent lattice defects created by the activated deformation processes by TEM.

Two commercial NiTi wires with different chemical compositions produced by Fort Wayne Metals - shape memory NiTi #5 wire (Ti-50.5 at. % Ni, 42 % cold work, diameter 0.1 mm) and superelastic NiTi #1 wire (Ti-50.8 at. % Ni, 42 % cold work, diameter 0.1 mm) were mainly used in these experiments. 30-50 mm long samples cut from single batch of cold worked wire supplied by the producer on textile compatible spools were prestressed to ~300 MPa, constrained in length and heated by short pulse of controlled electric power (pulse length 5-20 ms) [4,21]. Tensile stress, electric resistance and temperature were evaluated during the short electric power pulse (Fig. 1a). Depending on the applied pulse length (corresponds to maximum reached temperature), the heat-treated wires possessed either recovered or recrystallized austenitic microstructures with desired grain size (Fig. 1b,c). Fig. 1c links the power pulse length to the grain size. Since cold worked NiTi wires received the same degree of cold work reduction and the maximum temperature reached during the electric power pulse heat treatment was precisely controlled (Fig. 1a), the electro-pulse heat treatment method gives reproducible results in setting the desired austenitic microstructure (Fig. 1b) and mechanical properties (Fig. 1d) of heat-treated NiTi wires. Due to the short time thermal exposure during the electro-pulse heat treatment, diffusional processes in the heat-treated wires, such as precipitation or dissolution of precipitates, are suppressed.

Tensile thermomechanical loading tests on 0.1 mm thin NiTi wires were carried out using in-house-built tensile testers consisting of a miniature load frame, an environmental chamber (-30 °C to 180 °C), electrically conductive grips, load cell, linear actuator, and position sensor. The thermomechanical loading tests were performed in combined position, force and temperature control modes using low strain rates (~$10^{-4}$ $s^{-1}$) and low heating/cooling rates (~3-5 °C/min). Selected thermomechanical loading tests involving cooling down to -120 °C and heating up to temperatures above 500 °C were performed using a DMA 850 tester by TA Instruments using 10 mm long wire samples.

Thin lamellae for TEM analysis were cut from the subsurface layers of deformed wire (10 µm below the surface, typically with wire axis in the lamella plane) by focused ion beam (FIB) using a FEI Quanta 3D FIB-SEM microscope. Martensite variant microstructures in grains of deformed NiTi wires were analyzed by TEM using a FEI Tecnai TF20 X-twin microscope equipped with a field emission gun operating at 200 keV using a double tilt specimen holder. The recorded electron diffraction patterns were indexed using the lattice parameters of the B2 and B19' structure ($a_0$ = 0.3015 nm, a=0.2889 nm, b=0.4120 nm, c=0.4622 nm, β=96.8°) [53]. Martensite variant microstructures in grains of deformed NiTi wires were reconstructed by selected area electron diffraction with dark field (SAED-DF) method [25] and by the automated nanoscale orientation and phase mapping in TEM (ASTAR) [27].

The austenite and martensite phase fractions and textures evolving during tensile tests were evaluated by an in-situ synchrotron x-ray diffraction method [26]. The experiments were performed on the ID15A beamline at ESRF Grenoble using 64.2 keV beam energy (wavelength 0.19312 Å) and a beam size of 150 × 150 µm². A miniature thin-wire tester was installed on the beamline, and the gauge volume was placed in the wire center with the wire axis oriented perpendicularly to the x-ray beam. The wire samples were subjected to closed-loop thermomechanical loading tests, and 2D diffraction patterns were continuously recorded during the tests, taking advantage of the high intensity of the diffracted beams and the low exposure time (0.2489 s) of the used Pilatus3 X CdTe 2M detector. The Axial Direction Inverse Pole Figures (AD IPF) characterize the orientation distribution of the wire axis direction in B2 cubic and B19' monoclinic lattices. Radial symmetry of the recorded 2D x-ray diffraction data was assumed to determine the fiber textures of deformed NiTi wires [26].

## 3. Kwinking Deformation as the Mechanism of Plastic Deformation of B19' Martensite

The term "kwinks" was first introduced in experimental work [25], in which the authors tried to explain the observation of deformation bands in grains of plastically deformed B19' NiTi martensite (Fig. 2e), in which crystal lattices were either twinned or rotated with respect to the martensite matrix. The $(20\text{-}1)_M$-oriented twin-like deformation band was presented as a kwink band in Ref. [23]. The word kwinking arises from combining the words 'twinning' and 'kinking,' where the latter refers to deformation mechanism known to be responsible for plastic deformation of materials with strongly anisotropic plastic properties proceeding via coordinated plastic slip on a single slip system [54]. In B19' martensite, the plastic slip is indeed strongly anisotropic, with a single easy slip system being $[100](001)_M$ [55], and thus, the B19' lattice is prone to plastic deformation by kinking. At the same time, due to the low symmetry of the monoclinic structure and its ability to reorient through reversible twinning, the kink bands in B19' structure created by plastic deformation can possess twin-like relationships to the matrix (while kink interfaces are STGBs that display mirror symmetry of (001) planes across the interface, kwink interfaces display near mirror symmetry of crystal lattices across the interface). The kwink bands can therefore be interpreted either as deformation twins formed with assistance of unidirectional plastic slip, or, conversely, as plastic kinks initiated by deformation twinning.

The concept of kwinking (i.e., plastic straining via formation of kwink bands) has been elaborated in detail in Ref. [23] on both atomistic and continuum-mechanics scales, where it was shown that the experimentally observed formation of (20-1) kwink bands can be explained in terms of energy minimization. The full theoretical framework presented in [23] is, however, not necessary for understanding the kwinking deformation mechanism at a phenomenological level, as required for the scope of this paper.

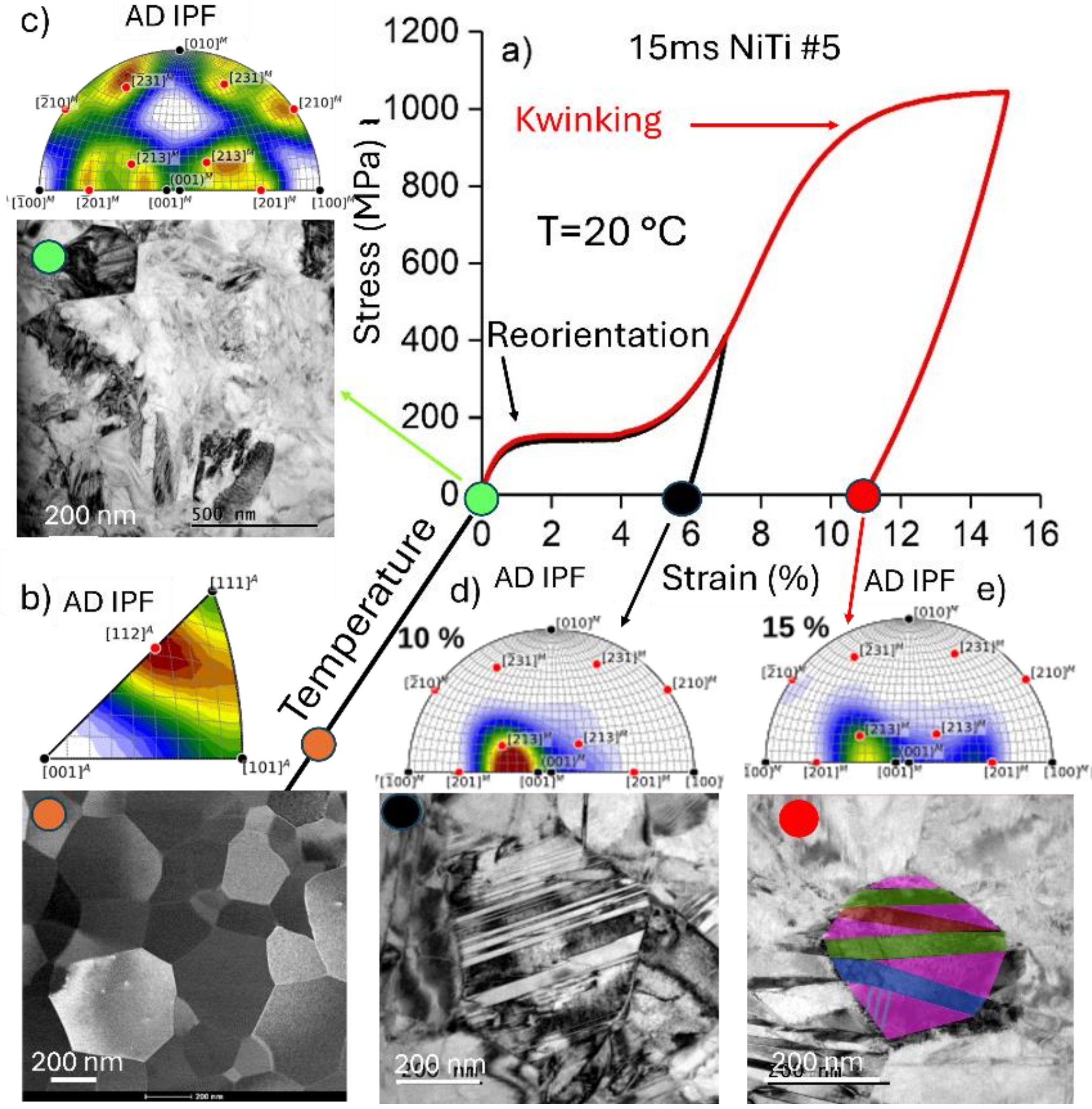

**Figure 2: Martensite variant microstructure and texture evolving during a tensile test on nanocrystalline 15ms NiTi #5 shape memory wire.** a) stress–strain–temperature curves recorded in tensile test at room temperature with denoted stages where martensite reorientation (7%) and plastic deformation via kwinking (15%) proceeds. The wire was unloaded, and TEM lamellae were cut and martensite variant microstructures in grains were reconstructed in four stages (solid circles) [25]. Figures (b–e) describe the martensite variant microstructures in grains and textures characterized by inverse pole figures of axial direction AD IPF evolving during the tensile test [26]. See Chapters 4, 6 and 7 and articles [24-27] for more detailed description and interpretation of these results.

The coupled plastic slip and reorientation of martensite can be sufficiently explained using a heuristic approach outlined in Fig. 3. Consider a grain inside a NiTi polycrystal subjected to unidirectional tensile loading containing B19’ martensite that is fully oriented, that is, the grain is filled with a single martensite variant with the maximum transformation strain aligned with the loading direction. For a strongly $<111>_A$ textured NiTi wire, for example, this means that the martensite variant in the considered grain is oriented with its $[10-1]_M$ direction along wire axis (Fig. 3a). This variant that complies best with the tensile loading is denoted as the blue V1 variant in Fig. 3. Now, we search for a deformation mechanism that may allow the grain to elongate in the tensile loading direction. As the martensite is fully oriented, the reversible twinning, that is, martensite reorientation within a single Eriksen-Pitteri neighborhood [51], which is otherwise the main pseudo-plastic deformation mechanism of B19' martensite, cannot fulfill this task. Indeed, as seen in Fig. 3b,c, both $(100)_M$ and $(001)_M$ compound twinning shorten the grain, having negative Schmid factors with respect to the tensile loading, and this holds true also for any other reversible twinning system (Type 1 and Type 2 [51]). The $(100)_M$ and $(001)_M$ twin bands contain the same crystallographic variant of martensite (red variant V2 in Fig. 3), which are, however, slightly rotated because of distinct twinning mechanisms. As a result, the grain is not only shortened along the loading axis, but also laterally sheared to the left (Fig. 3b) or to the right (Fig. 3c).

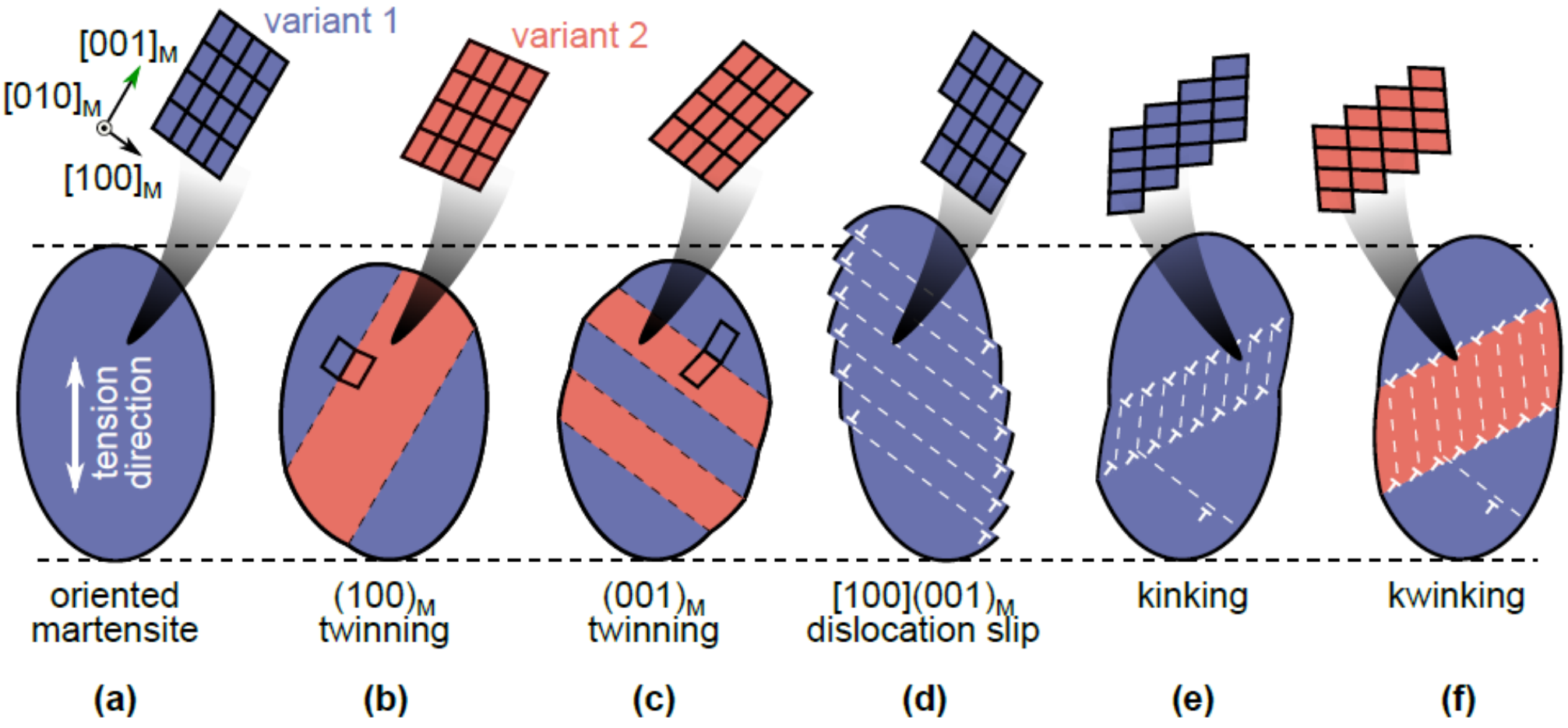

**Figure 3: Tensile straining of oriented B19' martensite through various deformation mechanisms** (a-f); the horizontal dashed lines outline the original length of the grain. While the reversible $(100)_M$ twinning (b) and $(001)_M$ twinning(c) shorten the grain, the plastic forming mechanisms including [100](001) dislocation slip (d), kinking (c) and kwinking (f) elongate it.

On the other hand, the only available $[100](001)_M$ slip system has a high and positive Schmid factor for the $[10\text{-}1]_M$ direction and hence can be easily initiated by the tensile loading, albeit producing also significant lateral shearing (Fig. 3d). However, conforming to the Von Mises law, a single slip system cannot allow for a homogeneous plastic deformation of a polycrystal, and thus, dislocation gliding on single $[100](001)_M$ slip system is restricted. In other materials with strongly anisotropic plastic properties, the lack of available slip systems is solved by creation of plastic kink bands, which is a mechanism outlined in Fig. 3e. The kink band spreads between two dislocation walls acting as tilt boundaries, while the material inside the kink band is effectively sheared and rotated with respect to the matrix. This results in a macroscopic strain (Fig. 3e) that is different from the plastic strain resulting from the coordinated $[100](001)_M$ slip in the matrix (Fig. 3d).

In other materials than NiTi, the kinking typically appears in wedge-like morphologies [56] that interlock each other and lead to significant work hardening of the material [57,58]. This, however, does not hold true in B19' martensite in NiTi. The reason is that the kinking in NiTi is complemented with reversible twinning, which brings additional degrees of freedom and allows the lattice to accommodate the compatibility stress concentration that the kinking induces. The simplest and most direct coupling between kinking and reorientation of martensite is outlined in Fig. 3f, in which the plastically slipped material inside of the band reorients, through reversible twinning, into variant V2. The kink band becomes a kwink band and the deformation process is called kwinking. By combination of dislocation slip in the martensite matrix, kwink band formation mechanisms and dislocation slip within the kwink bands, number of degrees of freedom increases and the plastic deformation of the polycrystal can proceed.

The interfaces between the matrix and kwink band are twin-like intermartensitic interfaces introduced by coordinated dislocation slip combined with twinning. The reorientation of the martensite by twinning reduces the tilt angles at the interfaces, allowing, thus, easier accommodation of the strains between neighboring grains, and, unlike the STGBs for which the lattice misorientation can be arbitrary, the kwink interfaces tend to align with certain crystallographic planes, which reduces their surface energy [23]. Indeed, the kwink bands form along $(20\text{-}1)_M$ planes, for which the interfaces are highly coherent (Fig. 4).

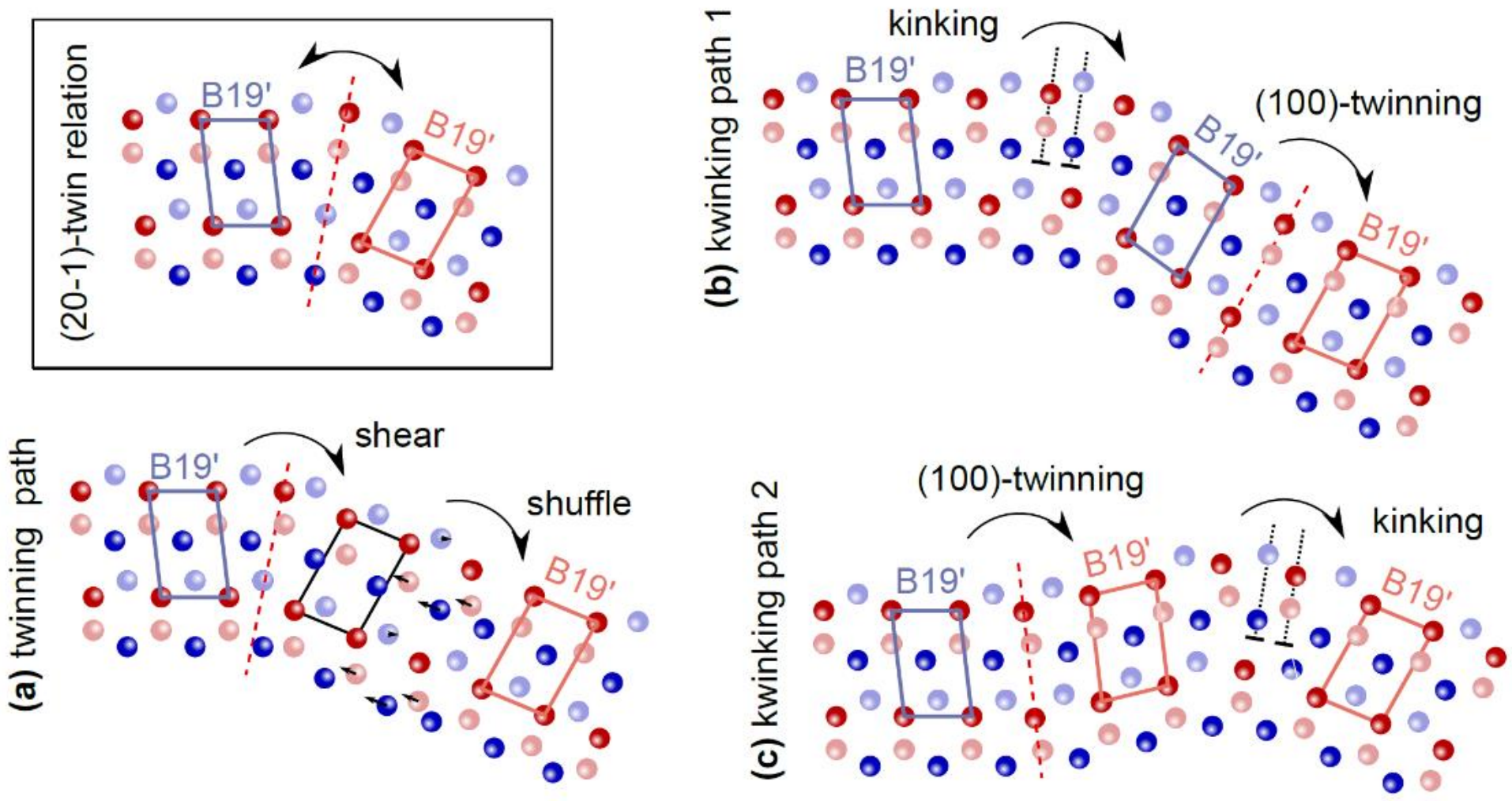

**Figure 4: Comparison of a deformation twinning path (a) vs. a kwinking path (b,c) towards (20-1) twin.** Assuming the twinning path (a), the monoclinic lattice must go through a complex deformation twinning pathway involving twinning followed by shear and shuffle [42] to reach the (20-1) twin configuration. Alternatively, the same configuration can be reached by kwinking path (b,c) involving combination of twinning and coordinated [100](001) dislocation slip based kinking [23].

The exact mechanism by which the kwink bands arise can be fully unraveled neither from their experimental observation by TEM nor by theoretical analysis. At the continuum mechanics level, the shape strains of kwink bands (the deformation gradients with respect to austenite) result from the combination of coordinated $[100](001)_M$ dislocation slip and $(100)_M$ compound twinning. In fact, the kinking and $(100)_M$ twinning commute, that is, exactly same strain field can be obtained by slipping the material first and applying the twinning afterwards (Fig. 4b), and by creating a twin band first and plastically slipping its interior afterwards (Fig. 4c). Both pathways towards the (20-1) kwink are energetically equally favorable.

In summary, plastic deformation of the monoclinic B19' martensite proceeds by kwinking deformation mechanism involving coordinated dislocation slip-based kinking combined with (100) deformation twinning. It can be understood as being a result of interactions between dislocation slip and twinning over a broad range of spatial scales. Macroscopically, it relaxes the external loading through pattern formation utilizing massive, coordinated dislocation slip, while it simultaneously reduces energy by creating coherent intermartensitic interfaces while relaxing the energy of the dislocation cores.

## 4. [100](001) Dislocation Slip in B19' Martensite

Results of TEM reconstruction of martensite variant microstructures in grains and martensite textures evolving during the tensile test on nanocrystalline 15ms NiTi #5 shape memory wire are shown in Fig. 2. The martensite variant microstructures in the wire stress-free cooled from 100 °C (Fig. 2b) to room temperature (below the $M_f$ temperature) contain multiple variously oriented domains of nanotwinned martensite (Fig. 2c). Due to the presence of (001) compound twins within the microstructure combined with peculiar electron diffraction contrast, individual grains and grain boundaries could be hardly distinguished in BF TEM images (Fig. 2c). However, when the wire was deformed beyond the end of the reorientation plateau (Fig. 2a), polycrystal grains became filled by single (001) compound twin laminates and grain boundaries became clearly visible (Fig. 2d). It shall be pointed out that this does not happen only in the selected grain that we analyzed (oriented in [010] zone in Fig. 2d), but that all polycrystal grains displayed these simple laminated microstructures when oriented in [h,k,0] zone. When the wire was loaded further beyond the end of the reorientation plateau (strain range 6-10% in Fig. 2a), this martensite variant microstructure did not change significantly. When tensile stress exceeded the yield stress for plastic deformation (kwinking stress), plastic deformation by kwinking started (strain range 11-15% in Fig. 2a) and multiple deformation bands (kwink bands) gradually filled all polycrystal grains (Fig. 2e). For further information on martensite variant microstructures see Chapter 6. Since the martensite lattices within newly formed deformation bands are misoriented from the matrix, texture of the martensite phase (characterized experimentally by inverse pole figures of axial direction in Fig. 2) evolved during tensile tests. For further information on texture evolution see Chapter 8.

The kwink bands are separated from the martensite matrix and between themselves by intermartensitic interfaces newly introduced by the kwinking deformation mechanism. The key point is that the strain with respect to parent austenite (sum of elastic, transformation, and plastic strain) must be compatible at these newly introduced interfaces (Fig. 2e). Although the most frequently observed (20-1) kwink interfaces display lattice misorientations and interface planes similar to the (20-1) twin, they are not deformation twins since they were not created by deformation twinning but by kwinking involving coordinated [100](001) dislocation slip combined with deformation twinning (Fig. 4b). kwinking deformation creates many intermartensitic interfaces which are not twins [24] (see Chapter 6).

Since the kwinking deformation proceeds via coordinated [100](001) dislocation slip in martensite combined with (100) deformation twinning, we pay special attention to these processes in this chapter. The (001) basal plane is the most densely packed lattice plane of the monoclinic B19' structure with largest lattice spacing. The [100](001) dislocation slip system has already been widely accepted in the field as the most preferable slip system in B19' martensite [41,55]. However, consideration of [100](001) is more theoretical concept then experimentally confirmed reality. Direct experimental support for the activation of the [100](001) dislocation slip in thermomechanical loading tests is limited. Experimental difficulties with TEM analysis of dislocation defects in martensite could be a reason. Another reason could be that the [100](001) slip dislocations gliding during the MT proceeding under external stress disappear within grain boundaries [13] and slip dislocations gliding during the kwinking deformation disappear within the newly introduced kwink interfaces. In direct confirmation comes from observation of peculiar dislocations in austenitic microstructures of NiTi wires that transformed under external stress [14,43]. The [100](001) slip dislocations in martensite are converted into <100>{011} slip dislocations in austenite by the reverse MT.

To understand the nature of the [100](001) dislocation slip in B19' martensite, one needs to consider the crystallography of the B2-B19' MT in NiTi. During the MT from cubic austenite to the lower symmetry monoclinic martensite, up to 12 Lattice Correspondence Variants (LCVs) can form, each defined by a characteristic lattice correspondence and strain gradient relative to the parent austenite. Each LCV may further undergo transformation twinning giving rise to wide range of twin interfaces experimentally observed in thermally induced B19' martensite (Fig. 2c). The lattice planes and directions between the B2 austenite and B19' martensite are mutually linked by the characteristic lattice correspondence of the B2-B19' MT. The lattice planes and directions are mutually inclined except for basal plane (001)M parallel to {011}A and (010)M plane parallel to {110}A plane (see Appendix B in [25]).

To explain the role of [100](001) dislocation slip and (001) compound twinning processes in kwinking deformation, the change of the crystal structure due to the B2-B19' MT can be characterized by 2D projections of B2 austenite and B19' martensite structures (Fig. 5) viewed perpendicularly to the (110)A austenite plane (parallel (010)M martensite plane). The monoclinic martensite lattice is distorted either leftwards or rightwards within the [010] zone giving rise to lattice correspondence variants V1 and V2 possessing common basal plane (001)M. The monoclinic structure can be easily sheared in the [100] direction on the (001) basal plane due to the week Ti-Ti bonds [60] (denoted by double black arrows in Fig. 5). This is the reason why (001) compound twin laminates consisting of two martensite variants V1 and V2 form during the forward MT in nanocrystalline NiTi. There are 6 such LCV couples derived from 6 equivalent {011} planes of the cubic structure. These 6 LCV couples, termed also Correspondence Variant Pairs (CVPs), form 6 variously oriented (001) compound twinned domains .

Martensite variant microstructures consisting of multiple CVP laminated domains appear in grains of nanocrystalline NiTi wires cooled into martensite stress-free or under low stresses (Fig. 2c). As the interfaces between various CVP domains are highly mobile, NiTi wires deform in tensile tests by martensite reorientation at low stress (100-300 MPa depending on temperature). The martensite reorientation creates single CVP laminate within each grain (Fig. 2d), but these laminates do not detwin during this process (if they do, then only partially). The MT proceeding under external stress gives rise to (001) compound twinned laminates filling whole grains that are significantly detwinned [10,25]. The observation of compound twinned laminates filling whole grains is essential for understanding the generation of incremental plastic strains by MT proceeding under stress and the role of dislocation slip in martensite in functional thermomechanical properties of NiTi [13].

When NiTi wires are deformed up to the end of the reorientation plateau, unloaded and subsequently stress-free heated above the $A_f$ temperature, the martensite laminates filling whole grains transform back to the parent austenite without generating plastic slip and permanent lattice defects. The reverse transformation temperature is, however, shifted ~30 °C above the $A_f$ temperature evaluated by DSC thermal cycling. This upward shift of the $A_f$ temperature by deformation is known as the "stabilization of martensite by deformation" [61]. It was often associated in literature with permanent lattice defects created by deformation. This is incorrect, no dislocation slip is needed to explain the experimentally observed upward shift of the $A_f$ temperature by martensite reorientation. It originates from higher energy required to form strain compatible habit plane interface between the oriented martensite (single laminate of (001) compound twins within grains) and austenite during the reverse MT of deformed NiTi (compared to stress-free thermal cycle) on subsequent heating [13,47].

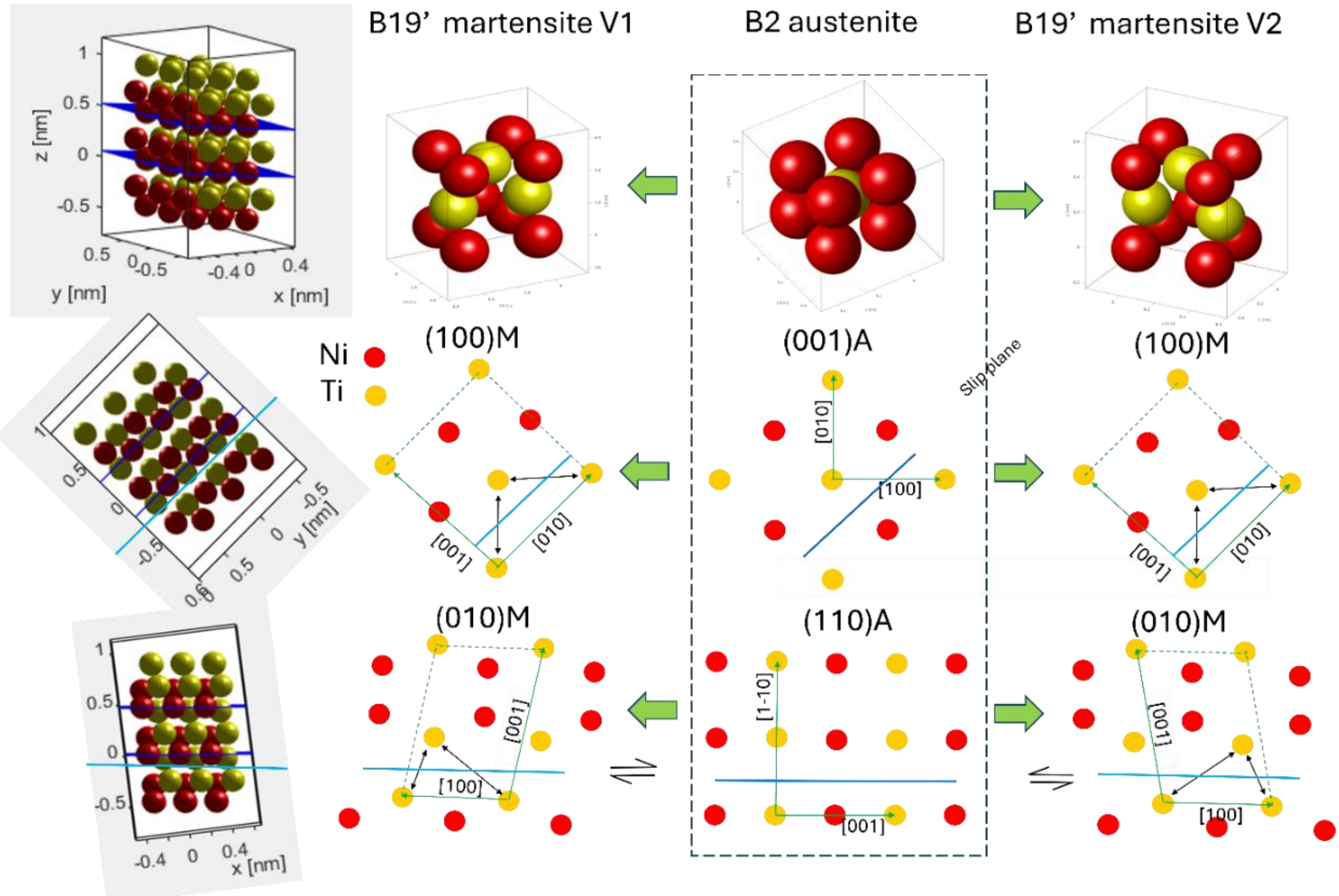

**Figure 5: Crystallography of the B2–B19' MT in NiTi (a) and intermartensitic interfaces involved in kwinking (b).** The sketch in (a) shows atom positions on $(001)_A$ and $(110)_A$ austenite planes prior to and after the martensitic transformation, turning them into lattice correspondent $(100)_M$ and $(010)_M$ martensite planes, respectively. Two layers of atoms are plotted together without differentiating between them. Mechanical stability of the B19' structure is significantly affected by weak Ti–Ti bonds [60] (denoted by double black arrows) which allow for easy shearing of the lattice in the [100] direction on the (001) plane. Notice that no mirror symmetry of atomic positions exists across the (001) twin plane (denoted by blue lines). The easy shear enables (001) compound twinning as well as dislocation slip on the (001) plane in the [100] direction. The [100](001) slip dislocations are inherited into austenite as <100>{110} slip dislocations.

In case of (001) compound twinned martensite in nanocrystalline NiTi alloys, there is a theoretical problem stemming from the fact that strain compatible habit plane interface cannot exist between austenite and single (001) compound twin laminate [62]. Habit planes between austenite and <011> type II twinned martensite were disregarded in this work since <011> type II twin laminates [3] were never observed in nanocrystalline NiTi. When B2-B19' MT in nanocrystalline NiTi proceeds under zero or small stresses, formation of multiple CVP domains behind and prior the propagating habit planes solves the problem of strain compatibility at the habit plane [13]. However, when the forward and reverse MT proceed under external stresses above certain characteristic thresholds, experiments show that grains are filled by single often significantly detwinned laminates. The strain compatibility at the habit plane thus must be attained differently - most likely with the help of elastic strain [49] and/or plastic strain on the martensite side of the habit plane interface [63]. In this case, however, incremental plastic strains [11] and permanent lattice defects [14] are generated by the MT proceeding under stress [13]

Another problem of MT in nanocrystalline NiTi wire proceeding under stress consists in that the (001) compound twin laminates filling whole grains are associated with large shape strains with respect to the parent austenite and these shape strains must be compatible at grain boundaries (GB). Since the detwinning of (001) compound twins requires extremely low energy [41,60] and yields a very large shear strain (11.2% [26]), it has been widely considered in literature to occur during the martensite reorientation. Despite that, (001) compound twin laminates were observed in reoriented martensite as well as in stress induced martensite [25]. For the same maximum stress, however, stress induced martensite is significantly more detwinned than the reoriented martensite. The higher is the stress under which stress induced MT proceeds, the more detwinned the laminates are.

The martensite laminates in oriented martensite loaded in tension can theoretically deform only via detwinning of (001) compound twin laminates and/or via dislocation slip on the (001) plane in the [100] direction yielding similar shear deformation. However, despite the very low stress required to activate both (001) dislocation slip and detwinning of (001) compound twins, these processes did not proceed (or proceed only marginally) in the studied nanocrystalline NiTi wire. We rationalize this by assuming the lack of independent slip systems – polycrystal cannot deform plastically via single deformation system in grains. The studied NiTi wires thus deformed elastically when loaded beyond the end of the stress plateaus regardless of the microstructure (grain size) and test temperature (Fig. 1d).

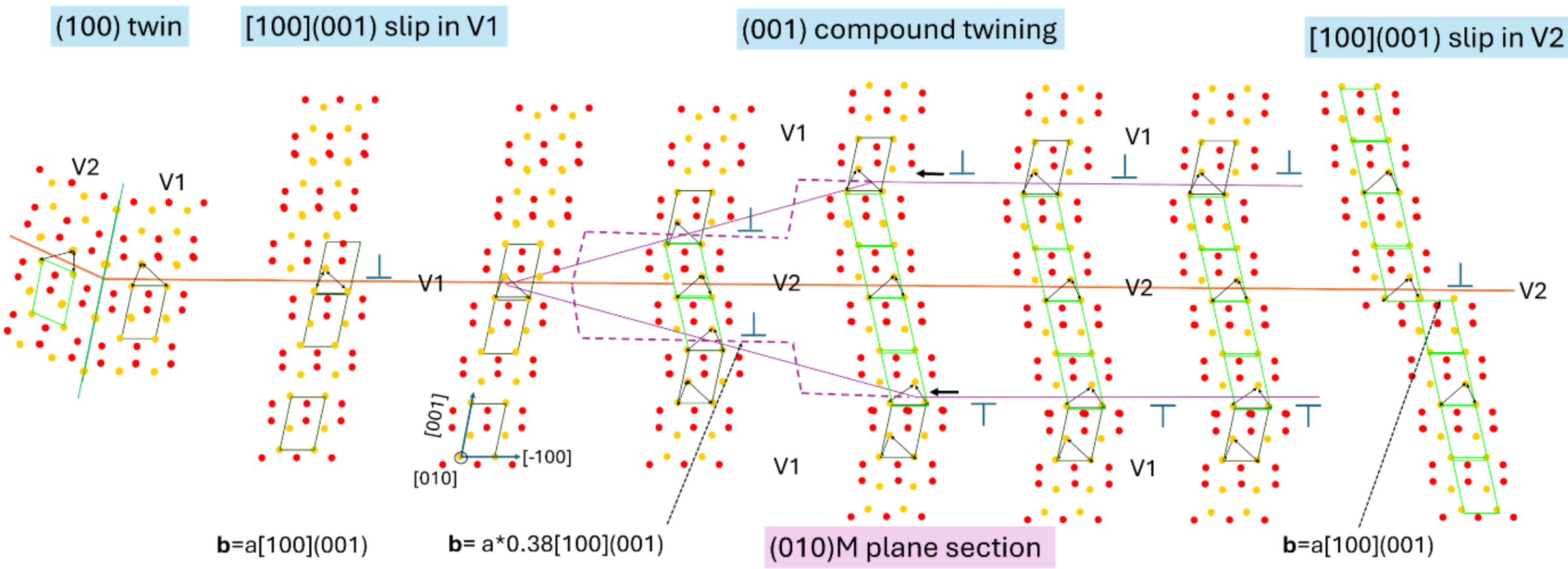

**Figure 6: Basic deformation processes in monoclinic martensite visualized in the [010] zone.** The instability of the B19' monoclinic lattice due to the easy shear on (001) plane in the [100] direction enables dislocation slip (**b**=a[100](001)) on (001) plane in the [100] shear direction in variants V1 and V2 and (001) compound twinning (mutual conversion between lattice correspondence variants V1 and V2 possibly occurring via partial dislocation slip (**b**=a*0.38[100](001)) on (001) plane in the [100] shear direction). For completeness, (100) deformation twinning, which reorients the martensite V1 into rotated variant V2 so that the (001) slip plane rotates away from the original orientation, in shown on the left side.

Nevertheless, when the forward and reverse MT proceed in tensile tests under stresses above the specific stress thresholds, the (001) dislocation slip and detwinning of (001) compound twins operate and alleviate shape strain incompatibilities arising among collectively transforming neighboring grains of the NiTi polycrystal cooled under external stress [13,14]. There are two essential differences between the detwinning of (001) compound twins and [100](001) dislocation slip - while shear strain due to detwinning yields crystallographically limited recoverable transformation strains, the [100](001) dislocation slip in martensite yields unrecoverable plastic strains. In other words, while the detwinning increases the recoverable transformation strains, the [100](001) dislocation slip keeps the recoverable strain unchanged but generates incremental plastic strain recorded in closed-loop thermomechanical cycle. The incremental plastic deformation is always linked to the MT proceeding under stress, meaning that the [100](001) dislocation slip is activated only when the forward and/or reverse MT proceeds under stress above the specific stress thresholds [13].

Results of TEM analyses of deformed martensite show that twinning (detwinning) of (001) compound twins proceeds via growth (disappearance) of thin (001) compound twin plates in the [100] direction. The question is whether this process should be treated as deformation twinning. We do not think so and, hence, we propose the following alternative mechanism. Assuming the weak Ti-Ti bonds in the (001) slip plane (Fig. 5), the atomic configuration of the (001) plane interface is inherent part of the crystal lattice either in V1 or V2. A partial shift of the V1 lattice on (001) plane in [100] direction (**b**=a*0.38[100](001)) flips the weak Ti-Ti bonds, which moves the interface and expands the V2 lattice at the expense of the V1 lattice (Fig. 6). Since coordinated partial dislocation slips on two parallel (001) plane interfaces driven by external stress convert the variant V1 into variant V2, the tip of the variant V2 propagating in Fig. 6 leftwards can be understood as a stress driven growth (shrinking) of the thin martensite plate.

Since the (001) compound twins were created by the forward MT on cooling and/or mechanical loading, they are transformation twins. However, are they really twins? Wherever the blue line in Fig. 5 is positioned in the monoclinic structure, no mirror symmetry across the interface can be established. This was called pseudo-mirror symmetry in literature [41,55]. Nevertheless, to avoid confusion, we use the term (001) compound twinning throughout this work.

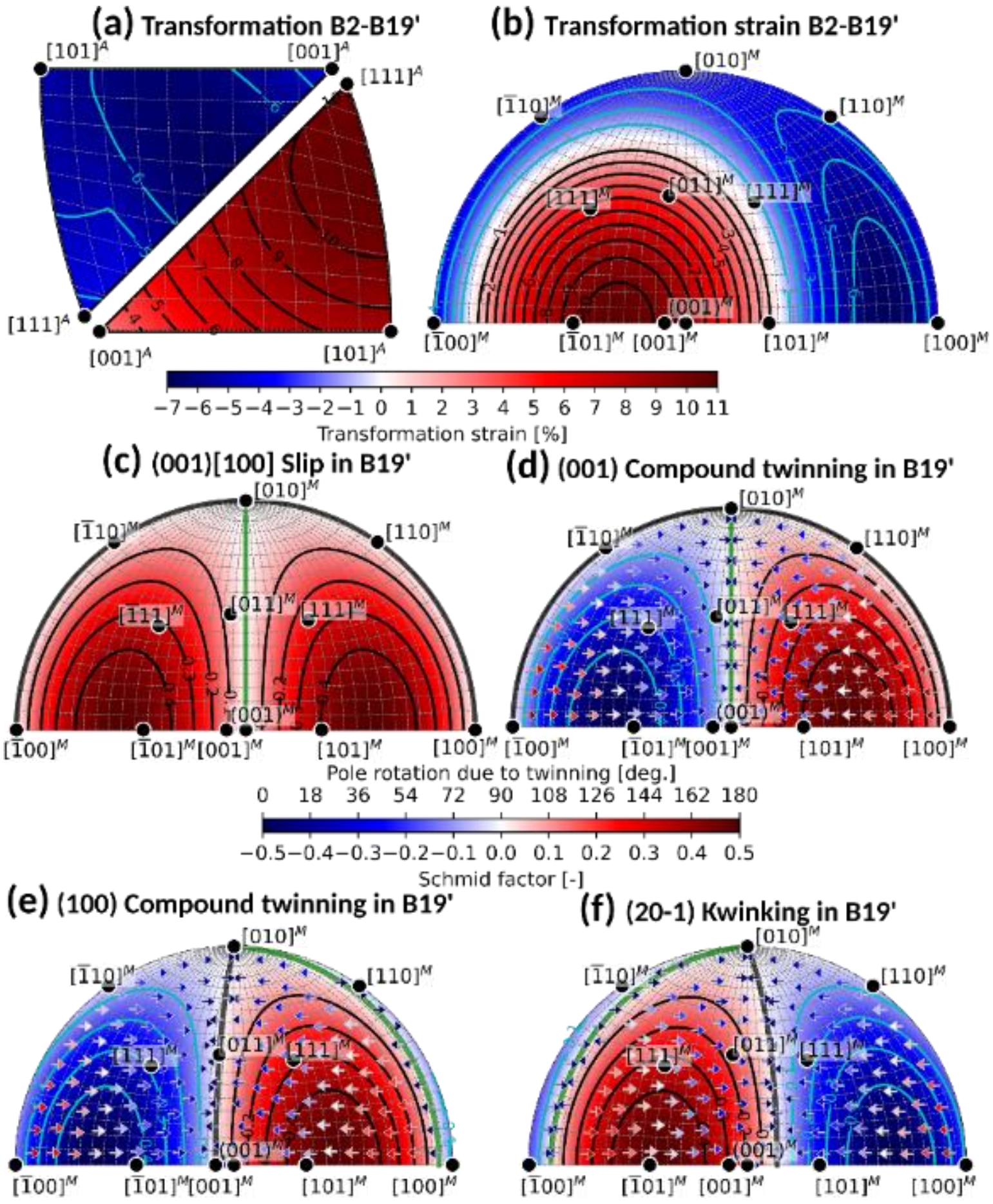

**Figure 7: Orientation dependence of MT and deformation processes in martensite relevant for kwinking deformation.** Orientation dependence of transformation strains of B2–B19' MT in NiTi in tension and compression plotted in cubic B2 austenite (b) and monoclinic B19'. Orientation dependence of the Schmid factor for: c) [100](001) slip system in B19' martensite, d) (001) compound twinning, e) (100) deformation twinning, f) (20-1) kwinking. Colored arrows in (d), (e) and (f) indicate directions of lattice reorientation with respect to the matrix due to the deformation process. Traces of slip planes or twinning planes are denoted by thick black lines, while traces of planes normal to the slip direction or twinning directions are denoted by thick green lines.

During the reverse MT upon stress-free heating, the reoriented martensite returns to the same parent austenite lattice it was created from, regardless of any (001) compound (de)twinning that may have occurred during the martensite reorientation under stress. If partial dislocation slip facilitated (de)twinning of (001) compound twin laminates occurs during the martensite reorientation, no plastic slip and no permanent lattice defects are created upon tensile straining followed by reverse MT upon heating. On the other hand, if complete dislocations **b**=a[100](001) glide through the grain during the martensite reorientation or stress induced MT proceeding under stress (Fig. 6), the cubic austenite lattice is restored upon heating, but incremental plastic strains, lattice rotations and slip dislocations remain within the austenitic microstructure.

The orientation dependent deformation mechanisms of MT, twinning, and plastic slip in B19' martensite (Fig. 7) give rise to elastic, deformation, and plastic anisotropies of textured nanocrystalline NiTi wires. When the $<111>_A$ textured NiTi wire is loaded beyond the end of the stress plateau in tension, most of the martensite variants in grains are oriented with the [-101] direction approximately aligned with the wire axis (see Fig. 7a,b). There is no transformation twinning system in martensite that can be activated when the oriented B19' martensite is pulled along the [-101] direction (Fig. 7). Despite their low Schmid factors, dislocation slip and detwinning processes are not activated (or only marginally) in nanocrystalline NiTi wire deformed beyond the end of the stress plateau. But there must be a way out of this trap since the oriented martensite eventually starts to deform plastically at the yield stress.

There is the (100) compound twinning (Fig. 7e, 3) that also converts the variant V1 into V2 but, in contrast to (001) compound twinning, it rotates the (001) slip plane (see Fig. 6 on the left). The (100) compound twining proceeds as true deformation twinning in mechanical loads [10]. It forms deformation bands observed in NiTi wires before they start to deform plastically by kwinking [10,12]. However, this is strange since Schmid factor for (100) twinning in tension is negative (Fig. 3b). Despite that, the experiments (see Fig. 13 in [10]) show that (100) deformation twinning becomes regularly activated in tensile tests, particularly at elevated temperature (stresses). When the (100) compound twinning takes place in a grain filled with (001) compound twin laminate), it reorients both martensite variants V1 and V2. The minority variant V2 becomes converted into the majority variant V1 within the twin and vice versa. This was nicely demonstrated by Zhang et al. [29] many years before the kwinking mechanism was discovered [23,24]. We deduce that the (100) deformation twinning becomes activated in tensile tests, because it is involved in kwinking deformation (Chapter 3), which is the only way the oriented B19' martensite can deform plastically.

In summary, due to the weak Ti-Ti bonds in its structure [60], the B19' monoclinic martensite in NiTi is prone to shear deformation on (001) plane in [100] direction taking place either via [100](001) dislocation slip and/or via (001) compound twinning. When the former is activated during the martensite reorientation and/or MT proceeding under stress, recoverable strains are not affected but incremental plastic strains are generated, and permanent lattice defects are left in the austenite phase in closed loop thermomechanical load cycles. When the latter is activated, recoverable strains increase and no plastic strains and no lattice defects are created. The (001) compound twinning (detwinning) can be alternatively viewed as stress driven motion of internal interfaces on (001) plane proceeding via glide of partial dislocations. Due to the availability of single slip system only, the B19' martensite is extremely plastically anisotropic.

## 5. Kwinking Stress as the Yield Stress for Plastic Deformation of Martensite

The onset of kwinking deformation is determined as the yield stress for plastic deformation evaluated in tensile tests on NiTi superelastic and shape memory wires at low temperatures at which the wires contain solely oriented martensite (Figs. 1,8). We call it "kwinking stress," because it characterizes the onset of kwinking deformation. The B19' martensite, however, can deform plastically also by incremental dislocation slip occurring at significantly lower stresses while the forward and/or reverse MT proceed in thermomechanical loading tests (the wire contains mixture of austenite and martensite) [8,11], which is a different plastic deformation mechanism. The kwinking stress (denoted $\sigma^Y_{M_kwink}$ in stress-temperature σ-T diagrams (Fig. 8)) characterizes the onset of massive plastic deformation of oriented martensite by kwinking - coordinated [100](001) dislocation slip based kinking combined with (100) deformation twinning proceeding in the absence of MT. When MT proceeds in a wire containing a mixture of austenite and martensite, kwinking deformation of martensite may be also activated under stresses that are lower than the kwinking stress while the forward and/or reverse MT proceed. This will be further discussed in Chapter 9.

Kwinking deformation can be activated both in tension and compression. We have systematically evaluated kwinking stresses $\sigma^{Y}_{M_kwink}$ from the results of thermomechanical loading tensile tests on superelastic NiTi #1 and shape memory nanocrystalline NiTi #5 wires having wide range of austenitic microstructures (grain size, $Ti_3Ni_4$ precipitates, permanent lattice defects etc.). NiTi wires were deformed in isothermal tensile loading and isostress heating experiments up to rupture [8,10,21]. The critical stress-temperature conditions for all deformation/transformation processes activated in tests on superelastic and SME NiTi wires having the same virgin austenitic microstructure are presented in σ-T diagrams (Fig. 8a,e) [13]. The remaining graphs in Fig. 8 show examples of stress-strain curves recorded in isothermal tensile tests in different temperature ranges. For examples of strain-temperature curves recorded in isostress heating tests see Ref. [8].

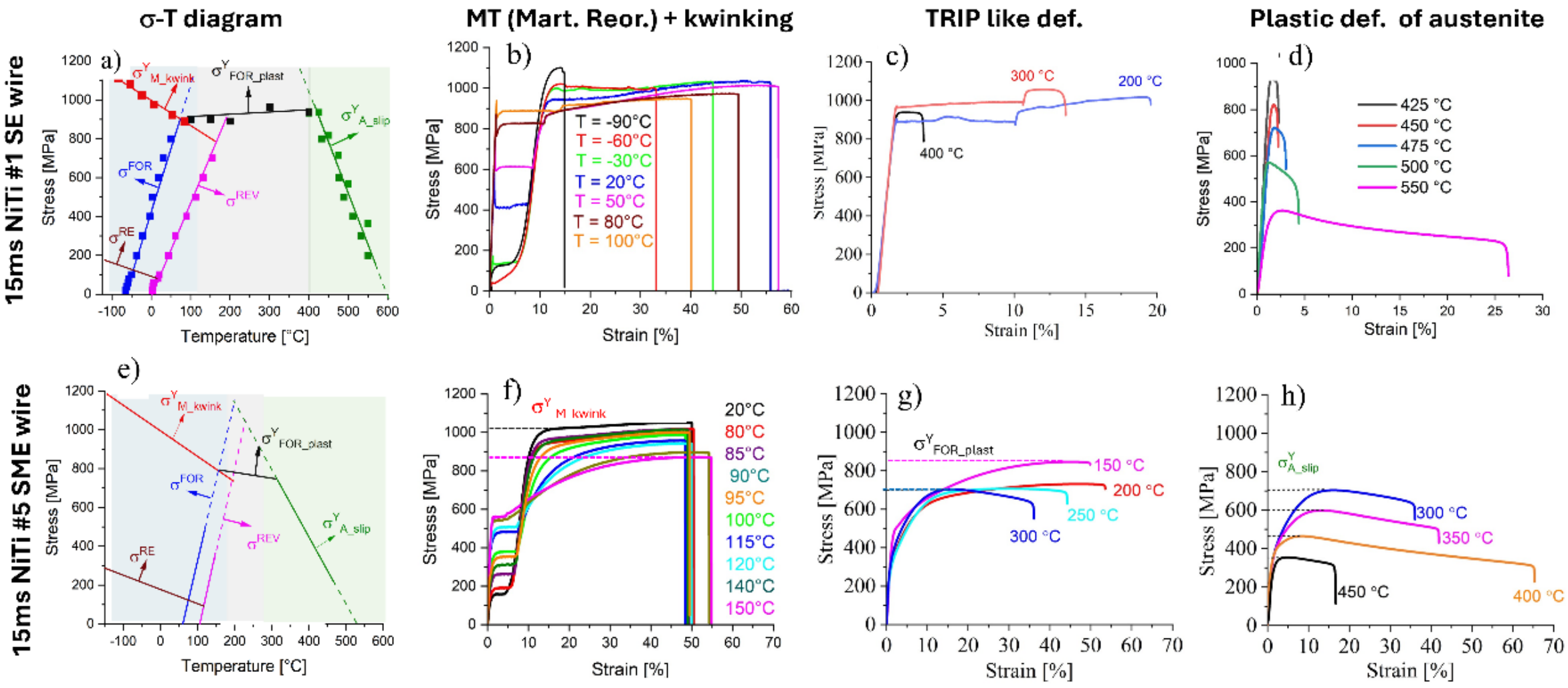

**Figure 8: Stress-temperature σ-T diagrams of a) 15 ms NiTi#1 superelastic wire and b) 15ms NiTi#5 (e) shape memory wire** provides information on critical stress-temperature conditions for activation of deformation/transformation processes in tension. The diagrams were constructed from the results of isothermal tensile tests (b-d,f-h) [11] and isostress thermomechanical tensile tests until fracture [8]. The wires have the same austenitic microstructure (grain size) but different chemical composition.

Notice that kwinking stress $\sigma^{Y}_{M_kwink}$ decreases linearly with increasing temperature (Fig. 8a,e). Moreover, it depends on the parameters characterizing the virgin austenitic microstructure of the wire, particularly on the: i) chemical composition (higher for superelastic NiTi wire than in the SME NiTi wire, ii) nanoscale chemical heterogeneity (Fig. 9), iii) coherent $Ti_3Ni_4$ precipitates (Fig. 9), iv) grain size (Fig. 1c), v) permanent lattice defects introduced by plastic deformation (Fig. 10) and vi) austenite texture (Fig. 8). While the parameters (i-iii) are related to the critical stress for [100](001) dislocation slip in martensite, parameters (iv-vi) concern mechanics of polycrystal deformation.

As an example of the former (i-iii), Fig. 9 shows the effect of low temperature aging at 250 °C for varying times 0-50 hours on tensile stress-strain response of superelastic 15 ms NiTi #1 wire in tensile tests at 20 °C. Kwinking stress increases and ductility decreases with increasing time of aging that introduces nanoscale chemical heterogeneity into the B2 austenite (Fig. 9c). We assume that the nanoscale chemical heterogeneity increases the critical stress for [100](001) dislocation slip in stress induced martensite, which in turn increases the yield stress for plastic deformation by kwinking recorded in tensile tests (Fig. 5b). The mechanism is not clear yet, most likely it is related to the effect chemical composition and heterogeneity on chemical bonding at the (001) slip plane (Fig. 5). It will be explained in chapter 7 that the wires strengthened by low temperature aging longer than 2 hours (Fig. 9a) fracture via instability of tensile deformation (necking) proceeding via kwinking deformation.

The effect of permanent lattice defects introduced by the plastic deformation on the magnitude of kwinking stress can be inferred from the results of series of tensile tests up to various maximum strains at various temperatures reported in Ref. [15]. The kwinking frequently starts with a short plateau (lower strain hardening in 3-30% strain range), where kwinking deformation massively proceeds (Fig. 10, 8b,f). The lower is test temperature, the longer this early kwinking strain range is. As the strain further increases in the tensile test, kwink bands fill the grains, the alloy strengthens against dislocation slip mainly due to the microstructure refinement by kwinking (Hall-Petch strengthening) and less by introducing dislocation defects (dislocations disappear within kwink interfaces).

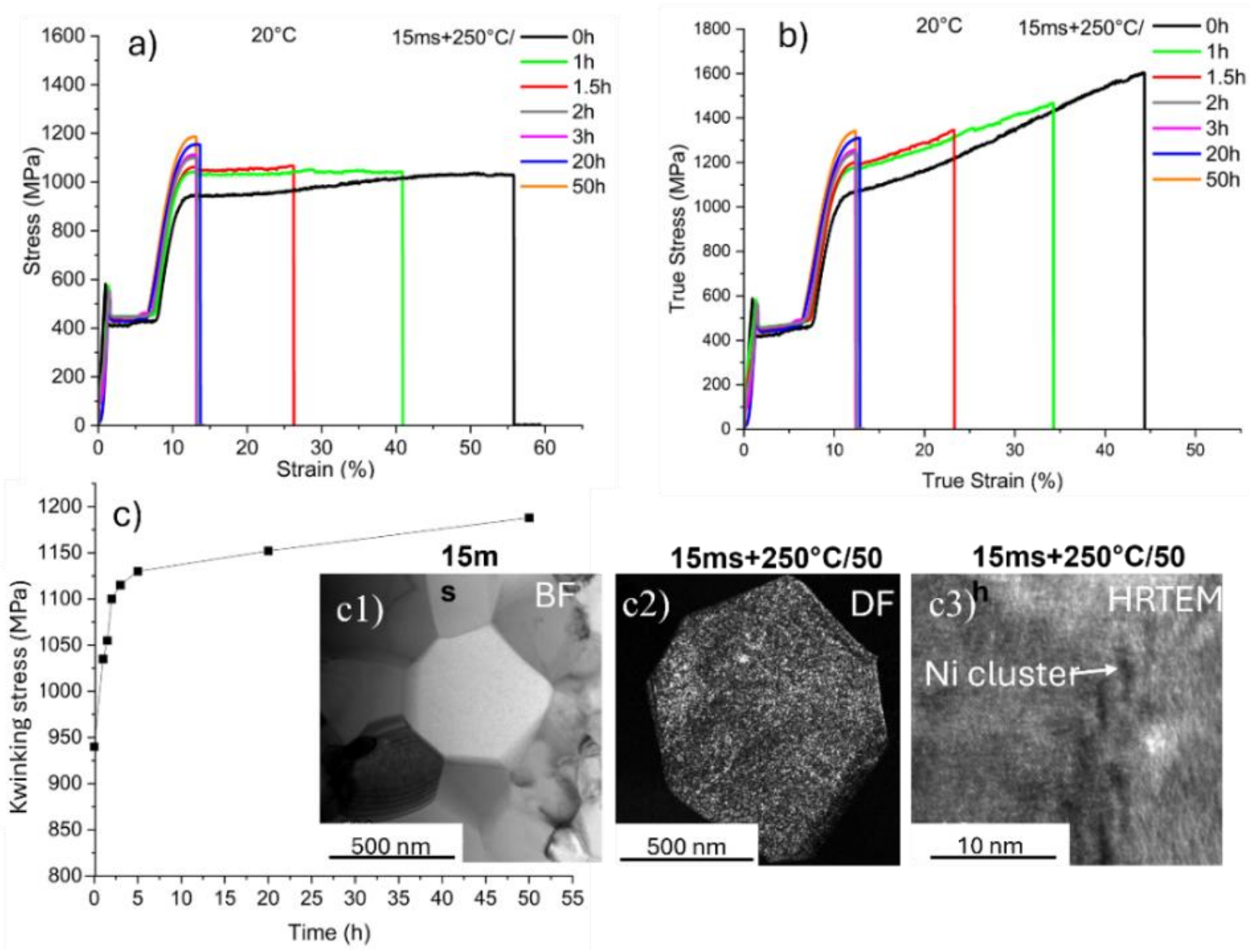

**Figure 9: Strengthening superelastic 15 ms NiTi #1 superelastic wire via low temperature aging** at 250 °C for various times 0-50 h affects microstructure and mechanical properties of the wires. a) stress-strain curves and b) true stress-true strain curves in tensile tests at 20 °C, c) dependence of kwinking stress on annealing time. BF, DF, and HRTEM images show nanoscale chemical heterogeneity within austenite grains (Ni clusters forming on (111) austenite plane) created by low temperature aging at 250 °C for 50 hours.

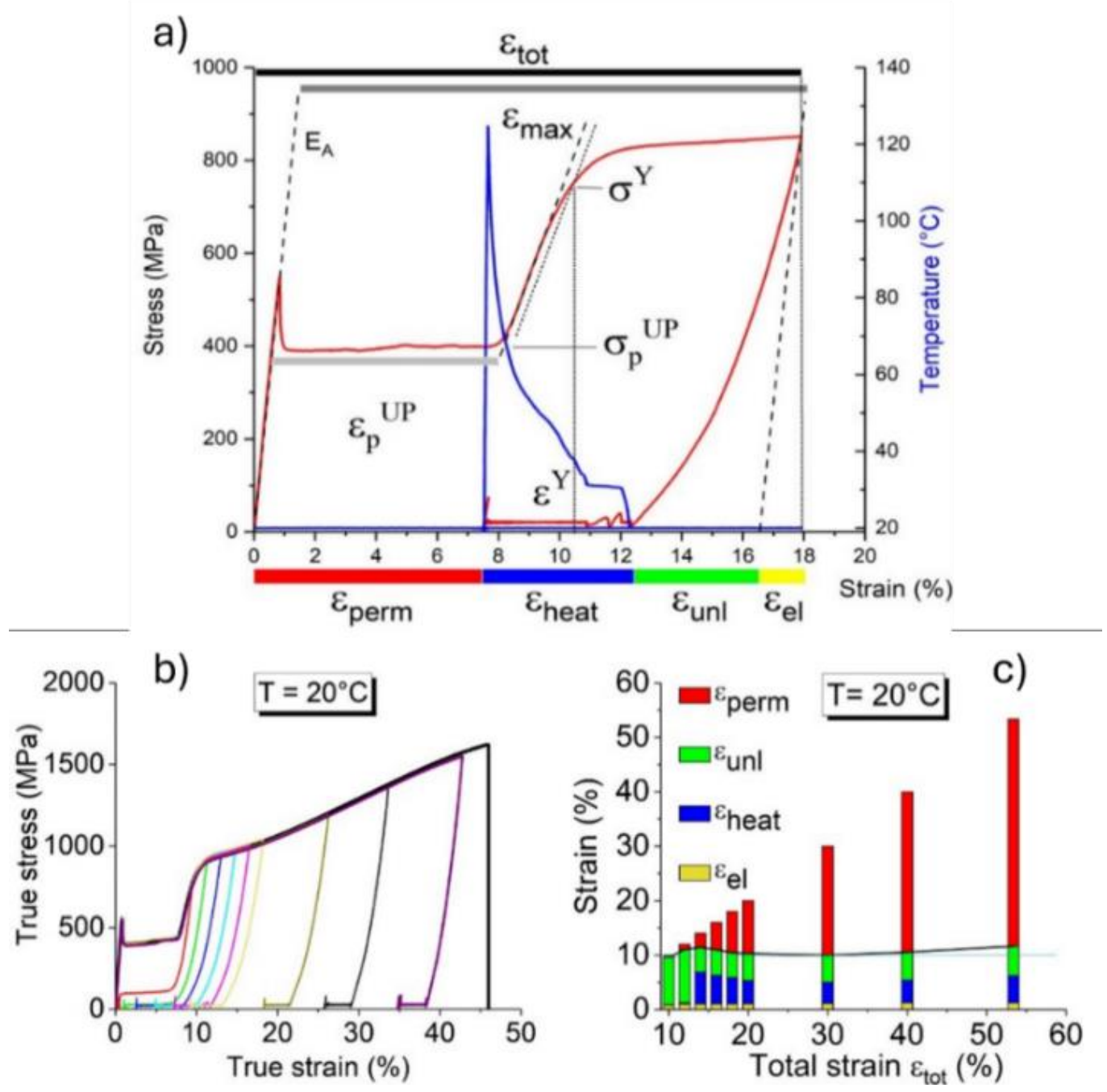

**Figure 10: Strengthening superelastic 16 ms NiTi #1 wire via plastic deformation** – strain hardening leads to the gradual increase of kwinking stress caused by the plastic deformation in tensile tests at constant temperature [15]. a) definition of strain components, b) true stress – true strain curves recorded in tensile test up to various maximum strains, unloading and stress-free heating the wire up to 200 °C, c) strain components evaluated in the tensile tests in dependence on maximum strain.

Recoverable and plastic strains introduced by the kwinking deformation in tensile test were evaluated from closed loop thermomechanical loading cycles with gradually increasing maximum strain (Fig. 10). Examples of true stress-true strain curves from tensile test at 20 °C and recoverable strains show that kwinking stress continuously increases (Fig. 10b) but recoverable strain (~10%) remains nearly constant up to rupture (Fig. 10c). The recoverable strains displayed by SMA polycrystals are believed to be governed by the crystallography of MT and texture [67]. From this point of view, the recoverable strains 10% evaluated in tensile tests at low temperatures (Fig. 10c) almost reach maximum transformation strain 11% of <111> oriented NiTi single crystal [3].

Recoverable strains evaluated from closed loop shape memory cycles on NiTi polycrystals are limited by the magnitude of reorientation stress (if it is too high, recoverable strains are low). Recoverable strains are also affected by the resistance to plastic deformation (if kwinking stress is too low, recoverable strains are low) [69]. To investigate that, we performed a closed loop shape memory test on superelastic 15ms NiTi #1 wire (Fig. 11). The results show that recoverable strains of NiTi evaluated from closed-loop shape memory tests is are significantly larger than superelastic plateau strains ~6% commonly evaluated in tensile tests. When the wire was loaded in the martensite state up to just below the kwinking stress, ~12% strain was completely recovered (Fig. 11a,b). When the wire was loaded further beyond the onset of plastic deformation by kwinking (Fig. 11c,d), recoverable strains became even larger (~13%), though unrecoverable plastic strain ~1% was generated in the closed loop shape memory test. These recoverable strains are larger than the theoretical B2-B19' transformation strain achievable in tensile tests on NiTi single crystal along <111>A direction [3]. This can be rationalized by considering the role of continuous anisotropic thermal distortion of monoclinic martensite in closed loop shape memory cycle in which the net recoverable strain is the sum of recoverable elastic, transformation, and thermal distortion strains. While conventional metallic wires elongate while being heated, the oriented martensite in NiTi wire gives rise to shortening (2%) due to anisotropic thermal contraction of the monoclinic structure along the [10-1]M direction (Fig. 11b).

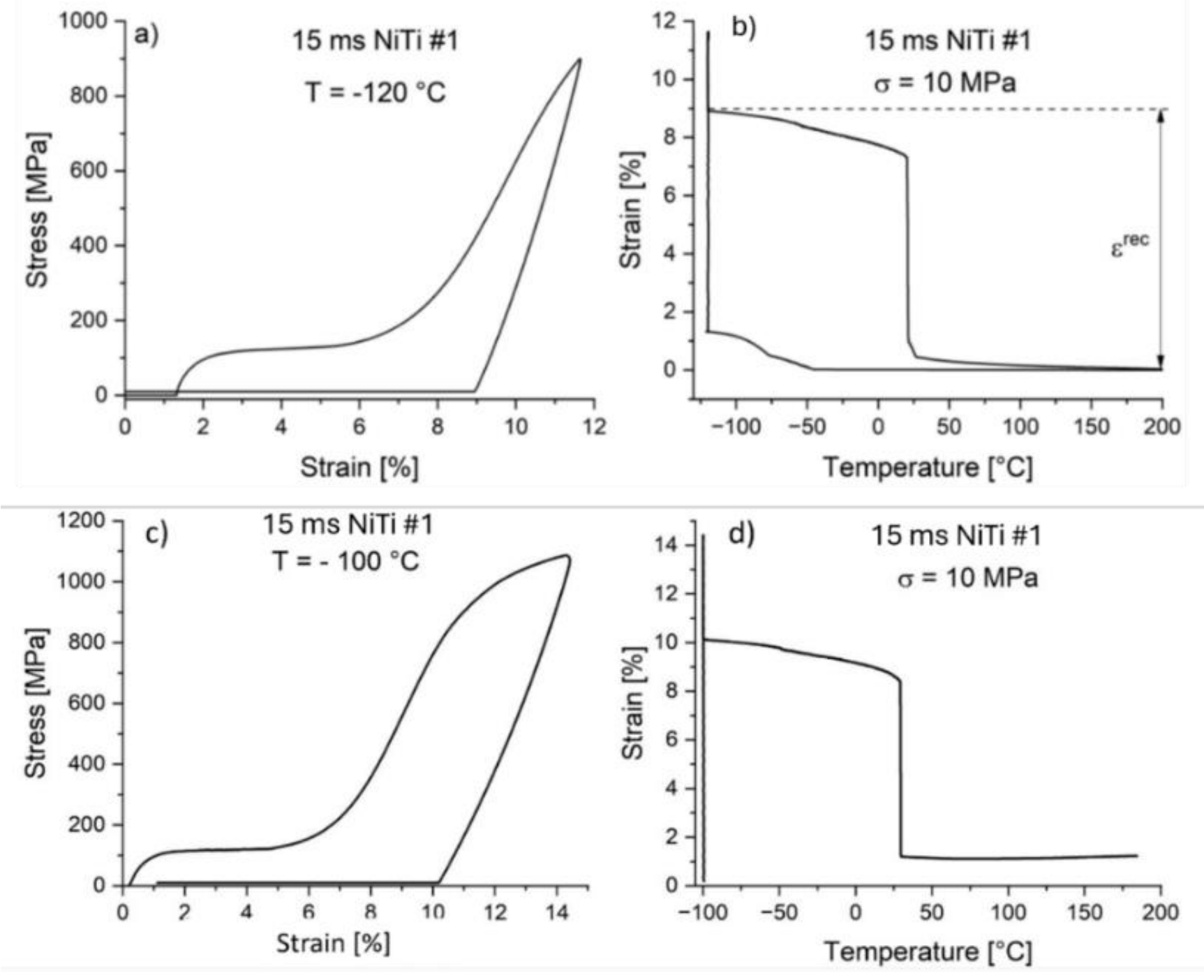

**Figure 11: Stress–strain–temperature response of superelastic 15 ms NiTi #1 wire in two shape memory tests** involving tensile deformation up to the kwinking stress (a,b) and beyond (c,d), unloading, and stress-free heating up to 200 °C. The imposed strain was fully recovered when the wire was deformed in the low temperature martensite state up to the kwinking stress (900 MPa) and up to 12% strain (a,b). When the wire was deformed beyond the kwinking stress (900 MPa; up to 14% strain (c,d)), recovered strain increased up to 13%, but plastic strain of 1% appeared in a closed-loop shape memory cycle. The recoverable strain is large because it involves anisotropic thermal distortion (~2%).

Another important result obtained from the experiment in Fig. 11 is that plastic deformation by kwinking at low temperatures does not limit the recoverable strains, it only generates irrecoverable plastic strain. When NiTi wires are deformed at elevated temperatures, plastic deformation limits recoverable strain [15] but this is not due to kwinking but due to a different TRIP-like deformation mechanism activated at elevated temperatures which may partially involve also plastic deformation of austenite.

Key questions are what defines the magnitude of kwinking stress, and what is the origin of its temperature dependence. Since kwinking deformation is a dislocation-slip-based process, the critical stress for activation of [100](001) slip in martensite naturally plays significant role. This explains why chemical composition, ternary alloying, local chemical heterogeneity and $Ti_3Ni_4$ precipitates affect the kwinking stress. The strong dependence of kwinking stress on grain size is via common Hall-Petch strengthening. Since kwinking stress is nearly strain rate independent, the observed near-linear temperature dependence of kwinking stress is unlikely to be due to thermal activation, as is common for conventional dislocation slip. The critical stress for [100](001) dislocation slip is, however, very low in NiTi [41] as it enables generation of incremental plastic strains accompanying MT under stresses below 500 MPa and the kwinking stress often exceeds 1GPa in superelastic NiTi wires [13]. So why the kwinking stress is so high and where its temperature dependence comes from?

As concerns the relation between the magnitude of kwinking stress and the critical stress for [100](001) dislocation slip in martensite, there are two problems. First is that we do not know how to evaluate the critical stress for [100](001) dislocation slip in martensite. Second is that kwinking stress is the threshold stress required to initiate highly localized deformation via coordinated dislocation slip that occurs due to shear instability in the B19′ lattice . Schmid law cannot be used to link the kwinking stress directly to critical stress for [100](001) dislocation slip in martensite. Similarly to the kinking stress, kwinking stress is likely controlled by the shear stress and the frictional influence of transverse stress. As concerns the temperature dependence of kwinking stress, results of our experiments show that it decreases with increasing temperature. We assume that this is because the critical stress for [100](001) slip in martensite depends on lattice parameters and elastic constants of the martensite (response of the lattice to shearing on (001) plane depends on temperature). The experimentally observed near linear temperature dependencies of lattice parameters [66] and elastic constants [65,68] of B19' martensite explain why kwinking stress decreases linearly with increasing temperature. There is a widespread empirical rule in the SMA field saying that NiTi shape memory elements must be deformed in martensite at temperatures as low as possible, if a fracture or plastic deformation shall be avoided. This empirical rule in fact describes strengthening of martensite via suppressing dislocation slip in it by decreasing temperature.

In summary, kwinking stress is the maximum tensile stress that can be applied in thermomechanical loads on NiTi wire without the danger of generating large plastic strains (in case of soft annealed NiTi wires) or without the danger of fracture (in case of strengthened NiTi wires). Plastic deformation by kwinking at low temperatures does not limit recoverable strains, only generates plastic strains. Kwinking stress can be increased by increasing the critical stress for [100](001) slip in martensite by adjusting chemical composition, ternary alloying or low temperature aging to control local chemical heterogeneity/$Ti_3Ni_4$ precipitates. Kwinking stress decreases with increasing temperature because the critical stress for [100](001) slip in martensite depends of lattice parameters (anisotropic thermal distortion) and/or elastic constants (resistance to shear on (001) plane) of B19' martensite that are temperature dependent. In addition, kwinking stress can be increased by manipulating the austenitic microstructure via post processing treatments such as: i) cold work/annealing to control grain size (Hall-Petch strengthening) and precipitates (precipitation strengthening), ii) plastic deformation to refine the microstructure (Hall-Petch strengthening) and introduce permanent lattice defects (strain hardening).

## 6. Martensite Variant Microstructures in Plastically Deformed NiTi

To investigate deformation mechanisms activated in NiTi polycrystals subjected to thermomechanical loading tests, we reconstructed martensite variant microstructures in whole grains created by MT proceeding under external stress [10,12] and/or by plastic deformation [24,25]. In addition, we also analyzed permanent lattice defects in grains of the austenite phase of plastically deformed and heated NiTi wires [14,45]. The TEM based methods used to reconstruct the martensite variant microstructures are described in detail in this chapter since they enabled the discovery of kwinking.

Since the grains in nanocrystalline NiTi wires are small (d<500 nm) and martensite variants form on nanoscale (LCV variants are 10-100 nm thin plates), martensite variant microstructures in deformed nanocrystalline NiTi wires cannot be analyzed by EBSD. Therefore, we have applied nanoscale orientation mapping in TEM to reconstruct martensite variant microstructures in grains [10,12,27] of deformed NiTi wires (Figs. 1a, 12,14).

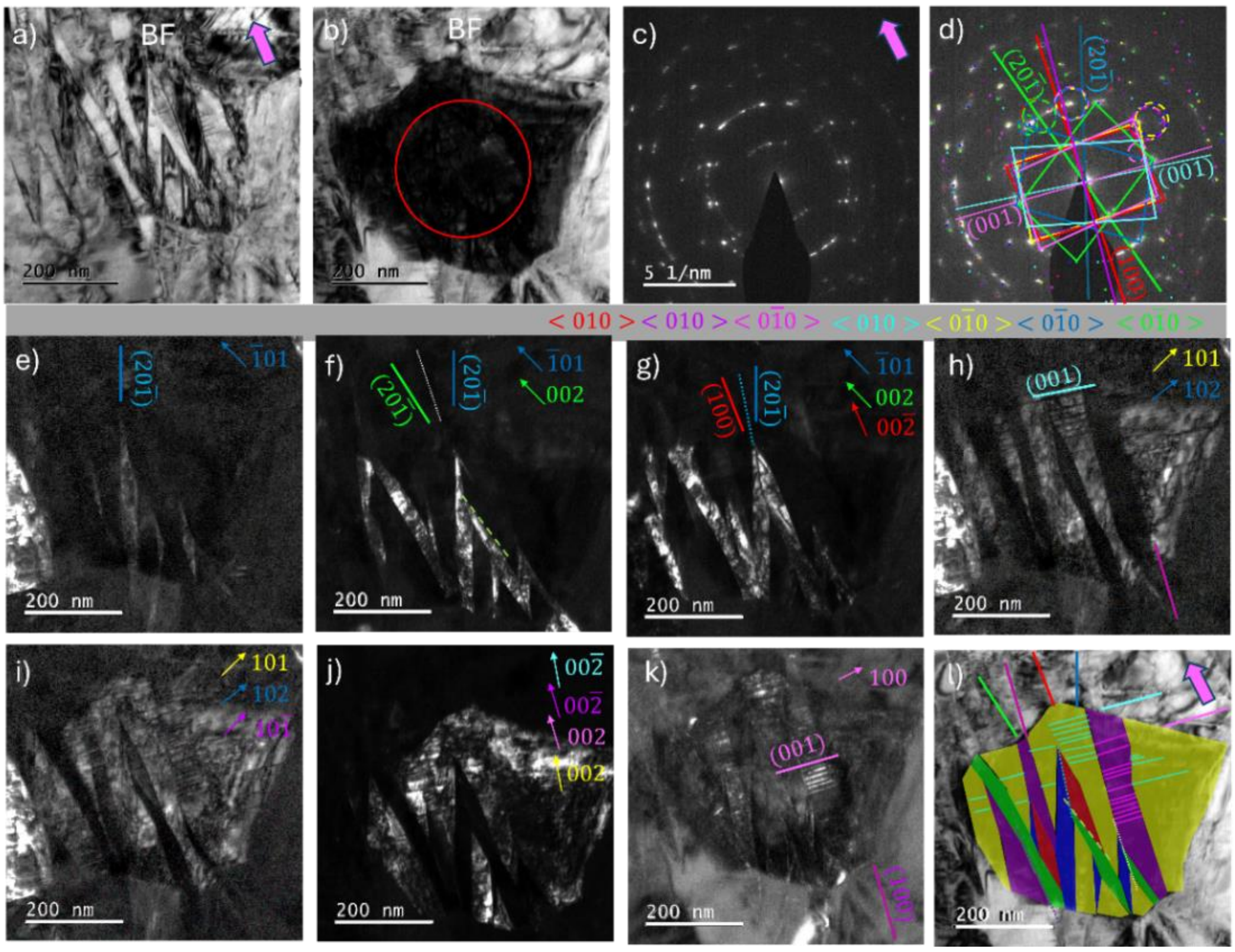

**Figure 12: Martensite variant microstructure in 15 ms NiTi #5 shape memory wire deformed in the martensite state at 20 °C up to 15% strain.** The martensite variant microstructure was reconstructed by SAED-DF method [27]. Although the selected grain is filled with deformation bands (a), it becomes completely dark when oriented into low index <010> zone (b). The electron diffraction pattern (c) was indexed to 7 martensite lattices oriented in the <010> zone denoted by different colors (d). The DF images (e,f,g,h,i,j,k) show the size and location of individual martensite lattices and traces of twin interfaces within the grain. The indexed composite diffraction pattern (d) and martensite variant map (l) represent the reconstruction of the martensite variant microstructure in the grain consisting of (001) compound twins (cyan, magenta), (100) deformation twins (red), (100) kwink band (violet) and (20-1) deformation twins (blue, green) within the martensite matrix (yellow). All denoted interfaces are parallel to the electron beam.

Due to the peculiar electron diffraction contrast from multiple variously oriented and frequently overlapping (001) compound twinned martensite variants, grain boundaries can be hardly distinguished in BF TEM images of martensitic microstructures in stress-free cooled wire (Fig. 2c). When nanoscale orientation mapping is employed (see Fig. 14), this becomes better, although individual martensite variants still cannot be resolved. However, when the wires were deformed beyond the end of reorientation plateau (Fig. 2d), the grain boundaries suddenly showed up, since individual grains were filled with single martensite laminates. As the laminates in neighboring grains are variously inclined to the electron beam, the TEM lamella still must be tilted to orient the grain of interest into low index zone [h,k,0] to resolve individual twins within the laminate (Fig. 2d).

When the kwinking stress was reached and the wire was deformed plastically, multiple deformation bands appeared in all grains (Fig. 2e). The bands were arranged in a special deformation geometry - all crystal lattices within a grain shared common [010] direction and all intermartensitic interfaces lied within the common [010] zone. Therefore, the microstructures in grains could be analyzed in grains oriented in the [010] zone [25,27]. This special deformation geometry is evidenced by dark BF images of grains containing many deformation bands oriented in the [010] zone (Fig. 12b). The composite electron diffraction pattern (Fig. 12c) in the [010] zone contains information on orientations of the crystal lattices in all martensite domains within the SAED area. DF images of the grain taken using various diffraction spots (Fig. 12e-k) are used to evaluate the size and location of individual martensite domains and traces of interfaces within the grain. Figs. 12d,l represents the SAED-DF reconstruction of martensite variant microstructures in whole grains of plastically deformed NiTi.

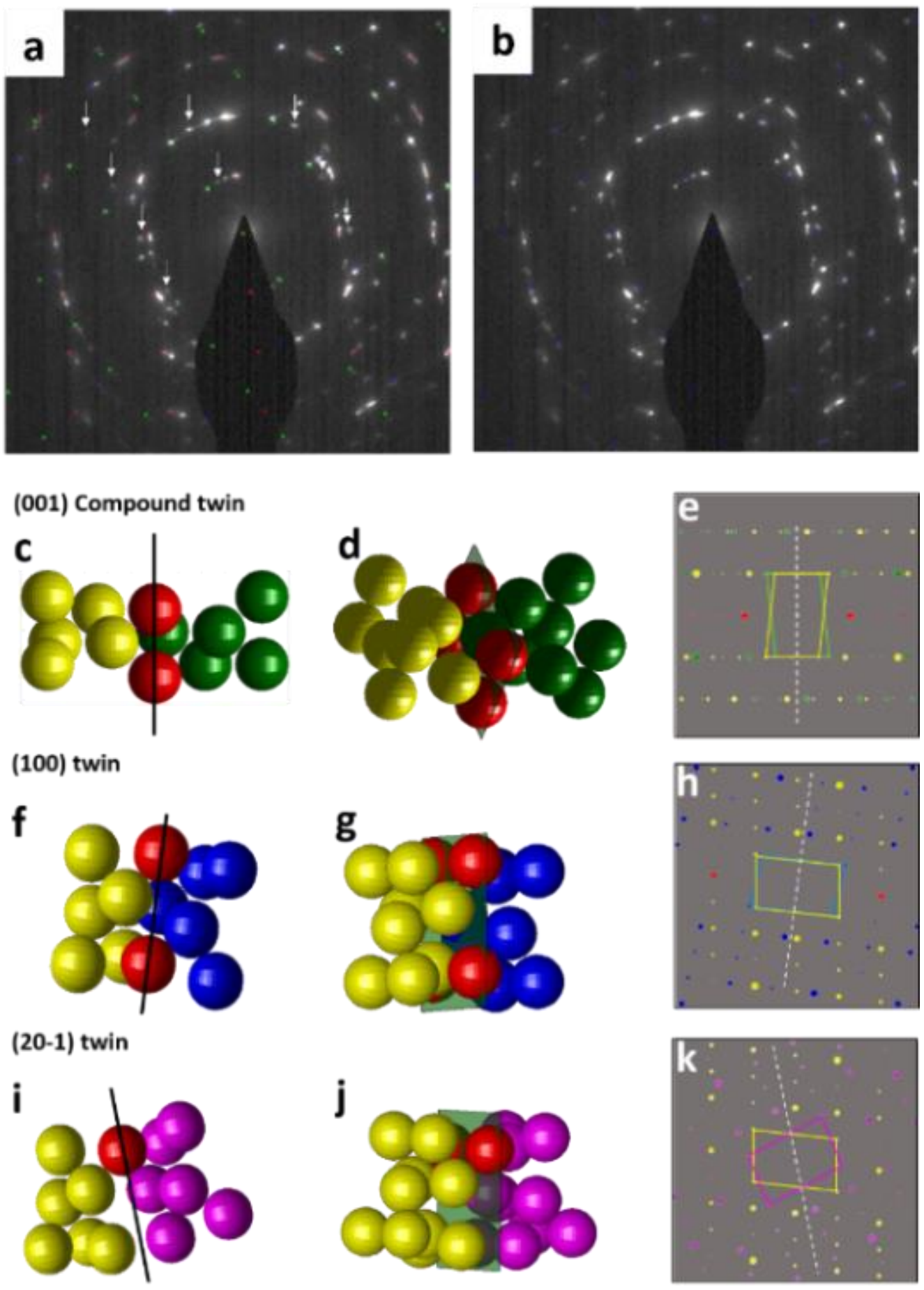

**Figure 13: Indexation of composite electron diffraction patterns from martensite variant microstructure in plastically deformed NiTi wire** [27] **using the software CrysTBox.** Up to 3 reciprocal lattices in the experimental electron diffraction pattern (a) can be automatically indexed using the Ransac multimodal settings of the diffractGUI tool of CrystBox (colored spots). Additional reciprocal lattices must be found by further manual assignment of additional Bragg spots to respective reciprocal lattices (additional spots denoted by arrows in (a) are colored blue in (b)). As the interfaces are parallel to the <010> zone axis, interface plane and misorientation can be evaluated. Three basic interfaces: (001) compound twin (c,d,e), (100) twin (f,g,h) and (20–1) kwink (i,j,k) observed in microstructures of the plastically deformed NiTi wire are visualized in 〈010〉 zone orientation (c,f,i) and in general orientation (d,g,j). Red spheres in c,d,f,g,i,j denote atoms shared by both adjoining lattices. The simulated composite electron diffraction patterns are shown for each interface in the <010> low index zone (e,h,k). Red spots in simulated diffraction patterns (e,h,k) denote the shared lattice plane normals - twinning plane.

A key step of the SAED-DF method is indexing the composite electron diffraction patterns from multiple martensite variants within grains. This is performed using the software CrysTBox [71], as demonstrated on Fig. 13. Up to three reciprocal lattices in the experimental electron diffraction pattern (Fig. 13a) are automatically indexed using CrysTBox (colored spots). Additional reciprocal lattices must be identified by manual assignment of further Bragg spots to the respective reciprocal lattices (additional spots denoted by arrows in Fig. 13a are colored blue in Fig. 13b). The interfaces within the grain are analyzed using CrysTBox as shown in Fig. 13c–j. See the methodological article [27] for more detailed information on the experimental methods used to reconstruct martensite variant microstructures in whole grains of plastically deformed NiTi.

However, tilting a generally oriented grain in the TEM lamella into the [010] zone (a single direction in the monoclinic structure) is not always possible. And even if grains are oriented, manual reconstruction of martensite variant microstructures in grains containing large number of kwink bands is laborious. This makes the SAED-DF method impractical for mapping martensite variant microstructures across multiple neighboring grains in plastically deformed NiTi wires. Therefore, we decided to employ nanoscale orientation mapping in TEM (Figs. 14,15,17), which is automatic, and which is in principle applicable to map martensite variant microstructures in generally oriented grains [27]. To demonstrate the application of this method to the investigation of the mechanism of plastic deformation of martensite by kwinking, we show below three examples of the ASTAR reconstruction of martensite variant microstructures in grains of 15 ms NiTi #5 shape memory wire plastically deformed at 20 °C up to 15% strain in tension (Figs. 14,15) and results of the reconstruction of deformation bands in austenite of superelastic 16ms NiTi #1 wire deformed at −30°C up to 18% strain, unloaded and stress-free heated up to 150°C to retransform all martensite to the austenite (Fig. 17).

First example (Fig. 14) shows martensite variant microstructures within multiple grains of plastically deformed NiTi wire. Notice that individual grains, which could not be resolved in BF images (Fig. 14b), can be clearly observed in ASTAR orientation maps of multiple grains (Fig. 14d,e,f). Since each grain in the wire deformed to 15% strain contains mainly martensite matrix (prevailing color in Fig. 14d,e,f) and only few scattered kwink bands, this is sufficient for the nanoscale mapping to evaluate grains orientation and GBs. Identification of individual kwink bands (Fig. 14g,h) and (100) twins (Fig. 14i) within grains is, however, not achieved.

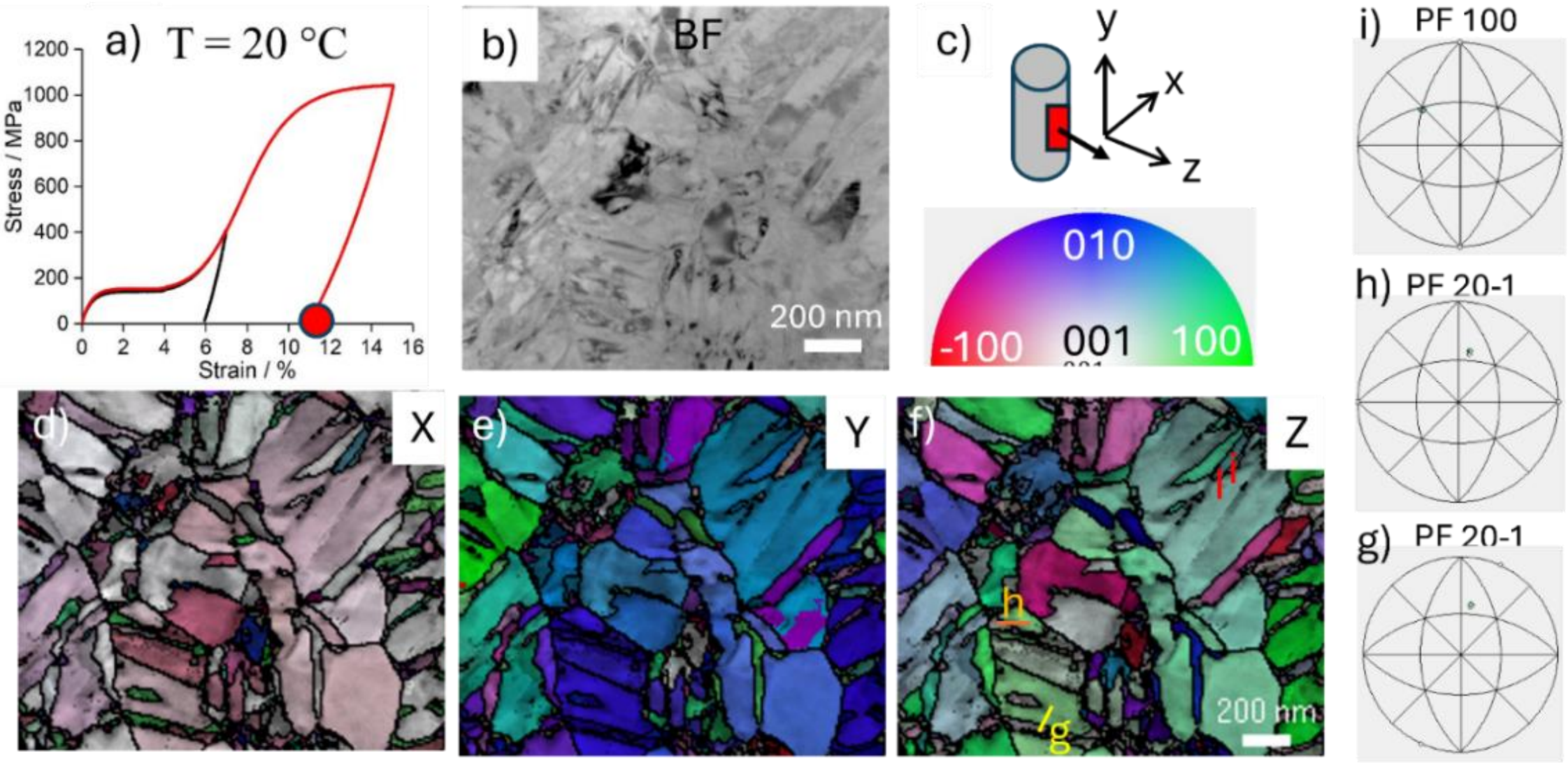

**Figure 14: Martensite variants in multiple grains of 15 ms NiTi #5 shape memory wire evaluated by nanoscale orientation mapping ASTAR.** The wire was deformed in a tensile test at 20 °C up to 15% strain. a) stress-strain curve, b) BF TEM figure of martensite variants in multiple grains, c) coordinate system and color scale for ASTAR mapping, ASTAR maps of x-(d), y-(e) and z-(f) directions. g-i) pole figures of deformation bands denoted in (f) are evaluated as (20-1) and (100) kwinks.

Second example shows results of the reconstruction of martensite variant microstructure within a single grain (Fig. 15) in the same wire. The three ASTAR orientation maps of x-,y-,z- sample directions (Fig. 15f,g,h) represent the reconstructed martensite variant microstructure. Note that the z-direction (Fig. 15f) is perfectly aligned with [010] crystal directions in all martensite variants due to the special deformation geometry. The other two maps of x- and y- directions (Fig. 15g,h) show (100) twin bands and (20-1) kwink bands in martensite matrix in different colors. The ASTAR dataset can be post-acquisition analyzed to characterize all interfaces by pointwise diffraction. virtual bright field (VBF) and virtual dark field (VDF) images visualize the martensite domains and intermartensitic interfaces. Results of the analysis of the microstructure by VBF and VDF imaging are shown in Figs. 15i-l (matrix), Figs. 15m-p ((100) twins), and Figs. 15q-t (20-1) kwinks).

Very important information on the kwinking deformation was obtained by HRTEM analysis of the newly created intermartensitic interfaces within these microstructures (Fig. 16) [24]. A grain containing large number of kwink bands (Fig. 16a) was oriented into [010] zone and the composite diffraction pattern was recorded and indexed (Fig. 16b). Individual diffraction patterns were associated with the kwink bands to characterize selected intermartensitic interfaces (Fig. 16c). HRTEM figures and FFT patterns (Figs. 16d-o) are used to characterize the boundary plane and lattice misorientation.

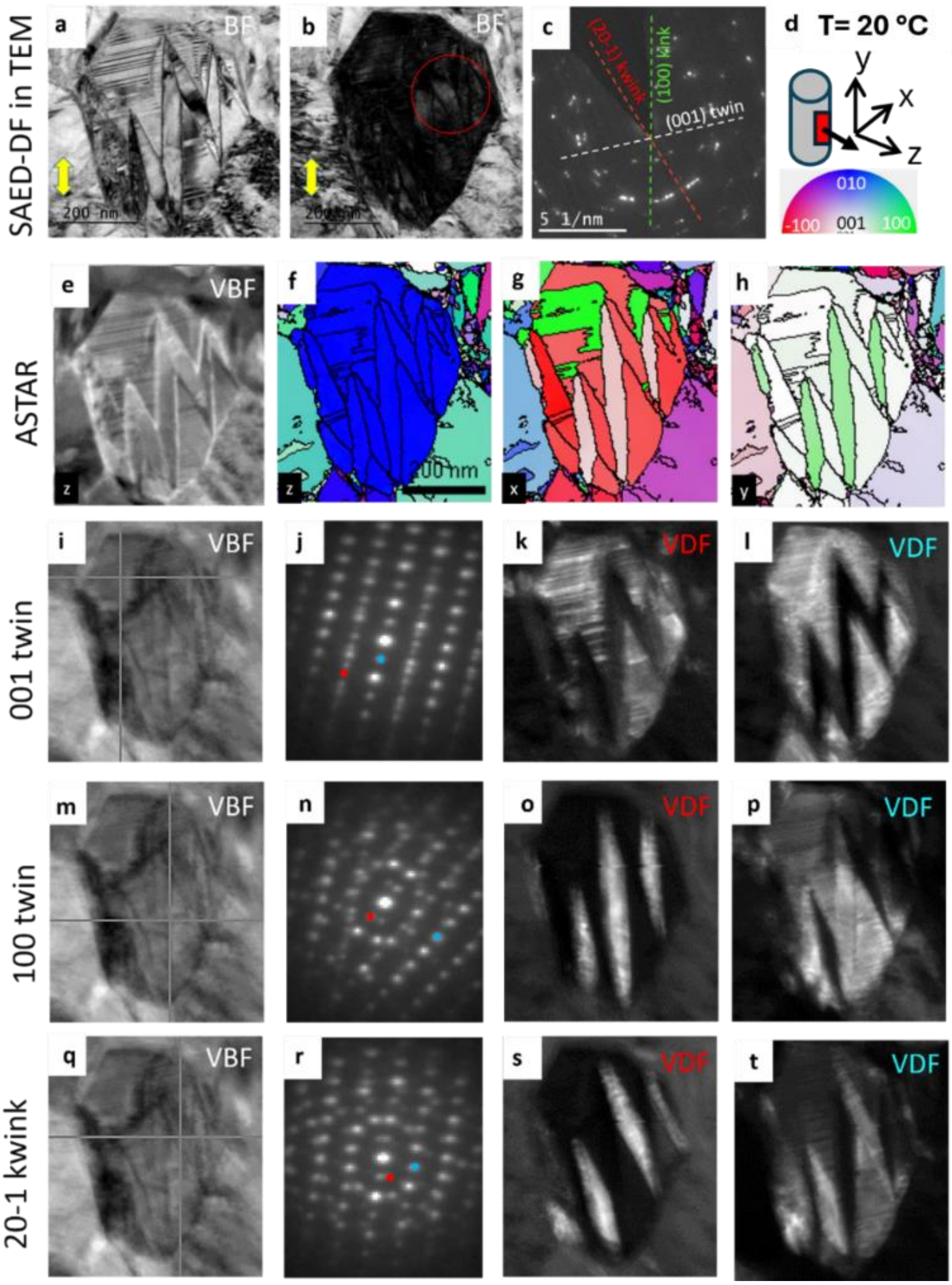

**Figure 15: Reconstruction of martensite variant microstructure within a grain of 15 ms NiTi #5 wire deformed at 20 °C up to 15% strain in tension by nanoscale orientation mapping ASTAR [27].** Upper row shows BF TEM image of the selected grain in general orientation (a) and in < 010> low index zone (b). Diffraction pattern (c) was taken from the area denoted in (b). Second row shows the ASTAR orientation maps of the grain along the z-axis (f), x-axis (g) and y-axis (h) aligned with the electron beam. The maps are colored according to the IPF scale in (d). Figures (i-t) show results of the point wise analyses of the ASTAR dataset revealing martensite variants and interfaces using VDF images taken using the reflections colored in diffraction patterns (j,n,r). The electron diffraction patterns (j,n,r) were taken from locations denoted by crosses in (i,m,q).

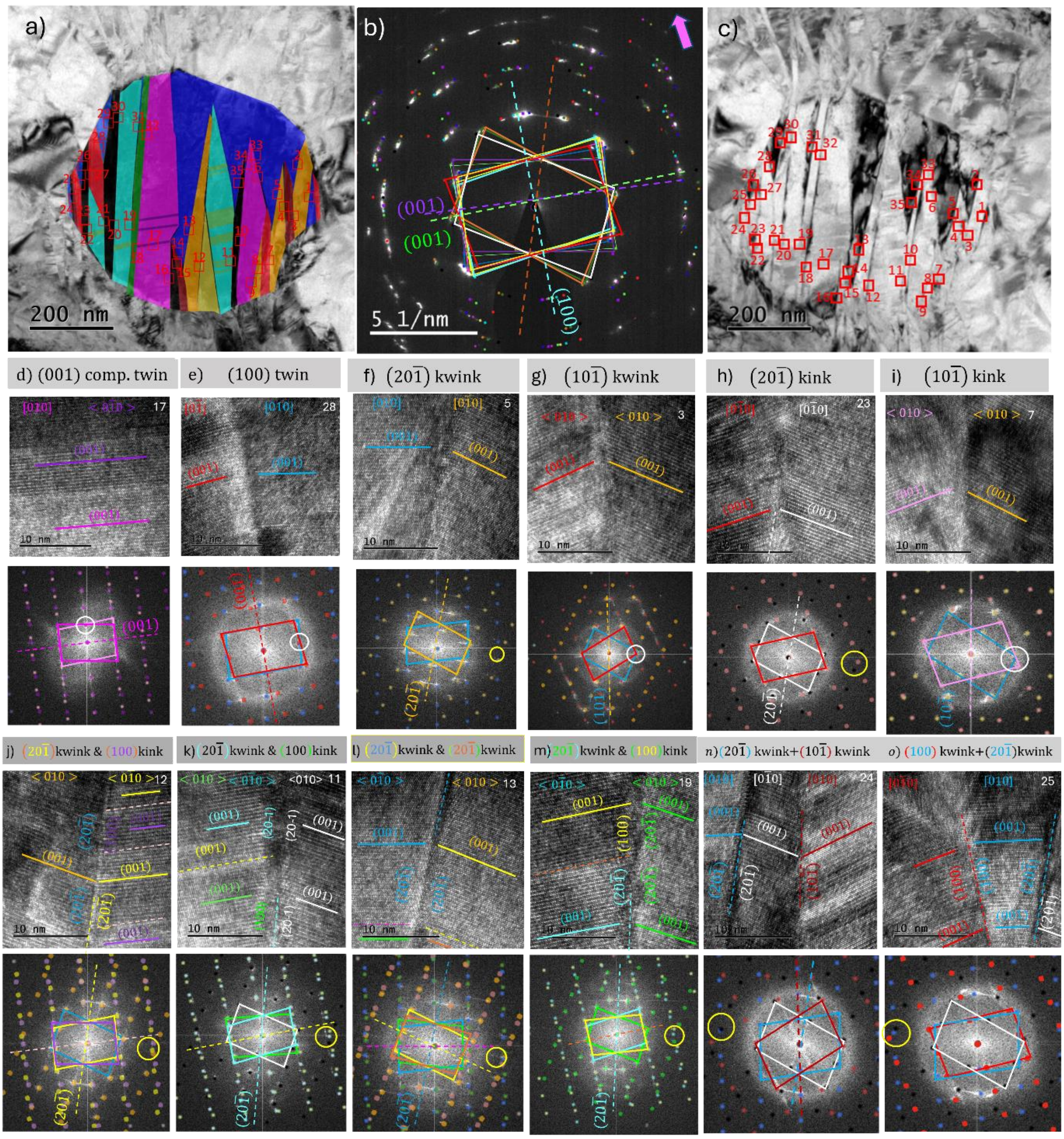

**Figure 16: Kwink gallery – examples of kwink and kink interfaces within the microstructure of a NiTi wire deformed in the martensite state at 20 °C up to 15% strain [24].** A grain with typical kwink band microstructure was oriented into the [010] zone (a), and a diffraction pattern (b) was taken from the SAED area covering the whole volume of the grain. Selected kwink and kink interfaces denoted in (c) were analyzed by HRTEM (d–o). The upper image shows the interface and the lower image the corresponding FFT patterns. As there are variously oriented lattices within some deformation bands, some interfaces are more complex (j–o). Color parallelograms in FFT patterns denote multiple misoriented crystal lattices. Solid lines in HRTEM images denote traces of (001) planes, dashed lines traces of interfaces, and circles around FFT spots denote interface plane normals. Note that the interfaces are not atomically sharp, do not lie exactly on twinning planes, and lattice misorientations do not correspond exactly to twinning misorientation. Individual kwink and kink interfaces are characterized by lattice misorientation and interface plane normals. Notice the symmetry of (001) lattice planes across the interface fulfilled by all interfaces as predicted by the kwinking theory.

The (001) compound twin (Fig. 16d) and (100) deformation twin (Fig. 16e) are basic interfaces formed by MT and deformation twinning, respectively. The other intermartensitic interfaces on (20-1), (10-1), (100) planes within the [010] zone are denoted as kwink interfaces (Fig. 16f,g) and kink interfaces (Fig. 16h,i). While kwink interfaces display a near mirror symmetry of adjoining crystal lattices, the kink interfaces separate crystal lattices that are rotated in such a way that only traces of (001) slip planes are mirror symmetric across the interface. This would be characteristic for kinking deformation mediated by dislocation slip on (001) plane. The interface planes are not atomically sharp, do not lie exactly on low index twinning planes and lattice misorientations do not correspond to twinning misorientations exactly. The kwink and kink interfaces are often combined (Fig. 16j-o), the traces of (001) slip planes are always mirror symmetric across interfaces, which is another indirect evidence that these interfaces are formed by plastic slip on (001) planes.

All this evidence brought us to the conclusion that these intermartensitic interfaces were not created by the martensitic transformation, conventional dislocation slip and/or deformation twinning in martensite but that they were created by the kwinking deformation mechanism involving coordinated dislocation slip based kinking combined with (100) deformation twinning described in Chapter 3.

When plastically deformed NiTi wires were unloaded and stress-free heated above the $A_f$ temperature, the martensite within grains transformed to the austenite. While the martensite matrix transforms to the same parent austenite from which it was created, the martensite within (20-1) kwink bands transforms into differently oriented deformation bands within the austenite phase (since crystal lattices are rotated, the band interfaces are STGBs) In addition to deformation bands, there is a high density of slip dislocations in austenite, gradients of lattice orientations and internal stress within austenite grains [14].

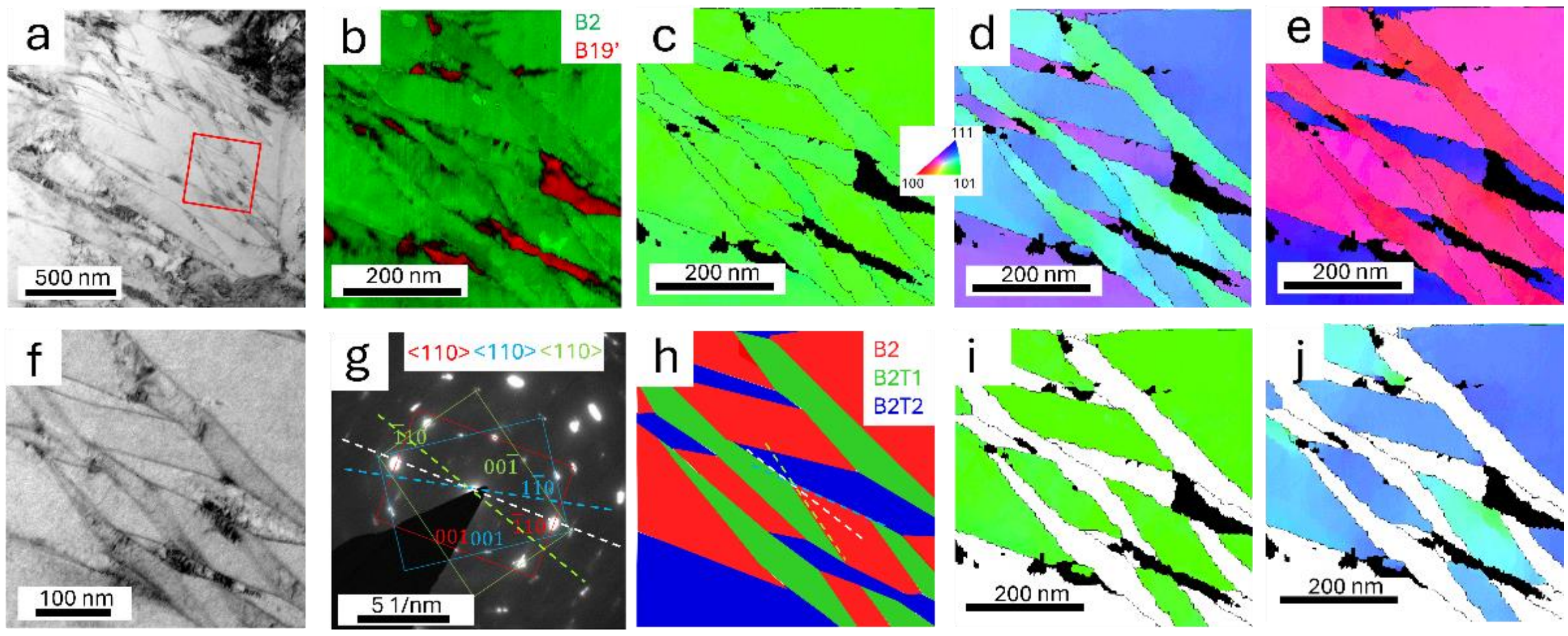

**Figure 17: Reconstruction of austenitic microstructure in superelastic 16ms NiTi #1 wire deformed at −30 °C up to 18% strain by nanoscale orientation mapping ASTAR [70].** The wire was deformed at −30 °C up to 18% strain, unloaded and stress-free heated up to 150°C, cooled to -30°C and heated to room temperature (Fig. 19h). a) BF image of a grain with multiple deformation bands. The red square area was selected for the reconstruction. The grain was oriented into [011] low index zone and orientation maps (c,d,e) showing orientations of the austenite lattice in z-direction (c), x-direction (d), and g) y-direction (e) were evaluated. The phase map (b) shows austenite with only few residual martensite particles. The BF image of the reconstructed area in general orientation (f) shows traces of interfaces denoted also in electron diffraction pattern (g) from the scanned area. The sketch of the austenitic microstructure (h) shows the B2 matrix and two {114} austenite twins B2T1 and B2T2. While the ASTAR orientation map of the matrix in z-direction (i) shows perfect alignment of the crystal lattice with the z-axis everywhere in the grain, the map in x-direction (j) shows orientation gradients.

Because the microstructures observed in plastically deformed superelastic NiTi were often misinterpreted in the past [15-20,22,30-35,37-39], we present below experimental results proving that the austenite deformation bands observed in plastically deformed superelastic NiTi originate from the kwinking deformation in martensite. The austenitic microstructures in plastically deformed superelastic NiTi wires were analyzed by nanoscale orientation mapping in TEM (Fig. 17), the band interfaces in austenite were analyzed by HRTEM (Fig. 18), and the inheritance between the martensitic and austenitic bands is discussed below using Fig. 19.

The key experimental fact is that deformation band arrangements in austenitic grains display a special geometry [70] which is derived from the special geometry of martensitic microstructures [27] described above. The austenite lattices within the matrix and deformation bands in each grain are rotated around a common [011]A axis (single [011] direction out of the six <011> directions available in the B2 structure). This particular [011]A direction is lattice correspondent and exactly parallel to [010]M direction, along which the martensite crystal lattices in plastically deformed martensite were aligned. The austenitic microstructures in austenite grain oriented in this [011]A zone can be reconstructed by the manual SAED-DF method as well as nanoscale orientation mapping (Fig. 17). Since the B2-B19' MT shortens the crystal along the [010]M direction, the sought [011]A zone axis is always oriented near perpendicularly to the wire axis. Therefore, TEM lamellae must be cut from deformed wires with lamella plane normals oriented perpendicularly to the wire axis.

Results of the nanoscale orientation mapping of a wedge austenitic microstructure oriented in [011] zone are shown in Fig. 17. The phase map in Fig. 17b shows that small amount of martensite kept in the austenitic microstructure after unloading and heating above the $A_f$ temperature. ASTAR orientation maps (Fig. 17c,d,e) show orientations of the austenite lattice in x-,y-,z-directions. The special geometry is like that of the martensite (Fig. 12). While the [011] austenite zone axis is perfectly aligned with the z-direction everywhere in the analyzed grain (Fig. 17c), the ASTAR maps of x- and y- directions (Fig. 17d,e) clearly show the mutually rotated crystal lattices within the matrix and deformation bands. The parent austenite grain is divided by two {114} austenite twins (B2T1 (green) and B2T2 (blue)) into multiple small crystal blocks of different orientations. The discontinuous lattice rotations lead to the refinement of the parent austenitic microstructure of the plastically deformed wire.

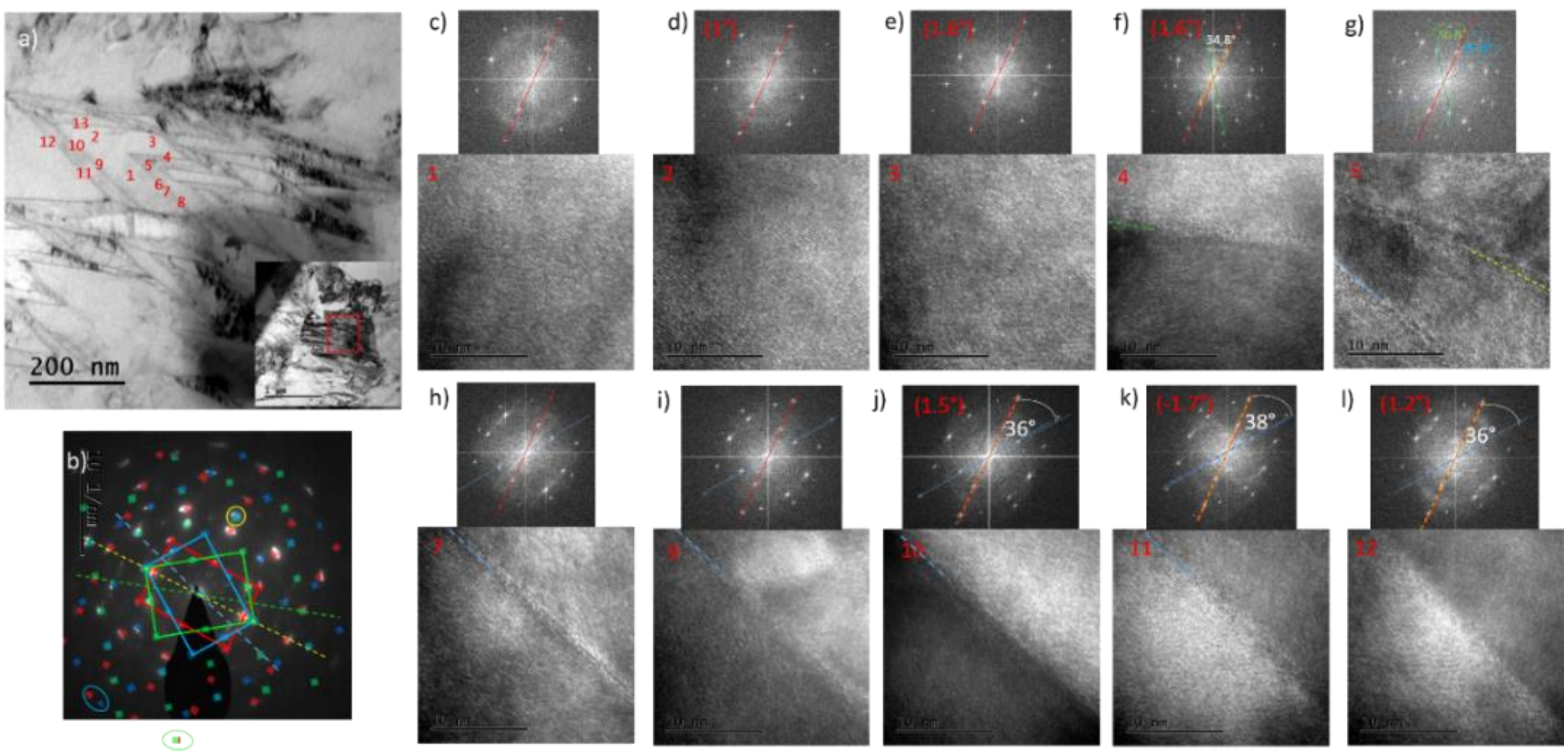

**Figure 18: Austenitic microstructure in 16 ms superelastic NiTi wire deformed at -30 °C up to 18% strain, unloaded and stress-free heated above the $A_f$ temperature** (Fig. 19h). a) BF TEM image of the wedge austenite microstructure containing {114} austenite twin bands. The grain was oriented into <011> zone and electron diffraction pattern (b) was taken and indexed as corresponding to the austenite matrix (red) and two austenite {114} twin bands (blue, green). HRTEM images (c-l) were taken from various locations numbered in (a). Red lines in FFT patterns denote orientations of the (110) plane of the matrix in given locations. Blue and green lines in FFT patterns denote orientations of the (110) plane with the twin. The orientation of the austenite matrix as well as lattice misorientation across interfaces vary from place to place within the microstructure. This disregards the possibility that the microstructure was created by deformation twinning in austenite.

While the crystal lattice of the matrix is perfectly aligned with the z-direction (Fig. 17i), it is locally misoriented in the x-direction (Fig. 17j) as well as in the y- direction. These added continuous lattice rotations were brought about by dislocation slip in the martensite (for more detailed information on the continuous lattice rotations see Figs. 12-15 in [14]). The refinement and rotations of crystal lattice in martensite caused by the kwinking in martensite (Fig. 12d) give rise to characteristic azimuthal broadening of austenite diffraction spots (Fig. 17g) regularly observed in TEM studies of plastically deformed NiTi [14,31-35].

To further support the view that the deformation bands were not created by {114} deformation twinning in austenite, we performed HRTEM analysis of austenitic interfaces (Fig. 18) in plastically deformed superelastic NiTi#1 wire. The obtained results clearly show that the observed interface planes are not sharp and do not always lie on {114} plane as typical for {114} deformation twinning and lattice misorientations across the interface do not exactly correspond to {114} austenite twin and vary from place to place within the analyzed microstructure. Nevertheless, all interfaces were STGBs (traces of interface planes bisect the angle between {011} planes).

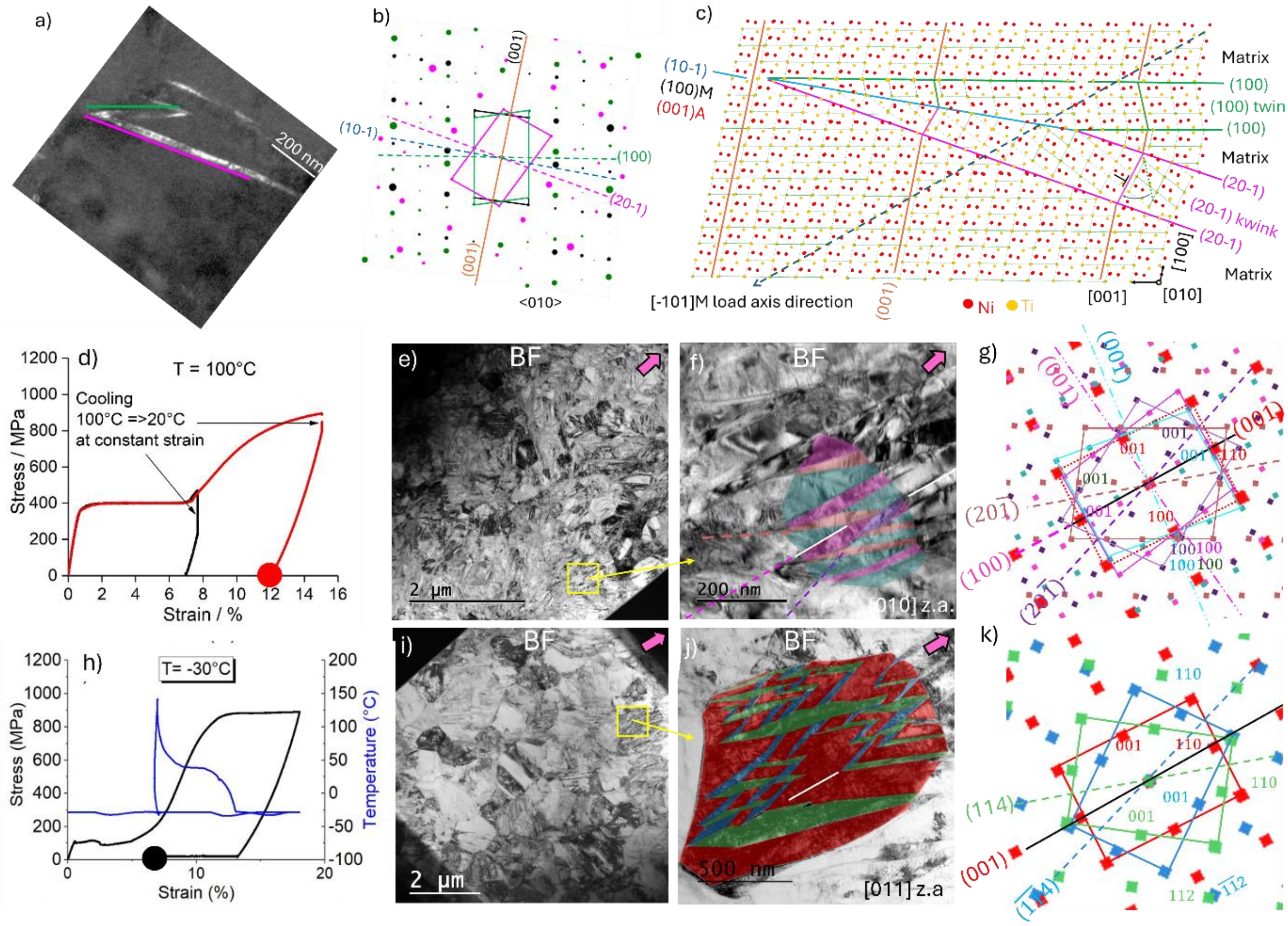

**Figure 19: Comparison of martensite variant microstructures (d-g) with austenitic microstructures (h-k) in plastically deformed NiTi wires.** The (20-1) kwink bands often form wedges with (100) twin bands. DF image of a single wedge (a), SAED pattern (b) and atomic model (c) describe the wedge configuration. The martensite variant microstructure in 15 ms NiTi #5 shape memory wire (d-g) was created by kwinking deformation in tensile test at 100 °C up to 15% strain, followed by cooling and unloading [25]. The austenitic microstructure in superelastic NiTi#1 wire (h-k) was created by kwinking deformation in tensile test at -30 °C up to 18% strain, unloading, heating up to 150 °C, cooling to -30 °C and heating to 20 °C [15]. The martensite variant microstructure (f) in a grain oriented in <010> zone (c,d) contains martensite matrix (cyan), (100) deformation twin (magenta) and two $(20\overline{1})$ kwink bands (brown and violet). The red reciprocal lattice (g) corresponds to the parent austenite lattice in the $[1\overline{1}0]$ zone from which all martensite variants within the grain were created. The austenitic microstructure (i) in grain oriented in [110] zone contains austenite matrix (red) and two sets of {114} bands (blue and green). The black lines in (g,k) denote the (100) martensite twin interface (formed by magenta & cyan matrix lattices) parallel to the (100) basal plane of the parent austenite. The comparison explains how the {114} deformation bands in austenite are created from two (20-1) kwink bands.

Finally, let us discuss the link between the martensitic and austenitic wedges often observed in martensitic and austenitic microstructures of plastically deformed NiTi. The (20-1) kwink bands in martensitic microstructures either stretch across whole grains or form wedges with (100) twin bands (Fig. 19a,b,c). The wedges are key elements of kwink band microstructures (see Figs. 12-16). Notice the rotation of the (001) plane through the wedge microstructure evidencing that the shape strain introduced by the (100) twinning is compensated by the shape strain of the (20-1) kwink.

The wedge microstructures in plastically deformed martensite and austenite are mutually compared in Fig. 19d-k. Since the martensitic wedges (Fig. 19f) are sharper than the austenitic wedges (Fig. 19j), the latter cannot be directly inherited from the former (another reason is that the lattice within (100) twin retransforms to the parent austenite lattice). When the reverse MT upon heating occurs, the martensite matrix (cyan) and the (100) deformation twin (magenta) transform to the parent austenite lattice they were created from, while the brown (20-1) primary twin and violet (20-1) secondary twin (Fig. 19f) transform into austenite twin bands (Fig. 19j) containing austenite lattice rotated from the austenite matrix (STGB on near {114} plane with lattice misorientation 38.9° around <110> axis called {114} austenite twins). The secondary (20-1) kwink band (violet) originates from the lattice within the (100) twin. The {114} austenite deformation bands thus appear in the austenitic microstructure because the heavily slipped and rotated crystal lattices within the primary and secondary (20-1) kwink bands (Fig. 19c) undergo reverse MT into rotated austenite lattices.

The {114} austenite planes are lattice correspondent to (20-1) martensite plane [25]. The inheritance (correlation) between the (20-1) kwink interfaces in martensite and {114} interfaces in austenite was proven experimentally by Ii et al. [30] who heated deformed NiTi alloy in TEM (see Fig. 2 in [30]). As the authors considered deformation twinning mechanisms (Table 2 in [30]), they arrived at the conclusion that twinning shear of (20-1) martensite twin (s=0.42) increases during the reverse MT up to the twinning shear of {114} austenite twin (s=0.707). Of course, this is highly unrealistic, kwinking deformation describing these interfaces as created by coordinated dislocation slip and twinning (Fig. 3b) (i.e., no {114} deformation twinning exists in NiTi) provides realistic explanation.

In summary, martensite variant microstructures in whole grains of plastically deformed nanocrystalline NiTi wires were reconstructed by nanoscale orientation mapping in TEM. The microstructures consist of deformation bands arranged in a special deformation geometry (twin misorientations and lattice rotations around $[010]_M$ axis, interface planes within the $[010]_M$ zone). Although (20-1) kwink bands were most frequently encountered in these microstructures, many of the observed interfaces were neither transformation nor deformation twins, did not show exact twin misorientations and twin planes, there was no mirror symmetry across the interface, crystal lattices were often only rotated. While deformation twin planes, as e.g. (100) deformation twin, are atomically sharp, kwink interfaces are nonplanar, atomically blurred on nanoscale and often consist of kwink and kink segments. When plastically deformed NiTi wires were unloaded and stress-free heated above the $A_f$ temperature, the martensite variant microstructures within grains transformed to plastically deformed austenite consisting of {114} deformation bands arranged also in a special deformation geometry (crystal lattices rotated around $<110>_A$ axis, STGBs as band interfaces). The analyzed interface planes were not sharp, did not always lie on {114} plane, lattice misorientations did not exactly correspond to {114} austenite twin and varied from place to place within the analyzed microstructure. Based on this experimental evidence, we have concluded that these microstructures (newly introduced interfaces) were not created by conventional dislocation slip in martensite or deformation twinning. They were created by kwinking deformation involving dislocation slip based kinking assisted by (100) deformation twinning. As there is no other alternative mechanism that can explain the experimentally observed special deformation geometry, kwink bands and new interfaces in martensite and STGBs in austenitic microstructure of plastically deformed NiTi, these features can be used as an indicator that the studied NiTi alloy deformed plastically via kwinking and not by alternative deformation mechanisms previously reported in literature.

## 7. Localized Plastic Deformation of NiTi in Tensile Tests

In contrast to conventional plastic deformation via dislocation slip, which tends to be homogeneous on the mesoscale, plastic deformation of B19' martensite by kwinking in NiTi is inherently localized on the nanoscale (kwink bands in Fig. 12 are bands of localized plastic deformation). In addition, plastic deformation of NiTi can also be localized on the macroscale—for example, in necks formed in strengthened NiTi wires during tensile tests [21], in shear bands in compression tests [72], in Lüders band fronts propagating via kwinking deformation in martensite in tensile tests [48], or in Lüders band fronts propagating alongside forward or reverse stress induced MT in tensile tests (see  Chapter 9).

Nevertheless, annealed nanocrystalline NiTi wires deform plastically on macroscale in a homogeneous manner with strain hardening (Fig. 1d) since plastic deformation of the oriented martensite via kwinking is inherently accompanied by strain hardening via refinement of microstructure (Hall-Petch strengthening). The magnitude of kwinking stress and strain hardening coefficient depend on the conditions for dislocation slip in martensite. In case of solution annealed coarse grain NiTi wires, in which [100](001) dislocations in martensite glide easily, kwinking stress is low and strain hardening coefficient is large. In strengthened nanocrystalline NiTi wires, in which the [100](001) dislocation slip in martensite is largely suppressed, kwinking stress is high and strain hardening coefficient is low ( see stress-strain curves in Figs. 2d, 9).

When annealed NiTi wire deforms plastically in tensile test (Fig. 2), kwinking deformation gradually fills polycrystal grains with kwink bands in the strain range 12-30%. Beyond 30% strain, plastic deformation continues mainly by secondary kwinking and dislocation slip within the kwink bands until fracture. Kwinking deformation in tensile tests results in characteristic evolution of martensite texture, accumulation of permanent lattice defects, and microstructure refinement, as will be further elaborated in Chapter 8. The evolution of microstructure and texture is most pronounced in the early strain range 12-30%, in which kwinking deformation proceeds massively. It was found [40] that plastic deformation of the superelastic 16ms NiTi #1 wire up to ~15% strain in tensile test at -30 °C, unloading and heating creates unique austenitic microstructure with optimized fraction of deformation bands giving rise to oriented tensile internal stress within the austenite matrix. The internal stress within the microstructure created by kwinking represents the origin of the large and stable TWSME observed during the subsequent thermal cycling of this wire (see Figs. 5-7 in Ref. [40]).

On the other hand, many of the strengthened cold drawn/heat treated nanocrystalline NiTi wires fracture systematically at ~12 % strain when the increasing stress reaches the kwinking stress (Fig. 1d). These systematic fractures were often considered in literature as evidence for tensile strength of NiTi. However, comparison of tensile stress-strain curves of NiTi wires having different austenitic microstructures (Fig. 1d) clearly proves that maximum tensile stress the superelastic wire NiTi #1 can withstand (true stress ~1800MPa) is nearly independent  on temperature and microstructure. NiTi wires having small, recrystallized grains (grain size ~10-100 nm), however, display fracture stress decreasing from 1800 MPa to 1000 MPa. NiTi wires with coarse recrystallized grains (grain size > ~100 nm), which deform plastically with strain hardening, fracture after large plastic deformation at ~1000 MPa tensile stress (at true stress ~1800 MPa), regardless of the austenitic grain size.

The nanocrystalline NiTi wires fracture via instability of tensile deformation (necking) caused by the activation of kwinking when the kwinking stress is reached [48]. This explains why the fracture stress of strengthened nanocrystalline NiTi wires varies from 1800 MPa to 1000 MPa (Fig. 2d). This instability of tensile deformation was investigated in dedicated tensile test at room temperature on 14 ms NiTi#1 wire  that fractured at ~14% strain (Fig. 20a). When one looks closely at the fractured wire, necking becomes clear. To reveal the deformation mechanism responsible for the plastic deformation within the neck, the experiment was repeated and the test was stopped before the wire fractured. After unloading and stress-free heating to 60 °C (Fig. 20b), macroscopic plastic strain ~ 2% due to the localized tensile deformation within the neck remained unrecovered (Fig. 20c). Four TEM lamellae were cut from the fractured wire, three from the neck region and one from outside the neck. The austenitic microstructures were analyzed. The density of {114} deformation bands increases from location 1 at the edge towards location 3 in the middle of the neck. This confirms that the NiTi wire fractured by necking caused by the activation of kwinking deformation when the kwinking stress was reached in the tensile test. Based on this observation, we realized that the fracture stress decreasing with increasing grain size (pulse time) (Fig. 1d) or with increasing aging time (Fig. 9a) reflects the grain-size (aging time) dependence of the kwinking stress. The fracture stress follows from the fulfilment of the Considère criterion for stability of tensile deformation [74], which states that strain localization (necking) occurs when the tensile stress equals the strain-hardening coefficient.

Besides the necking, localized macroscopic plastic deformation in Lüders bands propagating at ~1000 MPa was observed in tensile test on 15 ms NiTi#1 wire at 60 °C (Fig. 21a-d) [48]. The stress-strain curve displayed very long stress plateau, in which the wire deformed via propagation of Lüders band front with 40% localized strain. Tensile deformation was stopped at ~30% strain, wire was unloaded and TEM lamellae were cut from within the Lüders band (location B) and outside (location A). TEM analysis of microstructure from the Lüders band shows that the austenitic microstructure was refined by kwinking deformation within the propagating Lüders band front, as evidenced by the observation of ultrafine deformation bands and the amorphous ring diffraction pattern. Time evolution of stress and local strains evaluated by DIC (Fig. 21d) proves that the large plastic deformation by kwinking indeed occurs within the propagating Lüders band front. This confirms that, under suitable combination of austenitic microstructure (pulse length) and test temperature, NiTi wires may deform plastically via propagation of macroscopic Lüders bands fronts, in which the martensite deforms via kwinking up to 40% strain [48].

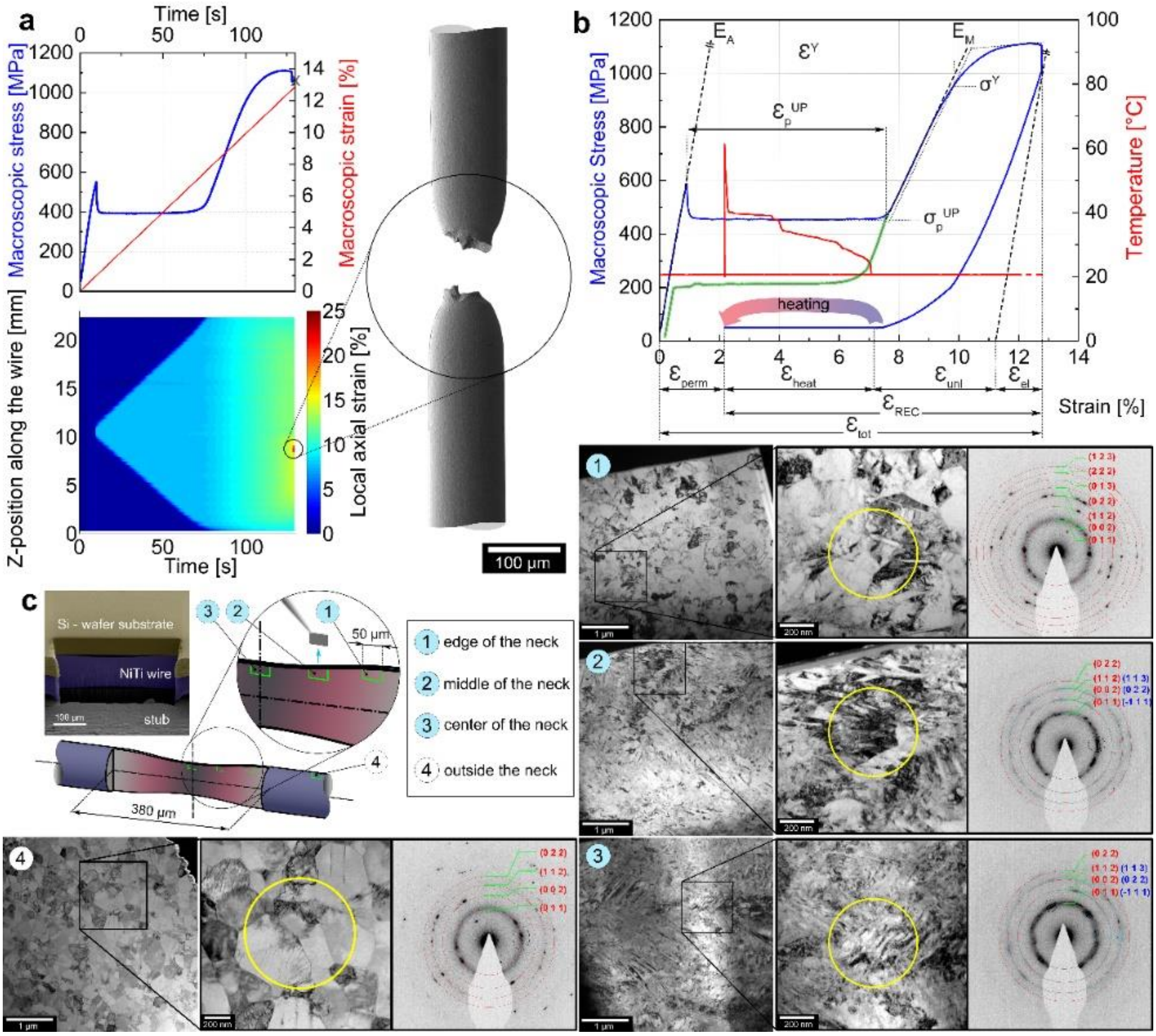

**Figure 20: Strain localization by necking in a tensile test proceeding via kwinking deformation [48].** a) Stress–strain curve of superelastic 14 ms NiTi#1 loaded in tension at T = 20 °C with DIC record evidencing strain localization. To prove that the wire fractures at 14% strain due to necking proceeding via kwinking deformation, tensile deformation was stopped when the neck started to form (b); the wire was unloaded, stress-free heated up to 60 °C, and cooled back to room temperature. c) TEM lamellae were cut from three locations in the neck and one location outside the neck. The analysis of austenitic microstructures in locations 1–4 shows {114} austenite twin bands, the density of which increases towards the center of the neck (locations 1 → 2 → 3 in (b)). This clearly proves that the wire fractured due to strain localization within the neck occurring via kwinking deformation.

Kwinking deformation gradually becomes suppressed in tensile tests on commercial cold worked/annealed NiTi wires with test temperature increasing above 100°C [9,10]. According to stress-temperature diagrams (Fig. 8a,e), the martensite shall deform plastically at temperatures higher than 100°C under stress lower than the transformation stress needed to induce it from the austenite. Obviously, this is not possible. We assume that the martensite stress induced from austenite at high temperatures deforms immediately by dislocation slip (i.e., plastic deformation proceeds massively already in the upper plateau range of superelastic stress-strain curve), which in turn prevents the stress induced MT to be completed at the end of the stress plateau. This gives rise to a TRIP-like plastic deformation, in which both austenite and stress induced martensite deform via dislocation slip and kwinking deformation becomes no more needed. When kwinking deformation in tensile tests gets suppressed with increasing temperature, ductility decreases (Fig. 8c,g). Although we have investigated this deformation mechanism by TEM [10] and x-ray diffraction methods [9], it remains unclear.

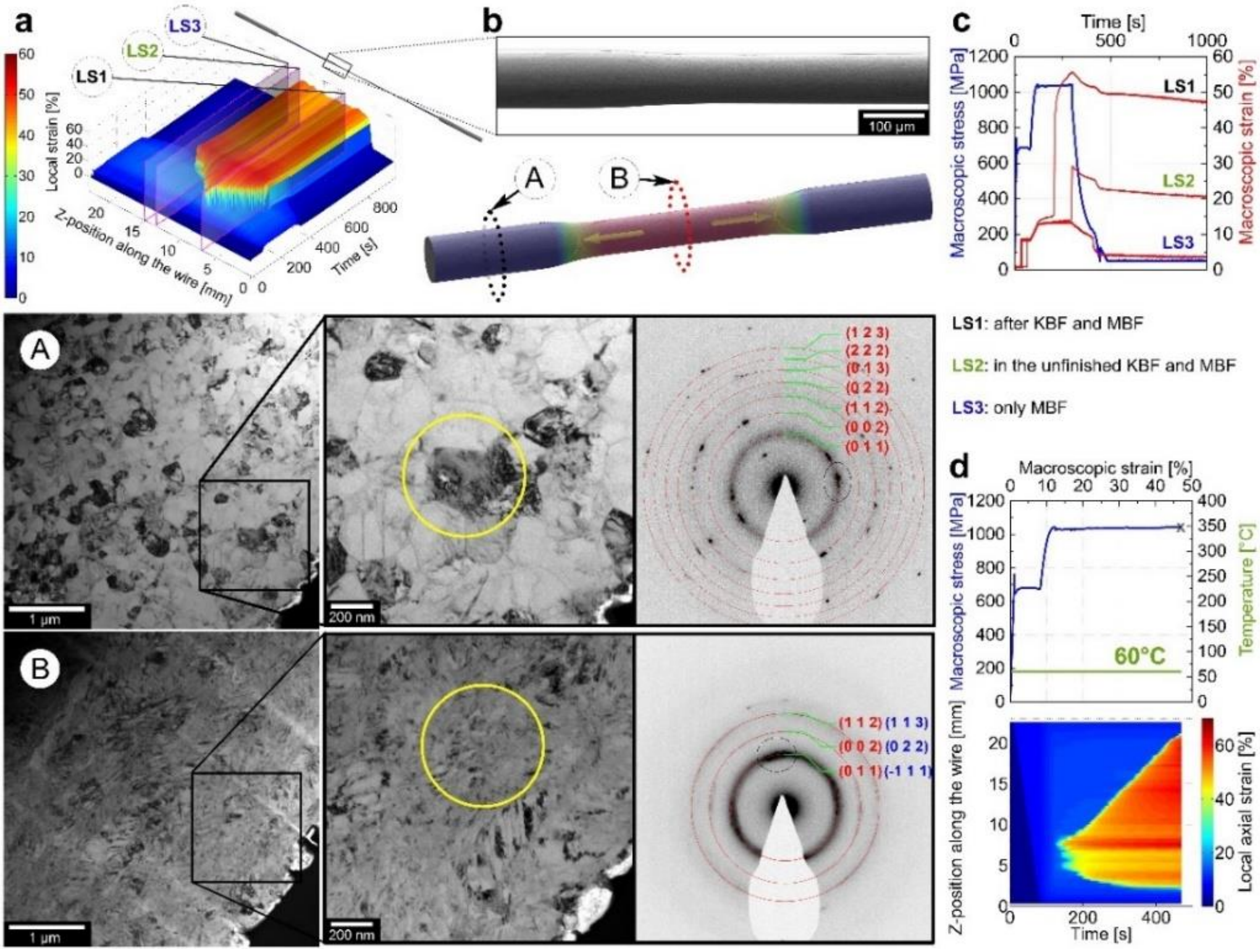

**Figure 21: Localized plastic deformation in superelastic 14 ms NiTi#1 wire deformed in a tensile test at 60 °C [48].** There are two stress plateaus on the stress-strain curve (d) in which tensile deformation is localized in Lüders band fronts propagating through the wire at constant stress. While only ~8% strain is localized in the first plateau due to stress-induced MT (680 MPa), very large strain ~40 % is localized (a) in the second plateau (1050 MPa) due to plastic deformation via kwinking. To confirm the activation of the kwinking deformation in the second plateau (b), tensile deformation was stopped at ~30% macroscopic strain, the wire was unloaded, TEM lamellae were cut from the deformed wire and lattice defects were analyzed in TEM (b). While the microstructure outside the Lüders band (A) contained only isolated dislocations, the microstructure within the Lüders band (B) contained high density of deformation bands and residual martensite. The microstructure was refined to near amorphous state (see the diffuse ring diffraction pattern in B). Time evolutions of macroscopic stress and local strains evaluated by DIC in locations LS1-3 across the Lüders band front (c) prove that the large plastic deformation ~40% indeed occurs within the propagating Lüders band front.

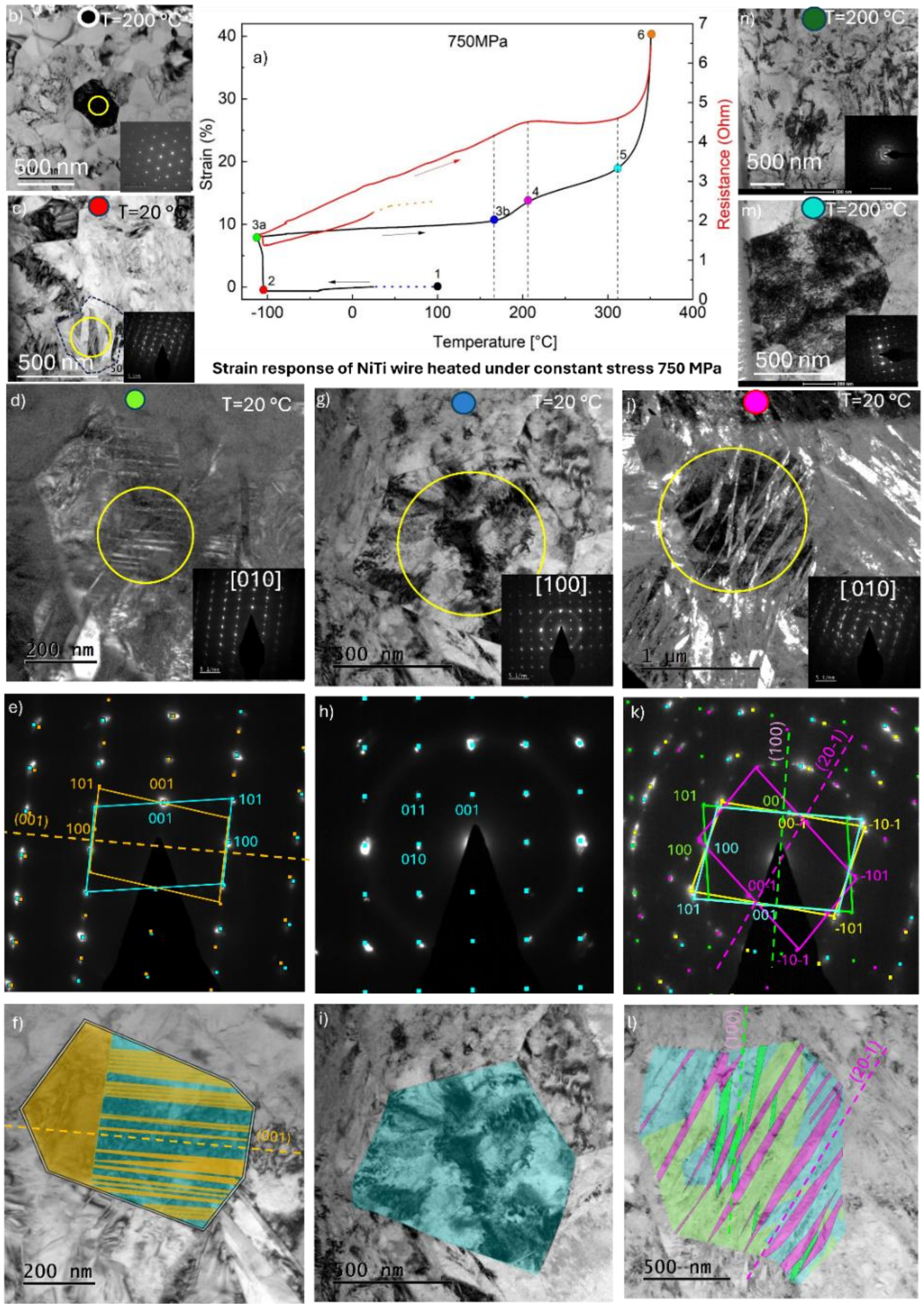

**Figure 22: Kwinking deformation activated during isostress heating under external stress** [8]. 15 ms NiTi #5 shape memory wire was deformed in the martensite state at -100 °C up to 9% strain (below the kwinking stress) and heated under constant stress 750 MPa up to the rupture at 350°C. To reveal deformation mechanisms activated in the heated wire, TEM lamellae were cut from the wire deformed up to gradually increasing maximum temperatures (stages 1-6). Except of stages 1,5,6, in which the wire contains austenite, there are martensitic microstructures in the wire in stages 2,3,4. When analyzing martensitic microstructures in strain states 3a, 3b and 4, the lamella was tilted into low index zones in the selected grain, diffraction patterns were taken from the yellow circle area covering large fraction of the grain interior and martensite variant microstructure in grains were reconstructed using SAED-DF imaging method. Dashed lines denote traces of martensite interfaces. Although the (001) compound twins are not visible in the [100] zone, the microstructure is the same in stages 3a (d,e,f) and 3b (g,h,i). High density of (20-1) kwink bands observed in stage 4 clearly prove that the heated wire deformed by kwinking between stages 3b and 4 (j,k,l). Upon further heating, the martensite transformed into austenite and the wire ultimately failed by plastic deformation of austenite via dislocation slip in stage 6.

The involvement of MT in this TRIP-like deformation mechanism is clearly evidenced by the yield stress increasing with increasing temperature (Fig. 8c,g). It shall be noted, however, that the temperature range, where this mechanism becomes activated, depends on the virgin austenitic microstructure of the NiTi wire. While solution annealed binary NiTi wires deform via this TRIP-like deformation already at room temperature, the strengthened nanocrystalline NiTi wires deform by it above 100 °C. In a sharp contrast, plastic deformation of NiTi wires via dislocation slip in B2 austenite is characterized by strain softening, rate dependence, decrease of yield stress and ductility with increasing temperature (Fig. 8d,h).

Finally, let us discuss plastic deformation by kwinking deformation that was observed to be activated when NiTi wire was cooled or heated under high external stress [8]. It can be homogeneous but also macroscopically localized. The localized deformation is facilitated by thermal gradients that are difficult to avoid in these experiments. There are two cases – either the kwinking deformation proceeds alongside MT (this will be discussed in chapter 9), or it is activated in the absence of MT. The latter occurs only upon heating and will be discussed below.

Results of an experiment, in which 15 ms NiTi #5 shape memory wire containing oriented B19' martensite was heated under high external stress up to rupture, are shown in Fig. 22. The wire was loaded at -100 °C up to 750 MPa tensile stress (400 MPa below the kwinking stress (Fig. 8e)) and heated under this stress held constant until it fractured. While the wire was heated from -100 °C to 170 °C, its length remained nearly constant (slightly increased), but when it was heated from 170 °C to 200 °C, it started to elongate. Tensile deformation continued upon further heating up to the rupture at ~ 350 °C.

Martensite variant microstructures in NiTi wires heated up to various maximum temperatures were analyzed by TEM (Fig. 22b-m). The results proved the presence of oriented martensite in the wire heated up to 170 °C (3b in Fig. 22a) and kwink bands were observed in the wire heated up to 200 °C (4 in Fig. 22a). We were curious what was happening with the oriented martensite exposed to high stress – high temperature conditions. We would like to analyze the martensite at these extreme conditions. The problem is, we saw only relicts remaining in the TEM lamella under no stress at room temperature. Nevertheless, based on the obtained results, we proposed that the heated oriented B19' martensite transformed into a long period modulated monoclinic structure (see Fig. 12 in [73]) before it deformed by kwinking and transformed into austenite (4-5 in Fig. 22a). The heated wire finally ruptured via dislocation slip in austenite at 350 °C (5-6 in Fig. 22a). When the wire was heated under stresses higher than 800 MPa, it fractured via kwinking deformation before the reverse MT could even start (see Figs. 3,4, in [8]).

In summary, although kwinking deformation in NiTi wires tends to proceed homogeneously with strain hardening, it may proceed also in macroscopically localized manner (necking, Lüders band, shear band). Whether kwinking deformation is homogeneous or localized, depends on the magnitude of kwinking stress and strain hardening coefficient – whether they fulfill the Considère criterion for stability of tensile deformation [74]. When the austenitic microstructure of NiTi wire is strengthened against dislocation slip in martensite, kwinking stress increases and strain hardening decreases making the alloy vulnerable against necking. The strengthened NiTi wires thus systematically fracture by necking proceeding via kwinking deformation activated when increasing stress reaches the kwinking stress at ~12% tensile strain. When NiTi wire loaded in tension at low temperatures up to stress below kwinking stress is heated under constant stress applied, it deforms by kwinking up to fracture when stress-temperature conditions for kwinking deformation are reached.

## 8. Texture Evolution and Refinement of Austenitic Microstructure by Kwinking in Tensile Tests

The mechanism of kwinking deformation (Chapter 3) was revealed based on the results of TEM reconstructions of martensite variant microstructures in grains of plastically deformed NiTi wires (Chapter 6), with a particular focus on the analysis of kwink bands and kwink interfaces. Although this is an excellent method for investigating deformation mechanisms in materials, it is not always clear whether the analyzed microstructures and lattice defects are representative of the investigated deformation mechanism (i.e., whether the conclusions derived from TEM analyses are statistically robust). Therefore, confirmation of the activated deformation mechanism by a complementary experimental method is needed.

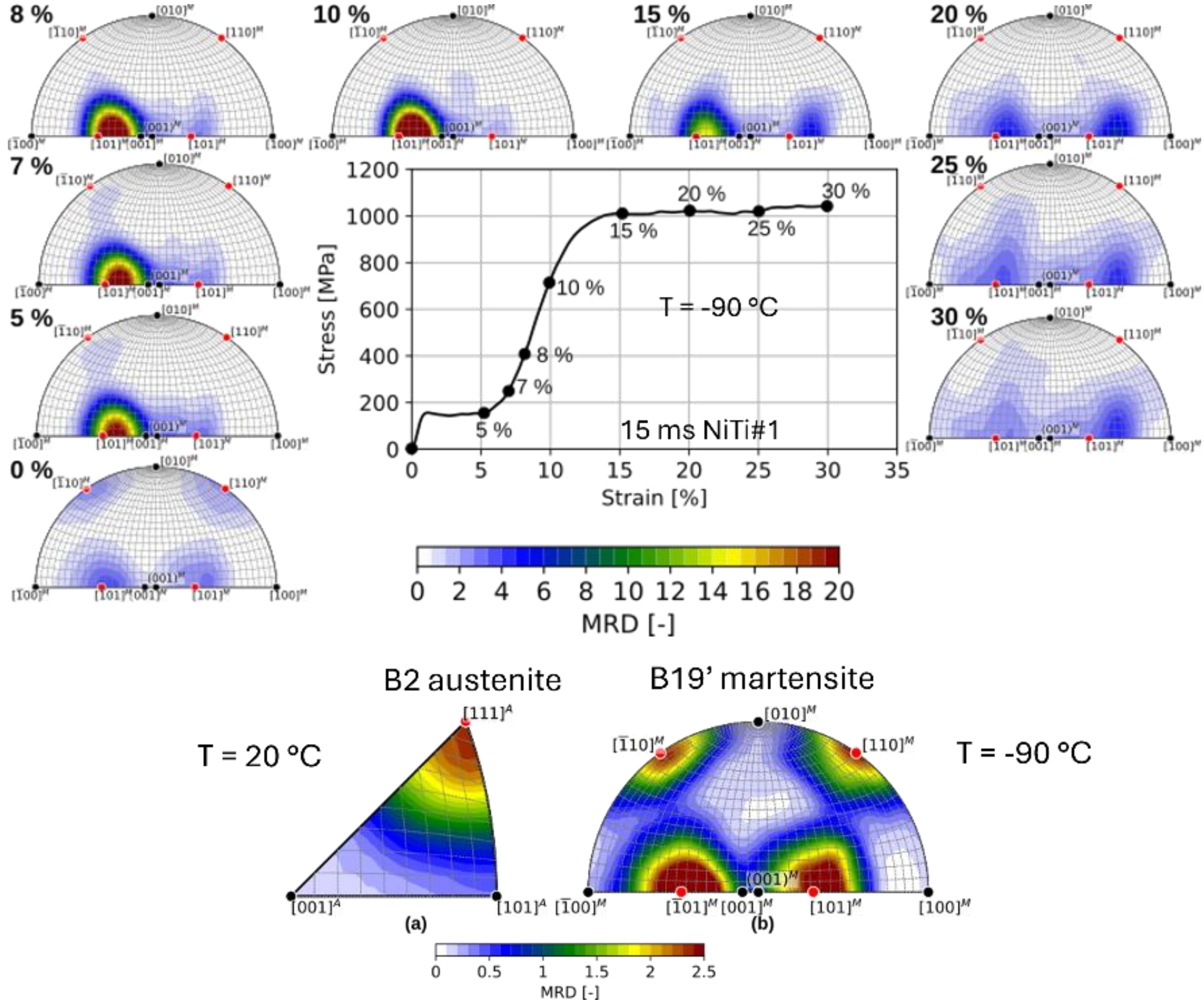

**Figure 23: Evolution of martensite texture of the superelastic 16ms NiTi#1 wire deformed in a tensile test -90 °C in the martensite state evaluated by in-situ synchrotron x-ray diffraction** [26]. The evolution of axial direction inverse pole figures AD IPF with increasing strain reflects the variation of crystal lattice orientations due to martensite reorientation (0-10% strain) and plastic deformation by kwinking (10-45% strain). AD IPF figures showing radially symmetric texture of B2 austenite and self-accommodated B19' martensite in a different scale are shown at the bottom. The red dots denote lattice correspondent directions in all martensite variants to the [111]A direction.

Since kwinking deformation rotates the crystal lattice in a defined manner, analysis of the evolution of martensite texture during a tensile test provides a complementary method with excellent statistical coverage. We therefore performed systematic synchrotron x-ray diffraction studies of texture evolution during tensile tests on both superelastic and shape memory NiTi wires. Results obtained in a tensile test on a 16 ms NiTi #1 superelastic wire deformed at −90 °C are shown on Fig. 23. The obtained results can be compared with the results from tensile test on 15ms NiTi #5 shape memory wire deformed at 20 °C (Fig. 24). Despite different (weaker) austenite texture of the 15 ms NiTi #5 SME wire, the texture evolutions are qualitatively very similar. Martensite reorientation occurs in the lower stress plateau and kwinking deformation in the higher stress plateau. The experimentally observed texture evolution was qualitatively interpreted based on the theoretical analysis of rotations of crystal lattices of stress induced martensite stress via (20-1) twinning [26], assuming $\langle 111 \rangle_A$ oriented austenite grains in the NiTi wire.

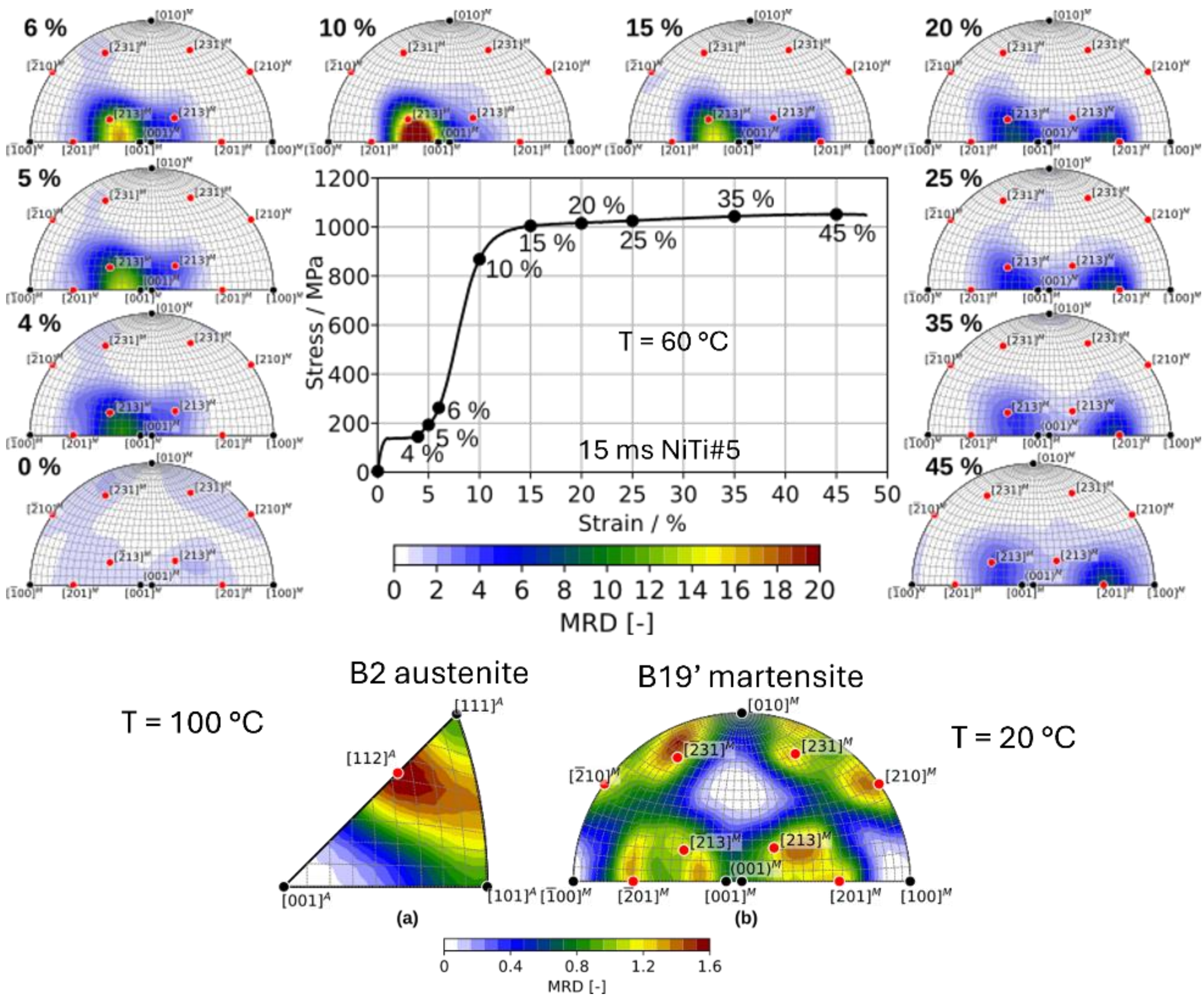

**Figure 24: Evolution of the martensite texture of the 15ms NiTi#5 shape memory wire deformed in a tensile test at 20 °C in the martensite state evaluated by in-situ synchrotron x-ray diffraction.** The evolution of axial direction inverse pole figures AD IPF with increasing strain reflects the variation of crystal lattice orientations due to martensite reorientation (0-10% strain) and plastic deformation by kwinking (10-30% strain). AD IPF figures showing radially symmetric texture of B2 austenite and self-accommodated B19' martensite in a different scale are shown at the bottom. The red dots denote lattice correspondent directions in all martensite variants to [112]A direction.

The texture evolution results (Figs. 23, 24) show that the intensity of the pole near [-101]M direction corresponding to the tensile stress induced martensite (Fig. 25a) gradually decreases while the intensity of the pole near [201]M direction on the right side of the AD IPF increases as the wires deform plastically beyond yield limits. This can be rationalized by assuming activity of the (20-1) kwinking in the tensile test (Fig. 25a,b). Besides the primary (20-1) kwinking, however, there is also the secondary (20-1) kwinking and dislocation slip in the matrix, which also affect the martensite texture and give rise to a kind of "rollover" evolution of the orientations of the martensite lattice during the tensile test (texture evolution is characterized by moving intensity between poles in IPFs (Figs. 25b,c) in directions of arrows). Quantitative interpretation of the observed texture evolutions would require comparison with texture evolution simulated by micromechanics model of NiTi polycrystal [74] which captures the activated orientation dependent deformation processes (B2-B19' MT, martensite reorientation, [100](001) slip in martensite, (100) twinning in martensite, and (20-1) kwinking).

Plastic deformation of oriented martensite by kwinking taking place in tensile tests on NiTi wires up to rupture at large strains reorients, refines, and hardens the virgin austenitic microstructure of the wire. The refinement of austenitic microstructure is a fascinating consequence of the kwinking deformation. It was investigated in tensile test on 16 ms superelastic NiTi#1 wire deformed at room temperature up to 53% strain. Recoverable and plastic strains were evaluated (Fig. 10) and permanent lattice defects in the wire deformed up to gradually increasing tensile strains, unloaded and heated were analyzed by TEM (Fig. 26). When the wire was unloaded and stress-free heated, the plastically deformed martensite transformed to plastically deformed austenite containing high density of dislocations and deformation bands, the density of which increased with increasing strain. Beyond the plastic yielding limit, as the stress induced martensite only started to deform plastically, the imposed strains recovered partially on unloading and partially on subsequent heating, but the imposed strain was largely recovered (Fig. 10). As the maximum strain increased (12-53%), plastic strain increased but recoverable strain remained large (~10%). The deformation bands became finer with increasing strain, and the microstructure became near amorphous at 53% maximum strain (Fig. 26). For more information on the analysis of recoverable and plastic strain components and permanent lattice defects generated by plastic deformation in tensile tests at various temperatures see Ref. [15].

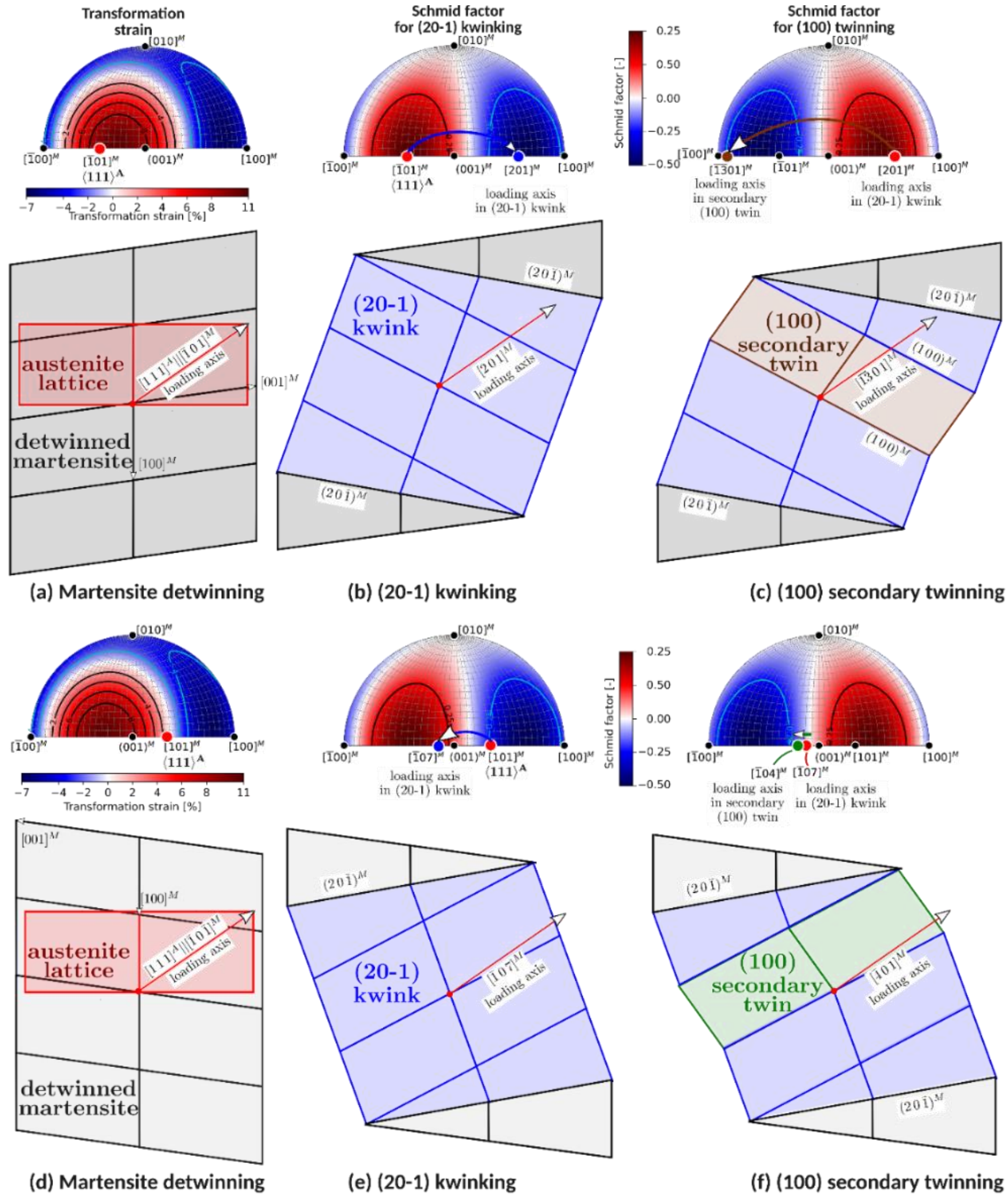

**Figure 25: The change of orientation of B19' lattice during tensile test due to kwinking.** The graphical construction illustrates how the orientation of crystal lattice ((101)A//(010)M projection) changes during tensile test on ideal <111>$_A$ fiber textured NiTi wire: a) stress induced MT in tension along the <111>A direction, b) primary (20-1) kwinking c) secondary (100) deformation twinning. Since (100) deformation twins were often observed in the microstructure of wires deformed in tension (despite negative Schmid factor), we included also graphical construction for d) stress induced (100) deformation twin, e) second (20-1) kwinking f) secondary (100) deformation twinning. The primary (20-1) kwinking followed by secondary kwinking (Fig. 19) leads to a kind of "rollover" deformation mechanism, which moves the intensity in the AD IPF figure (Figs. 23,24) around the [010] zone axis. This enables continuous plastic deformation of the wire by kwinking deformation in tensile tests which refines the martensite variant microstructure to nearl amorphous state.

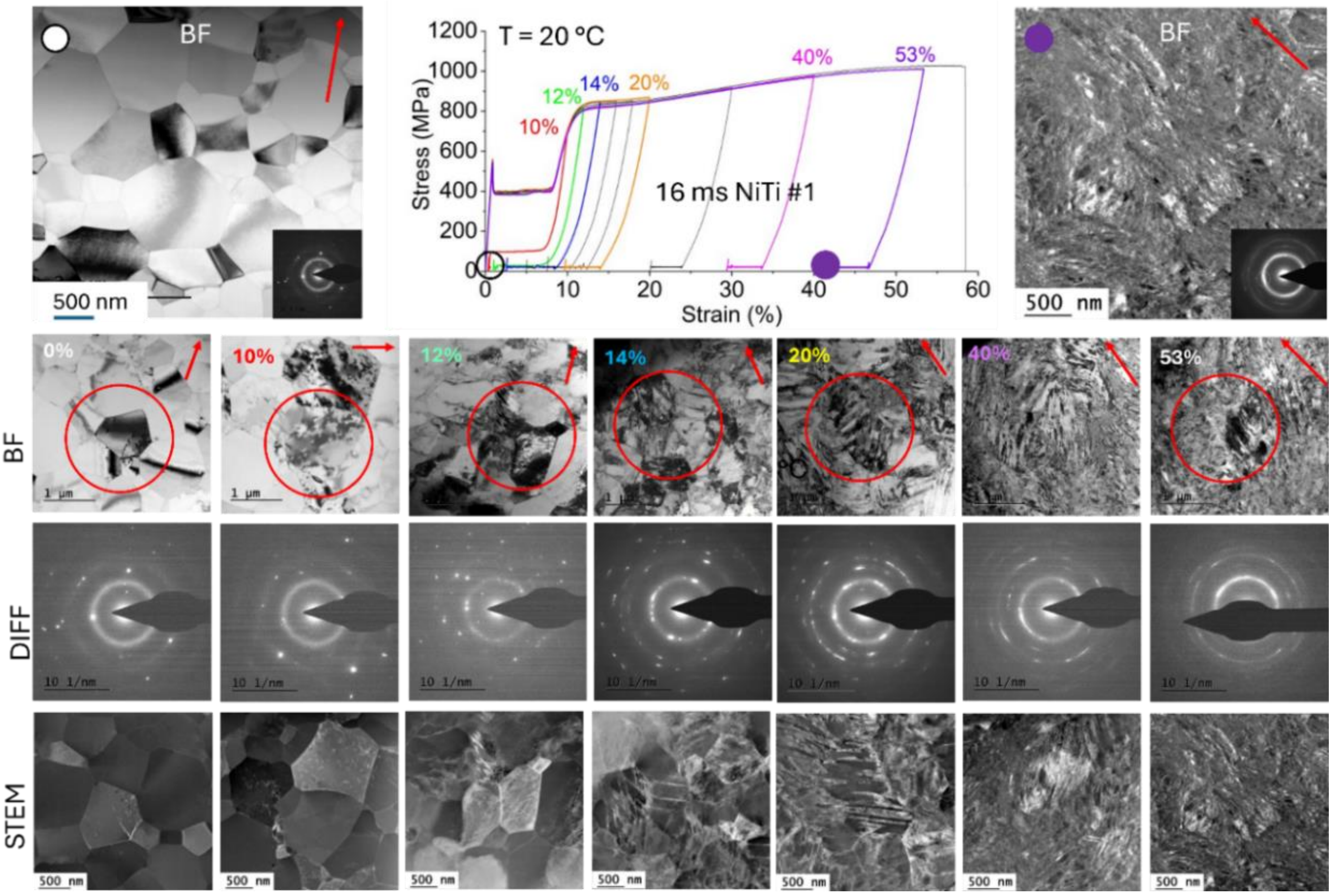

**Figure 26: Refinement of the virgin austenitic microstructure of superelastic 16 ms NiTi #1 wire deformed in a tensile test at 20 °C up to rupture** [15]. The refinement of austenitic microstructure is clearly evidenced by BF images, SAED patterns (taken from the large red circle areas) and STEM images from wires deformed up to gradually increasing strains, unloaded and stress-free heated up to 170 °C. The number of deformation bands in grains as well as number of diffraction spots in SAED patterns (same size) increases with increasing strain due to the plastic deformation of the B19' martensite by kwinking.

When writing the article [15], however, we did not know the kwinking deformation mechanism, we only knew that the peculiar (20-1) deformation twinning in martensite is partially responsible for the unrecoverable plastic deformation [22,36,52] and that it is somehow linked to {114} austenite twins observed in deformed austenite [33-35]. However, the explanations available in literature [32-34, 39, 50, 52] were not realistic and we were puzzled by the observation of the variety of internal interfaces (STGBs) in the austenitic microstructure of plastically deformed and heated NiTi wires. This motivated us to repeat the same experiment on shape memory wires [26,27] that remain martensitic after plastic deformation and unloading. The results of the analyses of martensite variant microstructures in plastically deformed NiTi wires lead to the discovery of kwinking and enabled to understand the origin of the high density of STGBs in austenite in plastically deformed superelastic NiTi (Figs. 18,19,26).

In summary, plastic deformation of B19' martensite in tensile tests on NiTi wires occurring via kwinking gives rise to characteristic evolution of the martensite texture that cannot be explained by conventional dislocation slip. The kwinking deformation leads to gradual refinement of the martensitic microstructure which provides strain hardening that enables stable tensile deformation at high stresses (~1GPa) up to large strains (~60%). The refined martensitic microstructure in plastically deformed NiTi wires transforms into plastically deformed austenite during the reverse MT when the wire is subsequently heated above the $A_f$ temperature. This way, the virgin austenitic microstructure of NiTi wires deformed plastically in tensile tests up to fracture at 60% strain becomes refined by the kwinking deformation in martensite to near-amorphous state. If recrystallization does not occur upon the heating, the austenitic microstructures in plastically deformed NiTi wires heated above the $A_f$ temperature give rise to very attractive mechanical properties [35, 75, 76].

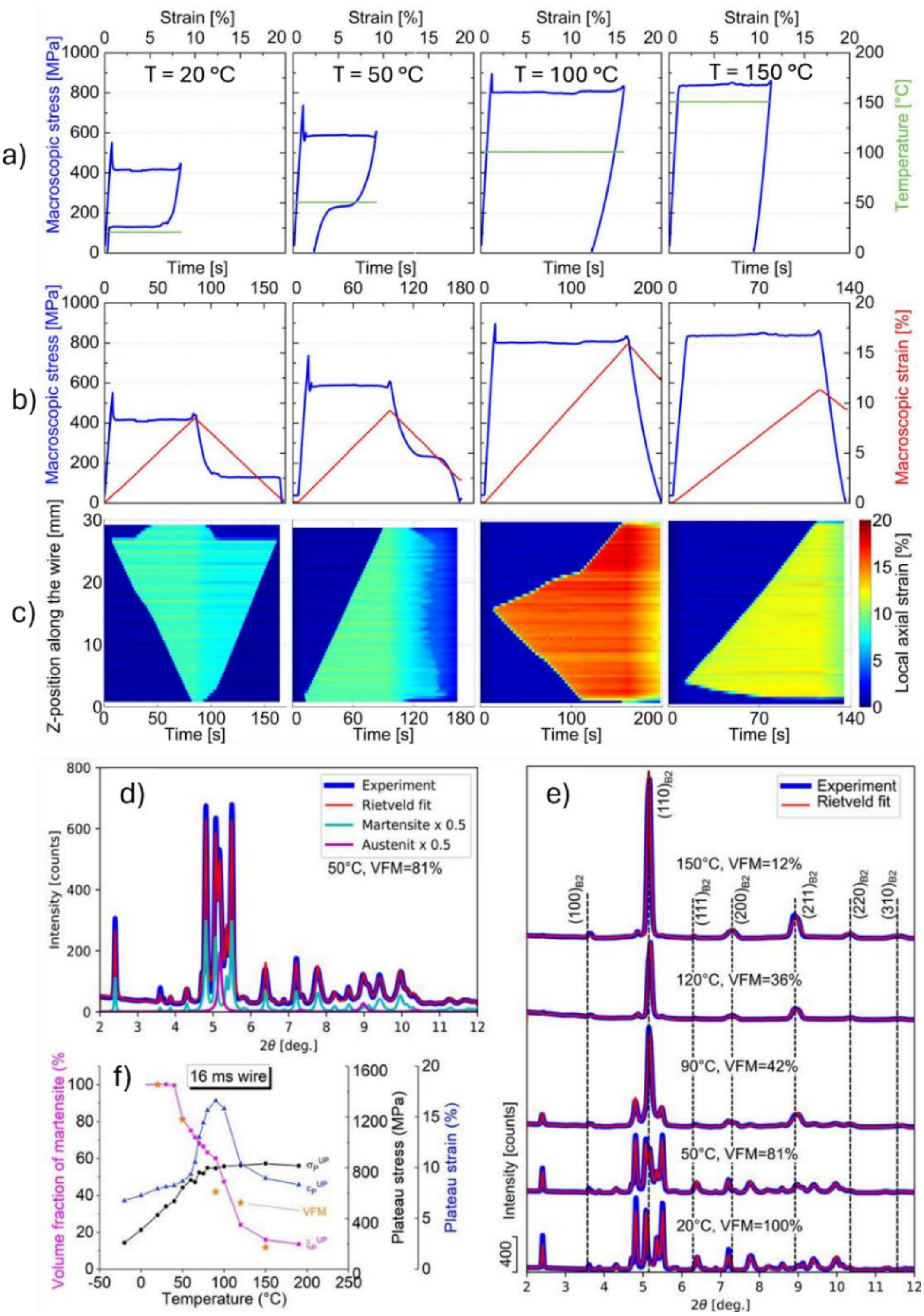

**Figure 27: Plastic deformation accompanying stress-induced MT in 16 ms NiTi #1 wire deformed in tensile tests at temperatures T=20 °C, 50 °C, 100 °C, 150 °C [28].** a-c) stress-strain-time responses and 1D DIC records showing localized deformation in Lüders bands due to stress-induced MT. c) 1D DIC strain maps inform about the evolution of the local axial strain during the tensile test in space-time coordinates. d) Rietveld fit of 360 degrees integrated 2D diffraction pattern recorded in in-situ synchrotron x-ray diffraction experiment during a tensile test at 50 °C. e) Fitted diffraction patterns measured at temperatures 20 °C, 50 °C, 90 °C, 120 °C,150 °C with denoted values of volume fraction of martensite (VFM) evaluated by quantitative phase analysis. f) VFM at the end of the stress plateau evaluated in tensile tests at various test temperatures plot together with plateau stress and plateau strain.

## 9. Kwinking Deformation in Cyclic Thermomechanical Loading Tests

The phenomena discussed so far concerned mainly plastic deformation by kwinking activated when a NiTi wire containing oriented martensite is deformed beyond the plastic yield limit (the kwinking stress). However, plastic deformation by kwinking may also be activated when the martensitic transformation proceeds under external stress below the kwinking stress. When this happens, incremental plastic strains suddenly become very large, often larger than recoverable transformation strains [10-13]. Of course, this makes responses of NiTi wires in closed loop superelastic and actuation tests no longer reversible. Let us briefly discuss three examples of thermomechanical loading tests, in which kwinking deformation was found to affect the cyclic stress–strain–temperature responses of superelastic NiTi #1 wires.

The first example concerns unusually long superelastic plateaus (>8%) accompanied by generation of irrecoverable plastic strains sometimes observed in tensile tests on NiTi wires (Fig. 27). These long upper stress plateaus have often been reported in literature without providing due explanation. To clarify their origin, superelastic NiTi #1 wires with different grain sizes were subjected to tensile loading-unloading tests at various test temperatures [28], lattice defects in deformed wires were analyzed by TEM, localized tensile strain was observed by DIC and martensite volume fractions at the end of the stress plateau were evaluated by in-situ synchrotron x-ray diffraction (Fig. 27).

It was found that tensile deformation is localized (Fig. 27a-c) and lengths of upper plateaus show maxima for specific combinations of grain size and test temperature (Fig. 27f). Simultaneously, TEM analysis of permanent lattice defects in wires exhibiting the longest upper plateaus showed presence of {114} austenite bands within the austenitic microstructures. These results suggests that stress-induced MT within the Lüders band front was accompanied by plastic deformation of stress induced martensite by kwinking. However, NiTi wires having various microstructures displayed this deformation mode at different temperatures.

To further explore this phenomenon, volume fraction of martensite existing in the wire at the end of the plateau was evaluated by in-situ synchrotron x-ray diffraction experiment [28]. It decreased with increasing test temperature across the temperature range, in which the long plateaus were observed (Fig. 27d,e,f). The localized deformation in Lüders bands in the tensile test is commonly associated with the stress induced MT. Despite that, the volume fraction of martensite at the end of the plateau decreased from 100% to 12% across the temperature range, in which the unusually long plateaus were observed (Fig. 27e). This was rationalized as follows. The upper-plateau strains increased with increasing temperature (stress) when the stress-induced martensite within the propagating Lüders band front started to deform plastically by kwinking because of the added plastic deformation. However, at the same time, less austenite transformed to martensite with increasing temperature. The observation of long upper stress plateaus at specific temperatures and microstructures (grain size) is thus due to trade-off between the stress-induced MT and plastic deformation of stress-induced martensite during the forward loading.

The second example concerns the role of kwinking deformation in generating large unrecoverable plastic strain during thermal actuation cycling of 15ms NiTi #1 wires under various external stresses (Fig. 28). The wires were subjected to single closed-loop thermomechanical loading cycles, in which either the forward MT (Fig. 28a,c) or the reverse MT (Fig. 28a,d) proceeded under constant stress. Cyclic strain–temperature responses were recorded (Fig. 28e–i), and permanent lattice defects created by the MT under stress in austenitic microstructures were analyzed by TEM.

It was found that the forward and reverse MTs proceeding under stress generate incremental plastic strains only when they occur under stress above certain characteristic thresholds. Wires thermally cycled under medium stresses (200–400 MPa) exhibited large recoverable strains (~8%) and only small incremental plastic strains. In contrast, wires cycled under high stresses (700–800 MPa), still below the kwinking stress of 900 MPa (Fig. 8a), showed very large plastic strains, and {114} deformation bands were found by TEM in the microstructure of the cycled wires. The incremental plastic strains reached up to ~8% (on cooling) and ~14% (un heating) under 800 MPa stress (Figs. 28c,d). This provides evidence that plastic deformation by kwinking occurred in the martensite induced on cooling (Fig. 28c) or reverse transforming to austenite on heating (Fig. 28d). The reason why kwinking deformation was activated under stresses lower than the kwinking stress is that kwinking proceeded alongside the MT.

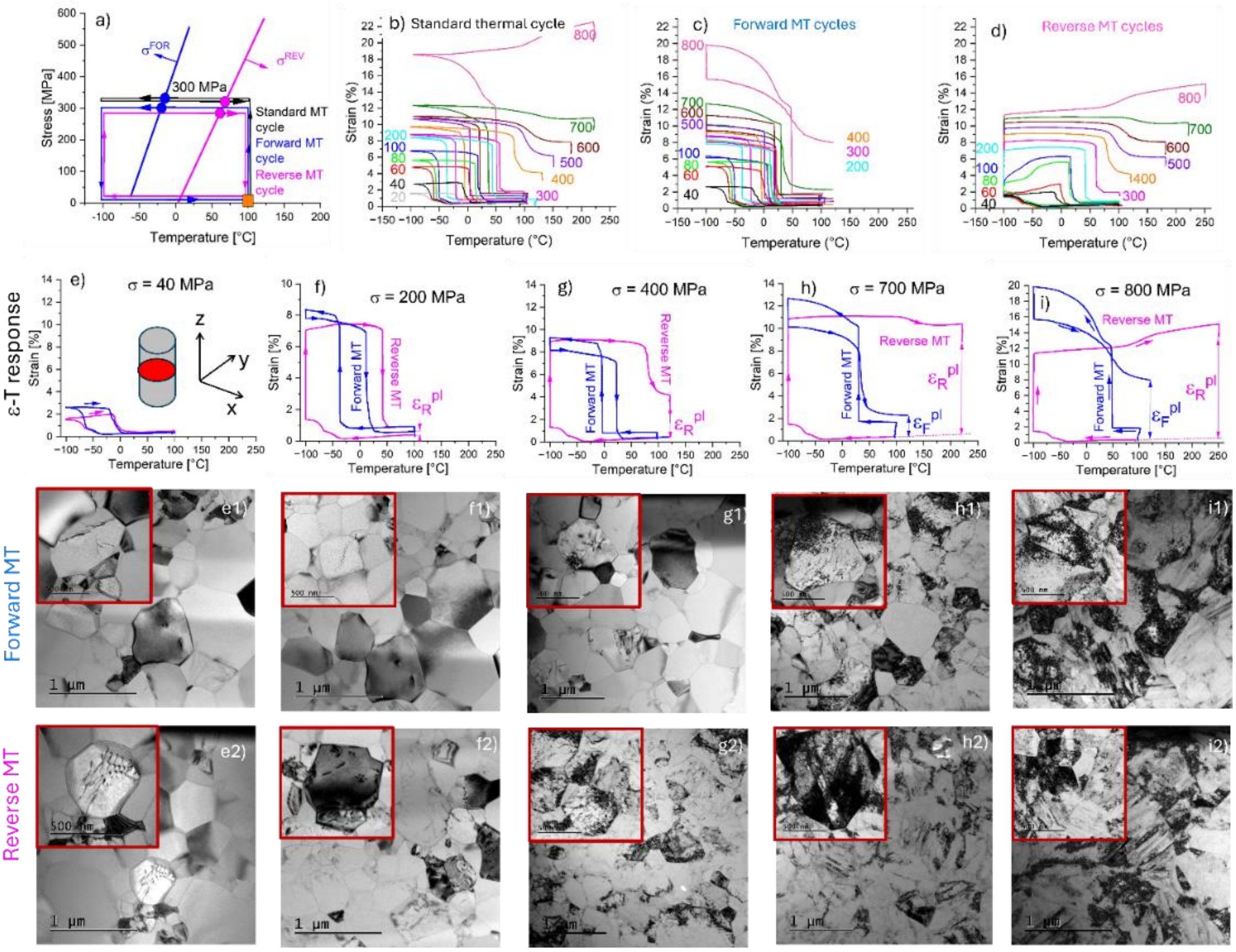

**Figure 28: Permanent lattice defects in austenite created during the forward and reverse MTs in 15 ms NiTi #1 wire upon cooling and heating under various external stresses (**40, 200, 400, 700, 800 MPa) [14]. a) Stress–temperature path of thermomechanical loading cycles. Strain–temperature responses recorded in b) standard thermal cycles (both forward and reverse MTs proceed under stress), c) forward MT cycles, and d) reverse MT cycles. Strain–temperature responses recorded in closed-loop thermomechanical loading tests (e–i), in which permanent lattice defects were created. The TEM lamellae were cut from NiTi wires perpendicularly to the wire axis. BF TEM images show slip dislocations, dislocation bundles and kwink bands within austenitic microstructures generated by forward MTs (e1–i1) and reverse MTs (e2–i2) under various tensile stresses. {114} deformation bands were found in the microstructure of the wire heated under 700 MPa (h2) and of wires cooled (i1) and heated (i2) under 800 MPa.

Finally, the third example concerns the role of kwinking deformation in shape setting of NiTi wire [4]. It is well known that NiTi wire can be shape-set into a spring by coiling it on a mandrel and heating it in the constrained shape above 500 °C. Strictly speaking, cold-worked NiTi wires should be used for shape setting. However, cold-worked NiTi wires are difficult to bend over a small radius, and researchers found empirically that already annealed NiTi wires can be shape-set by the same procedure as well, although the desired shapes are not set precisely due to significant spring-back. The mechanisms by which cold-worked and annealed NiTi wires are shape-set, however, are fundamentally different. When a cold-worked NiTi wire is shape-set, the new shape is established during recrystallization of the cold-worked microstructure around 500 °C. Since recrystallization occurs throughout the wire, the austenitic microstructure is uniform along the entire shape-set wire. By contrast, when an already annealed NiTi wire is deformed, constrained, and heated in the constrained shape, the oriented martensite transforming to austenite is responsible for setting new shape. However, the martensite transforms back to the same parent austenite everywhere except in regions where it is retained up to high temperatures by the stress locally increasing due to the imposed constraint. As the temperature increases. the oriented martensite in these regions, partially retransforms to parent austenite (raising the external stress) and partially deforms plastically via dislocation slip in martensite. This continues until the increasing stress reaches the kwinking stress; kwinking starts and the internal stress reaches its maximum. Activation of kwinking during constrained heating therefore prevents the internal stress from rising to levels that could nucleate cracks and cause fracture. Shape setting of annealed austenitic NiTi wires is thus safe due to kwinking deformation, although it is less precise (spring-back occurs) than shape setting of cold-worked wire. The key difference is that the austenitic microstructure in the LTSS shape-set NiTi wire is not macroscopically homogeneous, because it is modified only in regions where plastic deformation by kwinking occurred during shape setting.

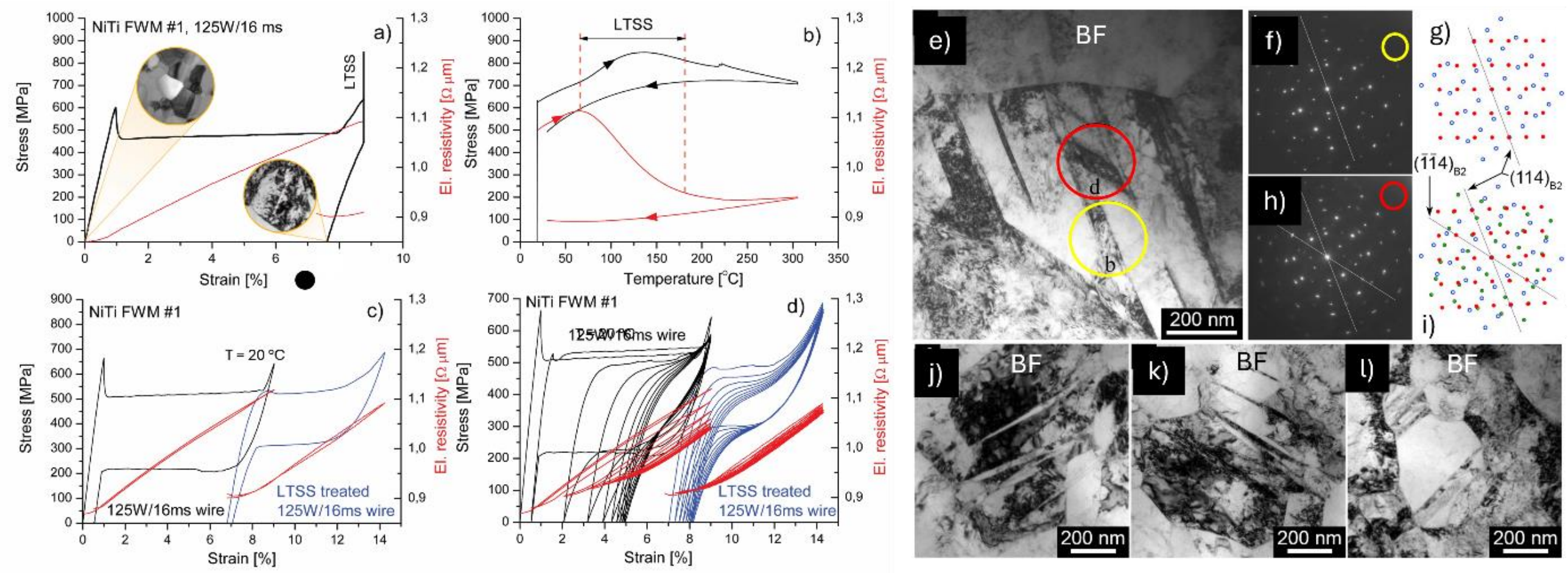

**Figure 29: Constrained heating (LTSS) experiment on superelastic 16 ms NiTi #1 wire simulating the shape setting treatment** (prestrain 9% at 20 °C, constrained heating/cooling up to 300 °C, unloading) [4]. a) Stress–strain response and b) strain–temperature response recorded in the test. c) Superelastic curves recorded before and after the test. d) Cyclic superelastic stress–strain curves recorded before and after the test. The austenitic microstructure in the wire changes during the constrained heating/cooling (b), as suggested by the insets in (a). Figures (e–l) show {114} austenite bands within the austenitic microstructure of the deformed wire observed by TEM.

To verify the above scenario experimentally, a dedicated closed-loop thermomechanical loading test [4] was performed on annealed 16ms NiTi#1 superelastic wire (Fig. 29). The wire was deformed in tension at room temperature up to the end of the superelastic plateau, constrained in length, subsequently heated up to ~300 °C, and cooled back to room temperature. Following the cycle, the wire was austenitic, shape-set (~7% longer) and displayed excellent superelasticity. The superelastic stress–strain response, however, differed from that of the virgin wire (Fig. 29c,d), indicating that the austenitic microstructure evolved during the cycle. A high density of slip dislocations and {114} austenite bands were found in the microstructure of the shape-set wire (Fig. 29e–l). This provides clear experimental evidence that the oriented martensite in the wire deformed by plastic deformation via kwinking while it was heated/cooled under constrained length. The maximum stress recorded during the shape-setting experiment (Fig. 29b) is observed at temperatures, at which kwinking deformation starts (increasing stress reaches the kwinking stress. NiTi wires with different virgin austenitic microstructures display different stress maxima at different temperatures (see Figs. 29 and 30 in [4]).

In summary, kwinking deformation may become activated in thermomechanical loading tests on NiTi wires when the forward and/or reverse MT proceeds under high external stress that is lower than the kwinking stress. When this happens, large unrecoverable strains generated by MT in closed-loop cyclic thermomechanical tests give rise to unexpected phenomena, such as long superelastic plateaus and/or large strains generated upon thermal cycling under constant stress. As a particular case, shape setting of annealed NiTi wires is controlled by kwinking deformation activated during the constrained heating of NiTi wire containing oriented martensite. The virgin austenitic microstructure of the heated NiTi wire changes to refined microstructure only in parts of the wire where oriented martensite resided during the constrained heating, while it stays unchanged elsewhere. Gradients of microstructure and functional properties are thus introduced by shape setting of annealed NiTi wires. The annealed NiTi wires are shape-set by the activation of kwinking deformation during the constrained heating up to relatively low temperatures (~300 °C), whereas shape setting of cold-worked NiTi wires proceeds via recrystallization upon constrained heating at higher temperatures (above ~450 °C recrystallization temperature) and same new microstructure is set everywhere in the shape set wire regardless of the applied constraint. These issues are very important for example for shape setting of NiTi stents from annealed laser-cut tubes and/or for shape setting of hybrid NiTi textiles with integrated NiTi wires [77] that cannot withstand high temperatures.

## 10. Kwinking Deformation in Constitutive Modelling of NiTi

**Plastic Deformation of Martensite in NiTi Modelling**

Constitutive models are mostly used for simulations of SMA components, where reversible functional responses are involved and the operating conditions are set to avoid irreversible plastic deformation. However, in case of NiTi, it has become clear that this is a counterproductive simplification, which makes the model predictions only qualitative, particularly as concerns cyclic thermomechanical loads. As already mentioned in the introduction, incremental dislocation slip in B19' martensite generated by MT proceeding under stress (Chapter 4) gives rise to functional fatigue, at the same time, however, it stays behind the enormous success of polycrystalline NiTi, since it prevents premature intergranular fractures frequently displayed by other SMAs.

As concerns plastic deformation of martensite, although several constitutive models already predict irreversible plastic deformation of NiTi [79-83], limited knowledge of the exact mechanism of plastic deformation makes predictions of general thermomechanical responses untrustworthy. The key problem is that plastic deformation, besides generating irrecoverable plastic strains, also modifies the virgin austenitic microstructure, which in turn significantly affects the functional thermomechanical responses. The kwinking deformation, while refining the virgin austenitic microstructure, renders NiTi the excellent strength/ductility ratio (tensile deformation up to 80% under stress ~1 GPa). These phenomena are beyond the scope of SMA models currently available in literature due to the limited awareness of kwinking in the field. Consideration of plastic deformation of martensite by kwinking in the formulation of advanced constitutive models of NiTi poses challenges for future development.

**Example of the Consideration of Plastic Deformation of Martensite by Kwinking in NiTi Modelling**

There is one specific area, where consideration of plastic deformation of martensite by kwinking is urgently needed in NiTi modelling. When fabricating NiTi medical devices, plastic deformation is inevitably introduced during the shape setting of already annealed NiTi components. Since this shaping process is usually carried out by heating to elevated temperatures of 400–500 °C while keeping the desired shape, it proceeds via the LTSS process involving kwinking deformation (Chapter 9, Fig. 29). The higher is maximum temperature applied in shape setting, the better is shape set quality and the softer the alloy becomes (displays functional fatigue). Therefore, one must search for optimized shape setting conditions based on the knowledge of kwinking.

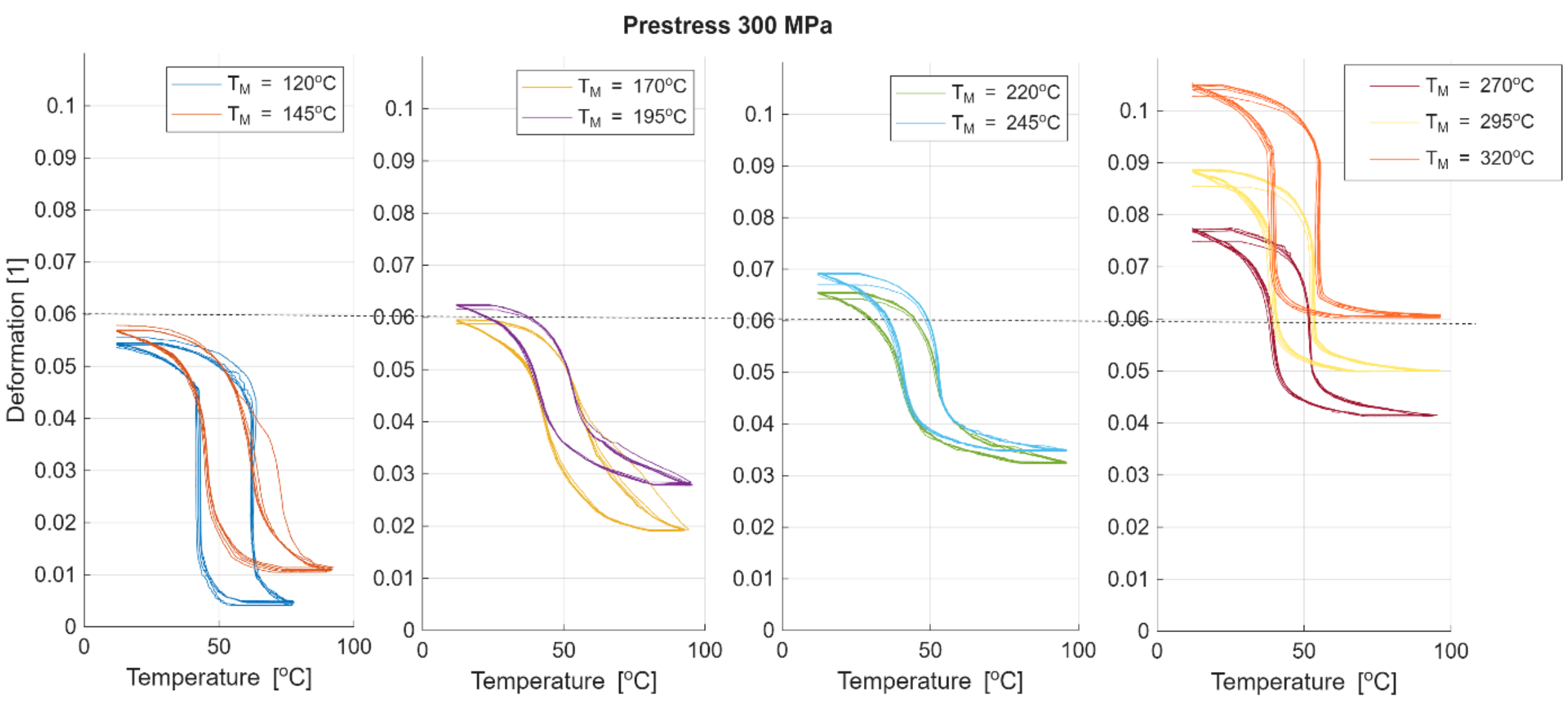

**Figure 30:** Strain-temperature responses of thermally cycled NiTi actuator wires subjected to low temperature shape setting treatments with maximum temperature $T_M$ gradually increasing from 120 °C to 320 °C (Fig. 29b).

As a demonstration of this, strain-temperature actuation responses of a commercial actuator wire (SmartFlex25, SAES Group) subjected to LTSS treatments up to maximum temperatures $T_M$ increasing from 120 °C to 320 °C [4] are shown in Fig. 30. The material state of the NiTi heated up to different maximum temperatures was adjusted by plastic deformation and recovery. For the purpose of the modelling, the microstructural states were parametrized by the plastic deformation occurring during the LTSS processes.

The results in Fig. 30 show that plastic deformation by kwinking during the LTSS treatment has two basic impacts on the functional response: i) it introduces irreversible plastic deformation but ii) it also changes the hysteretic strain-temperature actuation response. With increasing maximum temperature $T_M$, irreversible plastic deformation increases (the wire becomes longer). Simultaneously, the hysteresis and transformation strains exhibited by the NiTi wires heated up to intermediate temperatures are reduced and transformation ranges are broadened. However, the evolution of the functional response with increasing temperature (plastic deformation) is not monotonous. By heating above 170 °C the plastic deformation increases, but the functional response approaches the initial response of the virgin austenite. This non-monotonic evolution was successfully reconstructed by simulation using the constitutive model in [88]. It is thus conceivable that better understanding and modelling the kwinking deformation during the shape setting treatment can be used to improve shape setting procedures applied to NiTi components for medical devices and actuators. For example, special heat treatments can be designed to set the working conditions of antagonistic actuators, where the correct setting of the initial "slackness" is critical for the correct setting of the working stress, or to extend the temperature interval of the actuator for its easier control.

**Different Mechanisms of Plastic Deformation of Austenite and Martensite**

As explained in Chapter 5, polycrystalline NiTi alloys deform plastically at low temperatures only in the martensitic phase. Incremental plastic deformation of martensite occurs during the forward as well as reverse MT proceeding under external stress above certain thresholds [11,13,14]. Kwinking deformation tends to localize plastic deformation in the microstructure (kwink bands) and thus generates strong deformation heterogeneity. As discussed in Chapter 9, plastic deformation of martensite by kwinking is activated when the applied stress exceeds the kwinking stress or when the forward and reverse MT occurring under high external stress. In the absence of MT, the B2 austenite deforms plastically by dislocation slip only at very high temperatures above 300 °C (Fig. 8a,d,e,h) or at lower temperatures 100-300°C when it deforms alongside the MT as a part of the TRIP-like deformation mechanism (Fig. 8a,c,e,g).

So far, constitutive models of NiTi often consider plastic deformation of austenite, neglecting the obvious fact that the alloy is in martensite state when it deforms plastically. Since the mechanisms by which NiTi deforms plastically in martensite are significantly different from dislocation slip in austenite at elevated temperatures, the corresponding yield strengths and temperature dependencies of yield stress differ significantly (Fig. 8).

Plastic deformation of the B2 austenite is usually assumed to conform to classical plasticity theories (von Mises criterion) [83]. Plastic deformation of the highly anisotropic B19' martensite is very different. We assume that the martensite starts to deform plastically by kwinking when the distortion energy density reaches a critical value. However, since the experiments show a significantly different character from von Mises plasticity [84] and we did not perform yet dedicated experiments allowing for evaluation of the yield surface, we cannot offer more detailed information. In any case, independent descriptions of the plasticity mechanisms activated in the austenite and highly anisotropic martensite phases is necessary.

**Simulation of Functional Fatigue Related to Plastic Deformation**

Incremental plastic deformation in stress-induced martensite taking place during the forward as well as reverse MT proceeding under external stress above certain thresholds [11,13,14] (by dislocation slip not kwinking) affects the entire functional stress-strain-temperature response of NiTi in closed-loop thermomechanical load cycles. By introducing incremental plastic strains and related deformation heterogeneity into the virgin austenitic microstructure repeatedly during the thermomechanical cycling, the incremental plastic strains accumulate, transformation interval in temperature/stress widens and recoverable strain and hysteresis width decreases [13].

These phenomena, known as functional fatigue, originate from the plastic deformation of martensite by [100](001) dislocation slip accompanying the forward and reverse MT proceeding under stress [13]. Some of the constitutive models that capture the coupled MT and plastic deformation can simulate the functional fatigue [85,86,78,79,87]. Although these models partially capture the experimentally observed phenomena, such as an accumulation of plastic strains, a reduction in the hysteresis, an increase in the hardening during the cyclic loading, they suffer from the lack of understanding of the mechanism by which plastic strains are generated by the forward and reverse MT proceeding under stress. These models cannot capture plastic deformation by kwinking or the TRIP-like deformation at elevated temperatures.

**Simulation of TRIP-like Deformation at Elevated Temperatures**

Experiments on NiTi wires deformed in tension in wide temperature range show that, at temperatures above 100 °C, at which the kwinking stress becomes lower than the critical stress for stress induced MT (Fig. 8b,f), kwinking deformation becomes gradually suppressed with increasing test temperature [9,10]. At these temperatures, kwinking stress becomes lower than the yield stress for martensitic transformation (Fig. 8a,e), which is a nonsense. We assume that the martensite that is stress induced at high temperatures (100 °C - 400 °C) immediately undergoes plastic deformation which strengthens it, which gives rise to the microstructure consisting of mixture of austenite and martensite phases that deforms plastically via a TRIP-like deformation mechanism [9,10].

The question arises whether this additional inelastic deformation associated with the martensitic transformation should be included in the thermodynamic criteria for the onset of martensitic transformation. Analysis of the experimental results [88] suggests that the total deformation of martensite affects the resulting thermoelastic coupling. At elevated temperatures, the thermoelastic coupling coefficient decreases with a corresponding increase in martensite deformation (transformation + plasticity), and the onset of martensitic transformation is achieved at lower stresses than would be expected from thermo-mechanical coupling at lower temperatures. The forward martensitic transformation and the plasticity of martensite are in this case a fully coupled mechanism occurring simultaneously. Capturing this process with current constitutive models can be problematic: neither the activation condition for the phase transition nor the martensitic plastic deformation are achieved independently in this case, and the description of the coupled transformation and plasticity of martensite would require the introduction of an independent TRIP-like mechanism into the constitutive modeling.

An alternative approach [88] is to reformulate the constitutive model within the framework of generalized standard materials with internal constraints [89] and perform the optimization of the energy and dissipation functions simultaneously with respect to the internal variables describing both the transformation and plastic processes. It can be shown that optimization on a non-smooth energy surface (due to the rate-independent form of the dissipation function and the constraints on the internal variables) can lead to different results when performed independently on each internal variable or simultaneously (even in the strictly convex case). This is a mathematical representation of the possibility of two processes occurring simultaneously that would otherwise not be permissible separately.

**Simulation of Low Temperature Shape Setting of Annealed NiTi**

The experiment shows that plastic deformation of martensite plays key role in the reverse MT when it proceeds under external stress (Fig. 28). When MT proceeds under stress free conditions, reverse transformation temperatures are shifted upwards by plastic deformation, though not dramatically. This was implemented in constitutive modelling by further increase in dissipation [85, 90] additional to the stabilization occurring without any plastic deformation [47]. Although the applied stress increases $A_f$ temperature, it cannot prevent the reverse transformation of oriented martensite into austenite. It is less known, that this is enabled by the kwinking deformation activated by the high stress upon heating (Figs. 22,29). In other words, the temperature (stress) up to which the heated oriented martensite can survive in a constrained NiTi wire depends on the magnitude of the kwinking stress, which is a material parameter of the tested wire. When the kwinking stress is reached upon the constrained heating, the martensite transforms into austenite. Since the plastic deformation of martensite by kwinking occurs at the microstructural level in a highly heterogeneous manner (creates kwink bands and kwink interfaces), reverse MT proceeds into austenite containing deformation bands with rotated crystal lattice separated from the parent austenite lattice by STGB interfaces.

The consideration of the generation of plastic deformation by dislocation slip and kwinking during the LTSS process in modeling of NiTi is a very good example. When the constrained NiTi wire containing oriented martensite is heated in the LTSS experiment (see Fig. 29 and related text), an intriguing process of coupled plastic deformation in martensite driven by increasing stress and reverse martensitic transformation driven by increasing temperature takes place. Within the martensite phase, plastically deformed austenite begins to nucleate, thereby converting the transformation strain of martensite into plastic strain of austenite [4]. During this process, the overall plastic strain increases at the expense of transformation strain of the disappearing oriented martensite while the length of the wire stays constant (Fig. 29). Simultaneously, the virgin austenitic microstructure gradually evolves during the heating, affecting the functional response of the shape-set component (see Figs 15-17 in Ref. [4]). The microstructure is most severely affected when the wire is heated up to intermediate temperatures (end of the LTSS range in Fig. 29b). When the material is heated to maximum temperatures, its functional response becomes similar to the response of virgin undeformed samples (Fig. 30). This evolution of microstructure can be implemented in the constitutive model as a TRIP-like deformation mechanism [88]. Interestingly, this mechanism does not affect the total inelastic deformation (it only converts the transformation deformation into plastic deformation) and is therefore strictly temperature induced.

In summary, consideration of martensite plasticity in constitutive modeling of NiTi requires an independent description of the plastic deformation mechanisms activated in the austenite and martensite. Key issue is that the forward and reverse MT in NiTi proceeding under stress above certain thresholds tends to be accompanied by plastic deformation of martensite by dislocation slip and kwinking. This leads to cyclic instability of thermomechanical responses (functional fatigue). Constitutive models capable of simulation of functional fatigue that are currently available in literature [86,87] take advantage of the fact that the incremental plastic strains are small. These models cannot capture plastic deformation by kwinking or the TRIP-like deformation in tensile tests at elevated temperatures. When external stress exceeds the kwinking stress in a thermomechanical loading test (or the kwinking deformation proceeds alongside MT at stresses lower than the kwinking stress), plastic deformation in martensite via kwinking deformation assures strain compatibility at GBs. Kwinking deformation thus enables plastic deformation of NiTi polycrystal up to large strains but simultaneously modifies the virgin austenitic microstructure (characterized in SMA model by material parameters). Therefore, development of constitutive model capturing stress-strain-temperature responses of NiTi reaching beyond the plasticity limits implies updating not only internal variables, which is common, but also basic material parameters during the simulation. Such constitutive models do not exist yet, except of few models updating selected material parameters [85,88]. As a special case, we have discussed conversion of oriented martensite into plastically deformed austenite taking place during the low temperature shape setting experiment in which the virgin austenitic microstructure of the NiTi wire gradually evolves with increasing temperature, which affects the functional response of the shape set wire.

## 11. Kwinking Deformation in NiTi Technology

It is not clear whether kwinking deformation is unique to NiTi and some NiTi-based shape memory alloys which undergo B2-B19’ MT or whether it occurs in other SMAs as well. The available experimental results suggest that the presence of the B2-B19’ MT is the necessary precondition for the martensite to deform plastically by kwinking. However, the B2-B19’ MT itself does not warrant the activation of kwinking deformation. Various combinations of kinking, stress induced MT and deformation twinning were reported by researchers investigating deformation mechanisms in anisotropic hexagonal alloys such as titanium [91], zirconium [92], high strength nano metallic laminates [93] or high entropy shape memory alloys [94].

Considering all activated deformation mechanisms in NiTi, we realize that the austenite responds to mechanical straining by forming deformation bands containing crystal lattice associated with defined finite shape strains with respect to parent austenite. This inherently involves creation of new internal interfaces at which strain compatibility is fulfilled. These processes include martensitic transformation, deformation twinning (reorientation) and kwinking. Common features are the finite shape strains and newly introduced internal interfaces. In case of stress induced martensitic transformation, martensite variants form deformation bands and mobile habit planes separate the stress induced martensite from the parent austenite. In case of twinning, twin interfaces separating martensite variants move during the martensite reorientation which modifies the shape strains. In case of kwinking, kwink interfaces separate the martensite lattices plastically deformed by coordinated dislocation slip within kwink bands from the martensite matrix. While the SIMT and twinning that involves reconfiguration of atoms within the EP neighborhood of the cubic structure give rise to recoverable strains, kwinking deformation [23] enabled by the low resistance of the B19’ martensite to [100](001) dislocation slip in martensite moves the atoms out the EP neighborhood which results in unrecoverable plastic strain. In contrast, the shape strain achieved by coordinated dislocation slip within the (20-1) kwink band requires driving the atoms temporarily outside the EP neighborhood. Although, it is seemingly equivalent to the (20-1) deformation twinning analyzed by Ezaz et al. [42] as well as to the non-transformation pathway proposed by Gao et al. [52], the fact that kwinking proceeds via coordinated dislocation slip makes a big difference in that it rationalizes the wide range of phenomena discussed in chapters 5-10.

Activation of kwinking deformation in an alloy requires fulfillment of the following four requirements:
i) occurrence of martensitic transformation (enables twinning)
ii) plastic anisotropy of martensite (enables kinking)
iii) dislocation slip in martensite proceeding at low stress (enables coordinated dislocation slip)
iv) martensite reorientation proceeding at low stress (enables kwinking)

There are not many SMAs fulfilling these four conditions, though we did not perform any systematic literature search. The requirement for martensitic transformation (i) seems to be trivial but isn't. In fact, MT is required only because it enables (100) compound twinning, which makes the kwinking different from kinking (while (20-1) kwink bands in B19' martensite display defined shape strains, kink bands in Mg display any general shears). The requirement for plastic anisotropy of martensite (ii) is fulfilled in low symmetry monoclinic martensite but not in other SMAs transforming into orthorhombic or tetragonal martensite. The requirement for dislocation slip in martensite proceeding at low stresses is even more restrictive - majority of SMAs are highly resistant to dislocation slip in martensite. As proposed in Chapter 5, the resistance to dislocation slip in martensite (kwinking stress) depends on the elastic constants and lattice parameters of the B19' martensite. It can be further modified by varying chemical composition or via ternary alloying both of which affect lattice parameters and elastic constants. Superelastic NiTi wires are more resistant to dislocation slip in martensite than SME NiTi wires having slightly less Ni content [13]. Further increase of Ni content beyond 50.8 at. % results in further compositional strengthening [95]. While $NiTi_3Fe$ (at. %) alloys [96] are extremely prone to [100](001) dislocation slip in martensite, NiTiCu alloys [97] seem to be highly resistant to it. Further research is needed to investigate the effect of lattice parameters and elastic constants on the resistance of martensite in various SMAs to dislocation slip.

As concerns the requirement for low martensite reorientation stress (iv), this is also related to critical stress for dislocation slip in martensite controlled by the size effect (Hall-Petch strengthening). Higher stress is required to initiate dislocation slip in small CVP domains within grains than within laminates filling whole grains. The binary superelastic and shape memory NiTi wires undergo martensite reorientation in tensile test at low stresses 100-300 MPa, depending on the test temperature (Fig. 8a,e). Many SMA alloys, however, require much higher stress to reorient the martensite. These alloys display martensite variant microstructures consisting of multiple nanoscale CVP domains within grains in which the [100](001) dislocation slip is suppressed. Specifically, NiTiHf alloys transforming at high temperatures [98] display actuation under low stresses but undergo martensite reorientation and plastic deformation only under very high stresses. Researchers wondered why intermartensitic interfaces in NiTiHf do not move [99] when these alloys are mechanically loaded. As in the case of kwinking stress, lattice parameters, and elastic constants of the B19' martensite probably play significant role again and the size of the CVP domains [100] matters. Nevertheless, exact mechanism which controls the critical stress for martensite reorientation and its temperature dependence in NiTi based alloys [47] remains unclear. Why some alloys undergoing B2-B19' MT display kwinking and other alloys do not, remains to be investigated.

The key role of [100](001) dislocation slip in martensite in NiTi behaviors (Chapter 4) shall be kept in mind when developing NiTi based shape memory alloys. On one hand, [100](001) dislocation slip accompanying the forward and reverse MT proceeding under stress above certain thresholds facilitates strain compatibility at GBs which enables large recoverable strains, on the other hand, it gives rise to functional fatigue and affects the magnitude of kwinking stress. Small incremental plastic strains generated by MT proceeding under stress shall be tolerated but kwinking at low stresses shall be prevented. From another point of view, however, the coordinated [100](001) dislocation slip mediating the kwinking deformation [8] enables the excellent plastic deformability of NiTi.

The kwinking stress scales with functional fatigue [75]. The higher the kwinking stress is, the more stable is the stress-strain-temperature behavior in cyclic thermomechanical loads (functional fatigue). This can be understood by considering that both plastic deformation by kwinking and generation of incremental plastic strains during cyclic thermomechanical loads involve [100](001) dislocation slip in martensite. Because of this, the [100](001) dislocation slip in martensite must be carefully considered when designing NiTi alloys for cyclic loading applications [101], particularly to the coarse grain NiTi based alloys fabricated by novel adaptive manufacturing methods [102,103] that must be strengthened by other means than by decreasing the grain size. Adjusting chemical composition, ternary alloying or introducing chemical heterogeneities [104] are examples of such strengthening approaches. Since the B19' martensite is extremely plastically anisotropic (Fig. 5), textures of austenite and stress induced martensite (Figs. 23,24) is another way of strengthening that was found to play very important role in functional fatigue as well as in plastic deformation by kwinking.

In summary, polycrystalline NiTi can be strengthened against the [100](001) dislocation slip in martensite by:
i) adjusting chemical composition
ii) ternary alloying
iii) decreasing grain size
iv) introducing nanoscale chemical heterogeneity (or suitable coherent precipitates)
v) introducing permanent lattice defects
vi) introducing internal stress via thermomechanical training
vii) introducing texture into virgin austenite

The incremental plastic deformation via [100](001) dislocation slip in martensite accompanying MT proceeding under stress causes functional fatigue but prevents intergranular fractures and improves structural fatigue performance of NiTi [105,106]. Strengthening NiTi against the [100](001) dislocation slip in martensite suppresses functional fatigue and increases kwinking stress (broadens stress-temperature window in which NiTi displays functional properties but decreases ductility (promotes tensile instability via necking occurring when stress reaches the kwiniking stress) and degrades structural fatigue performance. Therefore, NiTi alloys must be strengthened against the [100](001) dislocation slip in martensite wisely based on the understanding of the mechanisms of functional fatigue of NiTi [13] and kwinking deformation [23].

## 12. Conclusions

Besides the functional thermomechanical properties, NiTi SMAs exhibit excellent plastic deformability in the martensite state due to the recently discovered kwinking deformation mechanism. Based on the understanding of kwinking deformation, we have overviewed and reinterpreted results of earlier studies providing experimental evidence on phenomena that originate from the kwinking deformation.

1. Plastic deformation of the monoclinic B19' martensite proceeds by kwinking deformation involving dislocation slip-based kinking combined with (100) deformation twinning. Kwinking deformation can be understood as being a result of interactions between dislocation slip and twinning over a broad range of spatial scales. Macroscopically, it relaxes the external loading through pattern formation utilizing massive, coordinated dislocation slip and nanoscopically, it reduces energy by creating coherent intermartensitic interfaces and relaxing the energy of the dislocation cores.
2. The extreme plastic anisotropy of B19' martensite in NiTi enabling the kwinking deformation originates from the low resistance of the B19' martensite to shear deformation on (001) plane in [100] direction. The shear deformation takes place either via [100](001) dislocation slip and/or via (001) compound twinning. The latter was proposed to proceed via stress driven motion of internal interfaces taking place via glide of partial dislocations on (001) plane.
3. The yield stress for plastic deformation of NiTi by kwinking (kwinking stress) is the maximum stress that can be applied in thermomechanical loads without the danger of generating large plastic strains in the case of soft annealed NiTi wires or rupture in the case of strengthened NiTi wires. Kwinking stress depends on temperature through its dependence on critical stress for [100](001) dislocation slip in martensite which depends on lattice parameters and elastic constants of the B19' martensite that are temperature dependent. Kwinking stress is the key material characteristics of NiTi that depends on the: i) chemical composition, ii) ternary alloying, iii) grain size, iv) nanoscale chemical heterogeneity, v) coherent $Ti_3Ni_4$ precipitates and vi) permanent lattice defects in the austenite microstructure.
4. The results of the reconstruction of martensite variant microstructures in grains and analyses of permanent lattice defects in plastically deformed NiTi by TEM prove that they originate from the plastic deformation of martensite by kwinking and not from alternative deformation mechanisms reported earlier. The martensite variants (deformation bands in austenite) in grains appear in special deformation geometries. These special deformation geometries and newly introduced interfaces (kwink interfaces in martensite and STGBs in austenite) could be created neither by martensitic transformation (or deformation twinning) nor by conventional dislocation slip. Their observation in microstructure of deformed NiTi serves as indicator that the alloy deformed plastically via kwinking.

5. Although plastic deformation of NiTi wires by kwinking tends to proceed homogeneously with strain hardening, it may also proceed in macroscopically localized manner (via necking, mobile Lüders bands, or shear bands). Whether the macroscopic deformation will be homogeneous or localized, depends on whether the magnitude of kwinking stress and strain hardening rate fulfill the Considère criterion for stability of tensile deformation.
6. The plastic deformation of NiTi wires by kwinking gives rise to characteristic evolution of martensite texture in tensile tests until rupture that cannot be explained by conventional dislocation slip. At the same time, the kwinking deformation leads to the refinement of austenitic microstructure to quasi-amorphous state. Taking advantage of that, NiTi materials with unique microstructures and properties can be prepared via carefully performed heat treatments of plastically deformed NiTi.
7. Kwinking deformation is responsible for unusual phenomena observed in tensile thermomechanical loading tests on NiTi wires, particularly to the long superelastic plateaus observed at specific temperatures, unusually large plastic strains generated upon thermal cycling under high stress and low temperature shape setting (heating of constrained NiTi). In these experiments, kwinking deformation was activated under stresses that are lower than the kwinking stress because kwinking deformation proceeded alongside MT.
8. The key implication of the results reported in this work for constitutive modelling of NiTi is that this alloy deforms plastically in martensite not in austenite when it is subjected to thermomechanical loading at low temperatures below ~100 °C. Since the mechanisms of plastic deformation of austenite and martensite are fundamentally different, this is very important. The plastic deformation of martensite by kwinking enables plastic forming without cracking (tensile deformation at ~1GPa stress up to ~ 80% strain). Simultaneously, it modifies (refines) the virgin austenitic microstructure (characterized in SMA models by material parameters). Therefore, constitutive models aspiring to capture stress-strain-temperature responses of NiTi reaching beyond the elasticity limits must consider the characteristics of the kwinking deformation (at least temperature dependent kwinking stress and strain hardening rate). Whenever plastic deformation occurs, the models must update not only internal variables, but also basic material parameters (transformation temperatures, Young's modulus, transformation strain, etc.). Constitutive models should be developed to cover the phenomena related to kwinking coupled with the martensitic transformation.
9. Whether kwinking deformation is unique to NiTi or whether it occurs also in other SMAs, is not clear yet. There are conditions that a metallic alloy must fulfill to deform plastically by kwinking. Among these conditions, the temperature dependent shear instability of the monoclinic B19' structure (easy shear on (001) plane in [100] direction giving rise to the low resistance to the [100](001) dislocation slip and (001) compound twinning) and temperature dependent lattice parameters and elastic constants of martensite are most important. They control functional fatigue as well as kwinking stress. Further research focusing on approaches towards controlling the shear instability, lattice parameters and elastic constants of martensite by adjusting chemical composition, or ternary alloying shall be undertaken while developing advanced NiTi-based SMAs.

## Data availability

Data from thermomechanical loading tests are available at https://doi.org/10.5281/zenodo.19730395 and other data will be made available on request.

## Acknowledgements

Support from Czech Science Foundation (CSF) project 25-16285S is acknowledged. P. Šittner acknowledges support from Czech Academy of Sciences through Praemium Academiae. MEYS of the Czech Republic is acknowledged for the support of infrastructure projects, CNL (CzechNanoLab LM2023051) and FerrMion (CZ.02.01.01/00/22_008/0004591).